\documentclass[traditabstract]{aa} 
\usepackage{graphicx}
\usepackage{subfig}
\usepackage{txfonts}
\usepackage{natbib}
\bibpunct{(}{)}{;}{a}{}{,} 
\usepackage{enumitem}
\usepackage{multirow}

\begin{document}
   \title{No Evidence for Significant Age Spreads in Young Massive LMC Clusters\thanks{Based on observations made with the NASA/ESA Hubble Space Telescope, and obtained from the Hubble Legacy Archive, which is a collaboration between the Space Telescope Science Institute (STScI/NASA), the Space Telescope European Coordinating Facility (ST-ECF/ESA) and the Canadian Astronomy Data Centre (CADC/NRC/CSA).}}

   \author{F. Niederhofer 
   			\inst{1,2}
   			\and
   			M. Hilker
   			\inst{1}
   			\and
   			N. Bastian
   			\inst{3}
   			\and
   			E. Silva-Villa
   			\inst{4,5}
            }

   \institute{European Southern Observatory, Karl-Schwarzschild-Stra\ss e 2, D-85748 Garching bei M\"unchen, Germany\\
              \email{fniederh@eso.org}
          \and
Universit\"ats-Sternwarte M\"unchen, Scheinerstra\ss e 1, D-81679 M\"unchen, Germany
          \and
           Astrophysics Research Institute, Liverpool John Moores University, 146 Brownlow Hill, Liverpool L3 5RF, UK
           \and
           Centre de Recherche en Astrophysique du Qu{\'e}bec (CRAQ) Universit{\'e} Laval, 1045 Avenue de la M{\'e}decine, G1V 0A6 Qu{\'e}bec, Canada
           \and
FACom-Instituto de Fsica-FCEN, Universidad de Antioquia, Calle 70 No. 52-21, Medell{\'i}n, Colombia
              }


  \abstract
{Recent discoveries have put the picture of stellar clusters being simple stellar populations into question. In particular, the color-magnitude diagrams of intermediate age (1-2 Gyr) massive clusters in the Large Magellanic Cloud (LMC) show features that could be interpreted as age spreads of 100-500 Myr. If multiple generations of stars are present in these clusters then, as a consequence, young ($<$1 Gyr) clusters with similar properties should have age spreads of the same order. In this paper we use archival \textit{Hubble Space Telescope} (HST) data of eight young massive LMC clusters (NGC~1831, NGC~1847, NGC~1850, NGC~2004, NGC~2100, NGC~2136, NGC~2157 and NGC~2249) to test this hypothesis. We analyzed the color-magnitude diagrams of these clusters and fitted their star formation history to derive upper limits of potential age spreads. 
We find that none of the clusters analyzed in this work shows evidence for an extended star formation history that would be consistent with the age spreads proposed for intermediate age LMC clusters.
Tests with artificial single age clusters show that the fitted age dispersion of the youngest clusters is consistent with spreads that are purely induced by photometric errors. As an additional result we determined a new age of NGC 1850 of $\sim$100~Myr, significantly higher than the commonly used value of about 30~Myr, although consistent with early HST estimates.

}

   \keywords{galaxies: star clusters: general - galaxies: individual: LMC - Hertzsprung-Russel and C-M diagrams - stars: evolution}
\titlerunning{LMC Clusters}
   \maketitle
%
\section{Introduction}
Globular clusters (GCs) have long been thought to be simple stellar populations (SSP), as their stars formed out of the same molecular cloud at approximately the same time. Therefore all stars in a GC should have (within a small range) the same age and the same chemical composition. However, recent discoveries have put this simple picture into question. Features like extended main sequence turn-offs (MSTOs), double main sequences (MS) and multiple subgiant and giant branches in the color-magnitude diagrams (CMDs) of GCs as well as abundance variations of light elements (e.g. C, N, O, Na, Al) have been found (e.g \citealt{Gratton12, Piotto12}). There are several attempts to explain the observed anomalies in GCs. 
The most common model implies the presence of multiple generations of stars inside GCs. In this model, the second generation of stars is thought to be formed out of enriched material from the first stellar generation. Proposed sources of the ejecta are AGB stars \citep{D'Ercole08}, fast rotating stars \citep{Decressin09} and interacting binaries \citep{deMink09}. A large drawback of this model, however, is that in order to produce the observed amount of second generation stars the cluster must have been 10 -100 times more massive in the past than observed today \citep{Conroy12}. 
An alternative scenario has been proposed by \citet{Bastian13a} where chemically enriched material from rotating stars and massive interacting binaries falls on to the accretion disks of low-mass pre-MS stars of the same generation and is finally accreted on the still forming stars. In this scenario all stars are from the same generation without significant age spreads among them.

Besides the anomalous features observed in old GCs which point towards a more complex scenario than a single stellar population, intermediate age (1-2 Gyr) clusters in the LMC show extended or double MSTOs (e.g. \citealt{MackeyNielsen07, Mackey08, Milone09, Goudfrooij09, Goudfrooij11a, Goudfrooij11b}). This might be a consequence of an extended period of star formation (100 - 500 Myr). Other theories relate the extended MSTOs to interacting binaries (e.g. \citealt{Yang11}) or stellar rotation (e.g. \citealt{BastianDeMink09, Yang13, Li14}). The centrifugal force in rotating stars decreases the effective gravity which causes a lower effective temperature and luminosity (e.g. \citealt{Meynet97}). However, \citet{Platais12} analyzed the CMD of the galactic open cluster Trumpler 20 and did not find evidence that rotation affects the CMD. Possibly, extended or multiple MSTOs in intermediate age clusters are associated with the presence of multiple stellar populations observed in old GCs \citep{Conroy11}. 

However, recent studies suggest that intermediate age LMC clusters, in contrast to galactic GCs, do not have spreads in chemical abundances. \citet{Mucciarelli08} analyzed the extended MSTO clusters NGC 1651, NGC 1783 and NGC 2173, and also the older cluster NGC 1978 and did not find chemical anomalies. Furthermore, NGC 1866 \citep{Mucciarelli11}, NGC 1806 \citep{Mucciarelli14} and NGC 1846 (Mackey et al. in prep.) also do not have abundance spreads. This calls into question the idea that the extended MSTO feature is due to age spreads, as such scenarios would naturally predict abundance spreads through self-pollution.
Thus, an understanding of the properties of intermediate age clusters can give valuable insight in the evolution of GCs.

If it is true that intermediate age clusters host multiple stellar generations then young ($<$1 Gyr) clusters with similar properties should display signs of age spreads inside the cluster as well. There are several studies that search for age spreads or ongoing star formation in young massive clusters (YMCs). \citet{Cabrera-Ziri14} analyzed the spectrum of a YMC in the merger galaxy NGC 34 and concluded that the star formation history (SFH) is consistent with a single stellar population. \citet{Bastian13b} presented a catalog containing more than 100 galactic and extragalactic young ($<$100 Myr) massive clusters and did not find any evidence of ongoing star formation in their sample. 

\citet{BastianSilva13} started a project to constrain possible age spreads in young massive LMC clusters. They analyzed the CMDs of the two young clusters NGC 1856 (281~Myr) and NGC 1866 (177~Myr) and fitted SFHs to the clusters' CMDs. They found no age spreads in these two clusters and concluded that both are consistent with a single burst of star formation that lasted less than 35 Myr. In this work, we continue the study by \citet{BastianSilva13} by searching for potential age spreads in eight more YMCs in the LMC.

\begin{figure}
\includegraphics[width=8cm]{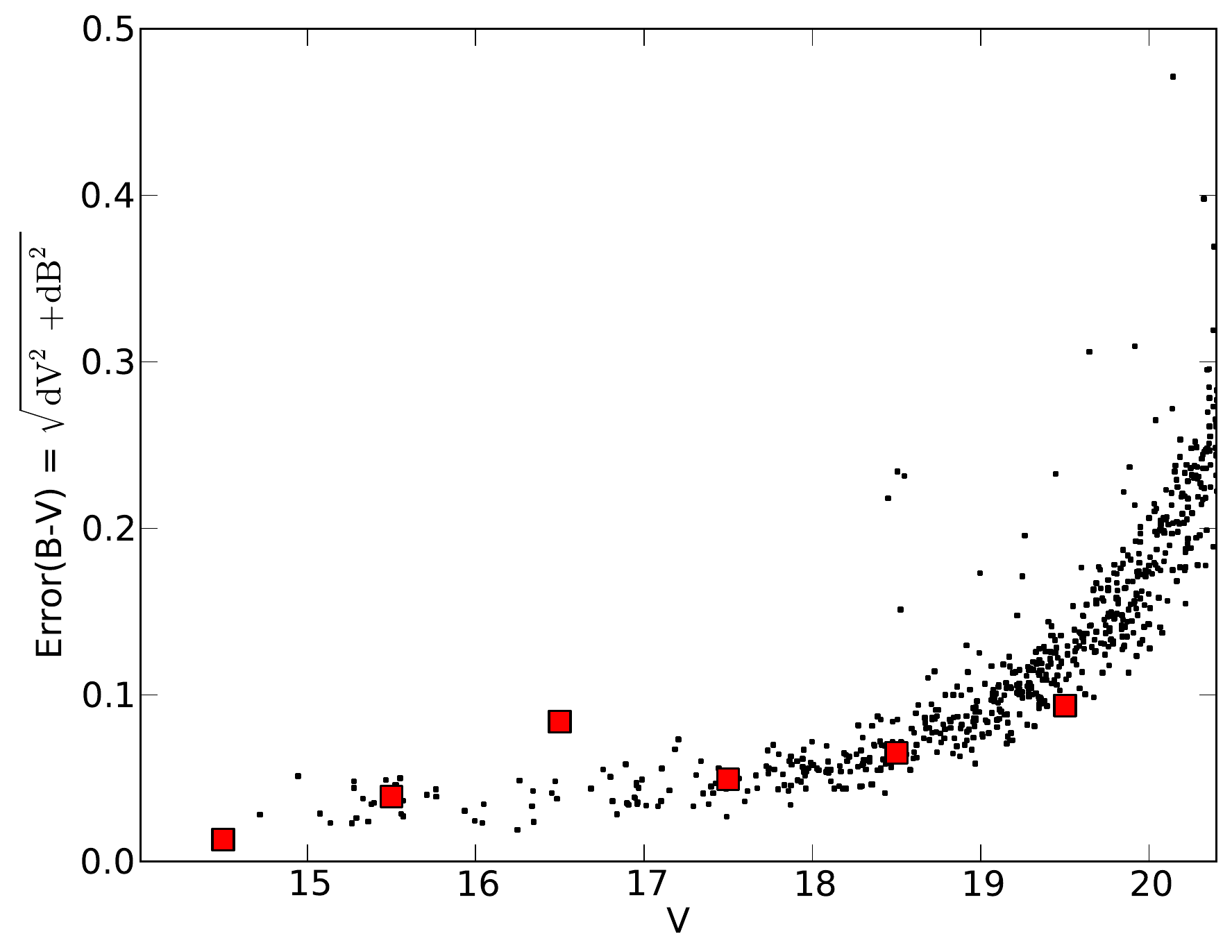}
   \caption{Color errors ($\mathrm{\sqrt{dB^2+dV^2}}$) of NGC 2136 as a function of the $V$-band magnitude . The filled (red) squares indicate the dispersion width of the main sequence.}
              \label{fig:ngc2136_error_vs_mag}
\end{figure}

We studied the CMDs of the clusters and fitted theoretical isochrones to them to determine their age, metallicity, extinction and distance modulus (DM). With these parameters we created model Hess diagrams of different ages and compared them with the observations to put initial constraints on any age spread that might be present within the clusters. In order to provide more quantitative results, we also fitted the star formation history of each cluster using the code \textit{FITSFH} (\citealt{Silva-Villa10} and \citealt{Larsen11}). This code creates a theoretical Hess diagram taking into account the photometric errors and some assumptions (e.g. metallicity). 
It searches for the best match between the data and a theoretical model which is a linear combination of Hess diagrams constructed from different isochrones. 
We do not take into account binaries and differential reddening (except for NGC2100 which has a considerable variation of extinction across the cluster) in the fitting of the SFH. Their effects on the CMDs will only increase any potential age spread that might be present in the clusters. Additionally, the archival photometric data tables that we use for this work contain only the standard deviation of the data reduction as photometric errors. Figure \ref{fig:ngc2136_error_vs_mag} shows, as an example, the errors in $B-V$ color of NGC 2136 as a function of the $V$-band magnitude. The filled squares (red) indicate the width of the MS, given by the standard deviation. The width of the MS is consistent with the overall trend of the errors, especially in the region 17 $\leq V\mathrm{[mag]} \leq$ 19 that is not sensitive to age. The real scatter of the errors around the global relation might be larger as the photometric standard deviations underestimate the real errors. However, underestimating the photometric errors will also lead to larger estimated age spreads. Therefore, all age spreads presented in this work are upper limits of real ones in the clusters. The aim of this work is not to quantify the exact extent of the SFH, but rather to test if significant (tens to hundreds of Myr) age spreads are present within the clusters. For all models and fittings in this paper we used the Parsec 1.1 isochrone set of the Padova isochrones \citep{Bressan12} and we assumed a \citet{Salpeter55} IMF.

The structure of the paper is the following:
Section \ref{sec:obs} describes the used data set and the further processing of the data. We perform tests with artificial clusters in Section \ref{sec:art_cluster_test}. The results of the fitting of the SFH are given in Section \ref{sec:res}. We discuss the results and draw final conclusions in Section \ref{sec:disc}.

\section{Observations and Data Processing\label{sec:obs}}
For our analysis we made use of data taken with the Wide Field and Planetary Camera 2 (WFPC2) on board of the \textit{Hubble Space Telescope} (HST). The data set is presented in detail in \citet{Brocato01} and \citet{Fischer98}. NGC 1850, however, is not part of these studies, we retrieved the data tables of this cluster from the HST Legacy Archive (HST Proposal 5475, PI: M. Shara). NGC 2157 is presented in \citet{Fischer98} whereas the \citet{Brocato01} data set covers the other six clusters. The images of the clusters were taken in the F450W and F555W filters (\citealt{Brocato01} data set and NGC 1850) and in the F555W and F814W filters \citep{Fischer98}. All the magnitudes in the HST filter system were transformed to standard Johnson \textit{BVI} magnitudes. We use the fully reduced data tables which consist of the stars' $x$ and $y$ pixel coordinates on the detector system, the magnitudes and their standard deviations in two filters each.

\begin{figure}
\includegraphics[width=8cm]{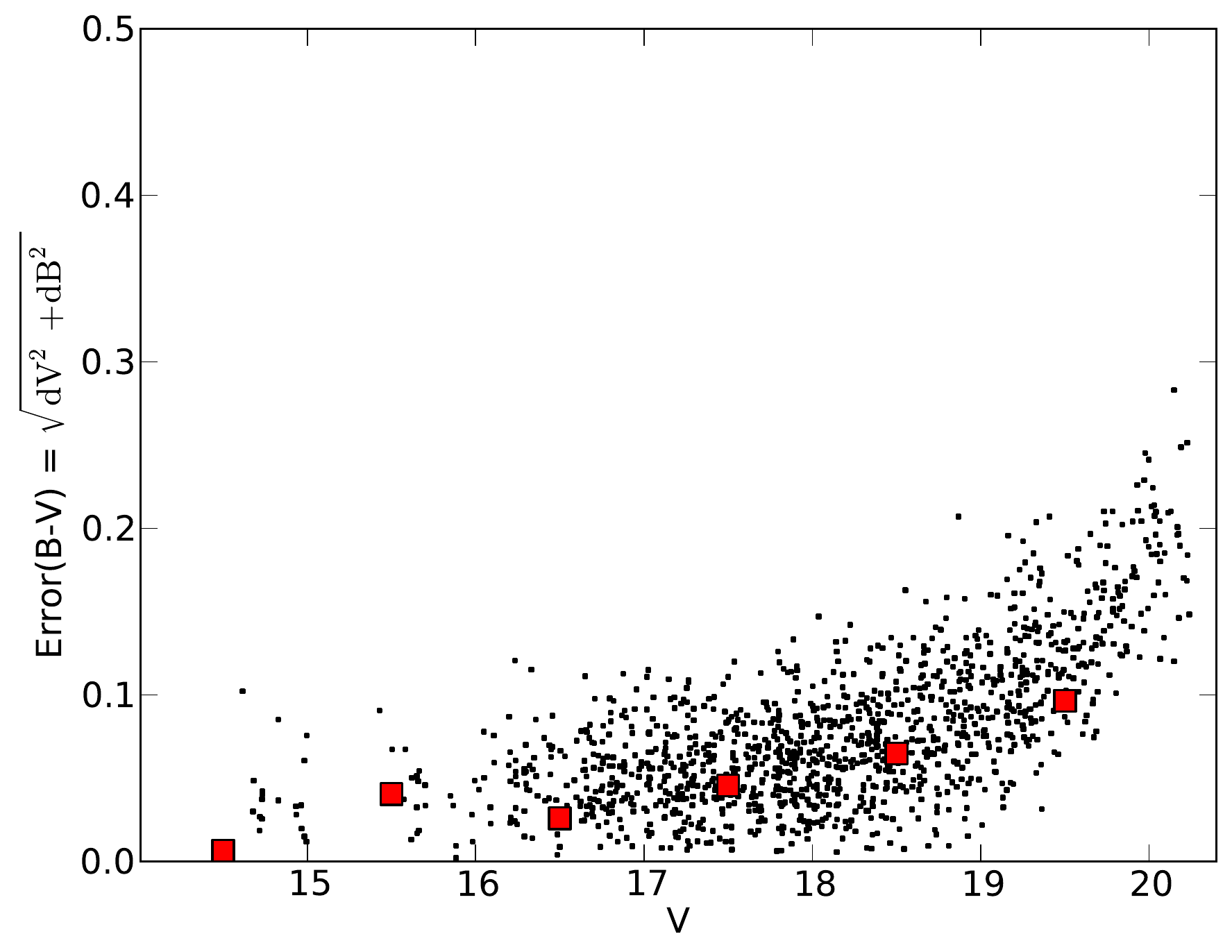}
   \caption{Modeled color errors ($\mathrm{\sqrt{dB^2+dV^2}}$) of NGC 1850 as a function of the $V$-band magnitude. The filled (red) squares indicate the observed dispersion width of the main sequence.}
              \label{fig:ngc1850_error_vs_mag}
\end{figure}

The HLA catalog of NGC 1850 does not include photometric errors. So we created artificial errors that were modeled from the \citet{Brocato01} data set. We took these clusters from the data set that have the same exposure time (40s)  as NGC 1850 as a reference and we simulated the errors such that they follow the same exponential increase with fainter magnitudes as the errors from the reference clusters. Figure \ref{fig:ngc1850_error_vs_mag} shows the artificial errors in $B-V$ color as a function of the $V$-band magnitude. The filled (red) squares indicate the width of the MS (standard deviation) of NGC 1850. It is comparable with the mean error in the interval 17 $\leq V \mathrm{[mag]} \leq$ 19 where the MS is not affected by age effects. We are aware that the scatter of the modeled errors could also be underestimated as the errors were modeled from the standard deviation of the photometric reduction of other clusters (see previous Section).

Additionally to the photometry of NGC 2157, \citet{Fischer98} provide data of a background field which is located 26\arcsec east and 110\arcsec north of the cluster for subtraction of field stars. For our analysis we restrict ourselves to the inner regions of the clusters. 

As the clusters in the \citet{Brocato01} data set and NGC 1850 are not centered on any of the four WFPC2 detector chips we had to determine their cluster centers in the $x$,$y$ CCD coordinate system. We did this by first creating artificial blurred images of the clusters by convolving the flux weighted spatial positions of the stars in each cluster with a Gaussian. We chose Gaussians with a $\sigma$ ranging between 40 and 60 pixels (changing from cluster to cluster) in order to get a smooth flux distribution in the center of each cluster. Afterwards we fitted elliptical isophotes to the created images using the IRAF\footnote{IRAF is distributed by the National Optical Astronomy Observatories, which is operated by the Association of Universities for Research in Astronomy, Inc., under cooperative agreement with the national Science Foundation.} task \texttt{ellipse} and used the center of the innermost isophotes as the cluster center. Only those stars which are located inside two times the core radius $R_{\rm core}$ (radius at which the density is half the central density) given by \citet{Brocato01} are used for further analysis. $R_{\rm core}$ is between 2.0 and $\sim$4 pc (see Table \ref{tab:Cluster_Param}), at an assumed distance of 50 kpc \citep{deGrijs14}. NGC 2157, however, is approximately centered on the PC chip of the WFPC2. The stellar density is highest on this chip with a steep fall off towards the other three chips. For this cluster we use just the stars located on the PC chip which has 800 $\times$ 800 pixels with a pixel scale of 0$\farcs$046 per pixel. Assuming a distance of 50 kpc to the LMC the PC chip covers an area of 8.9 $\times$ 8.9 pc. This area is comparable with the area given by the projected half-light radius of NGC 2157 (5.4 pc, \citealt{McLaughlin05} assuming \citealt{King66} models).  

\begin{figure}
   \resizebox{\hsize}{!}{\includegraphics{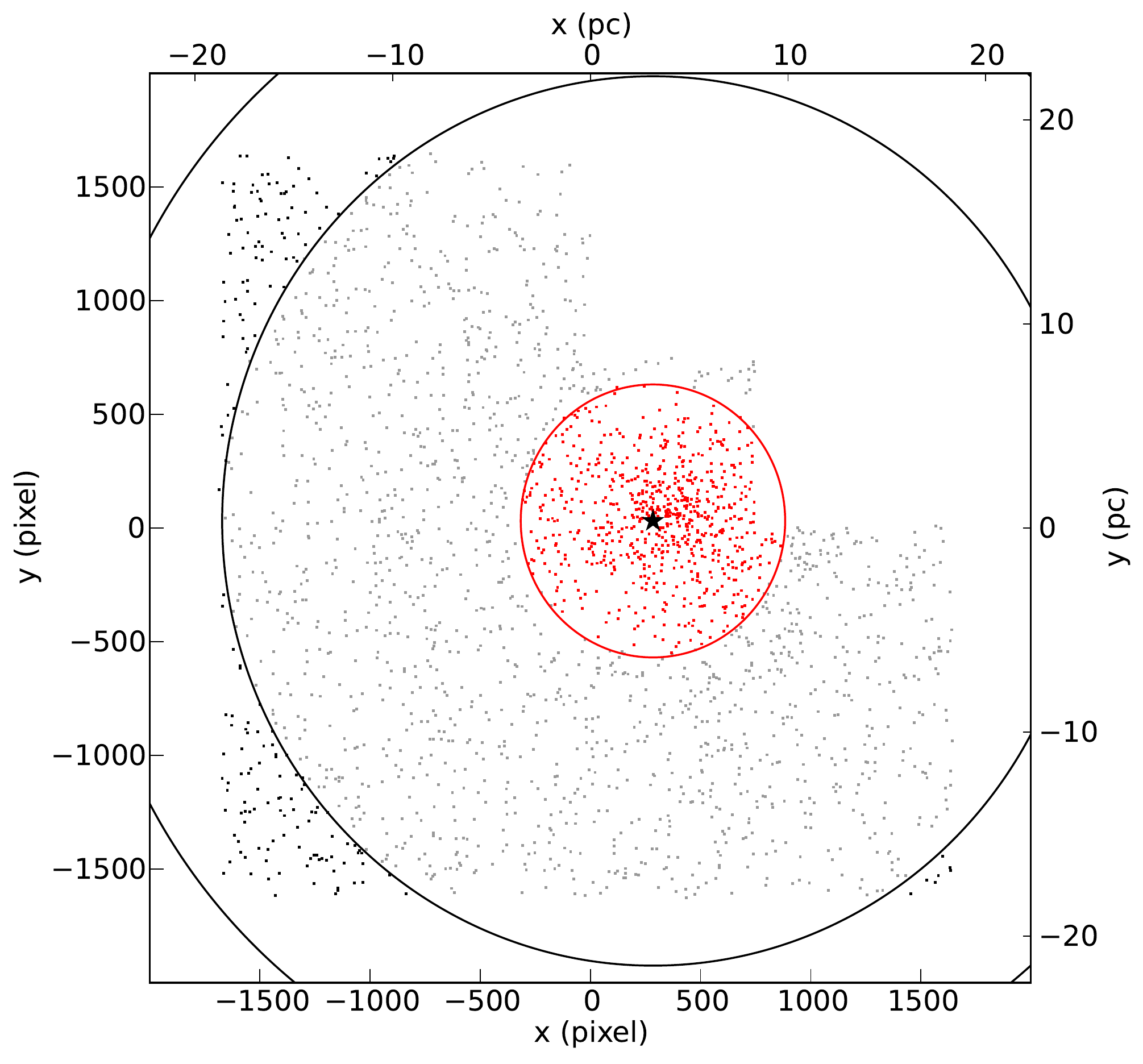}}
   \caption{NGC 1847; Positions of the stars on the WFPC2 chip. The inner circle (red) corresponds to two times the core radius of the cluster. The intersection of the outer two circles forming an annulus with the corners of the WFPC2 field of view is the area which is used for field star removal. The asterisk symbol marks the center of the cluster as determined by us.}
              \label{fig:ngc1847_xy}
\end{figure}

For our purpose we need to further analyze the data sets. The first step is the subtraction of the field star contamination which is a combination of Galactic foreground stars and LMC field stars. For all the clusters, except for NGC 2157, we do not have extra field exposures so we have to use the cluster images themselves for the field star subtraction. To minimize the contribution of cluster stars we constructed areas that are as far away from the center of the cluster as possible. We chose annuli centered on the respective cluster that intersect the corners of the chips such that the intersection area of the annulus and the chip is approximately the same as the area where we want to subtract the contaminating stellar population (see Figure \ref{fig:ngc1847_xy} as an example). After defining the areas we created CMDs for the stars in both regions and subtracted for every star in the field CMD the star in the corresponding cluster CMD that has the closest geometrical distance in color-magnitude space. However, we have to bear in mind that also in this outer regions a certain fraction of cluster stars is still present that is subtracted as field stars. This becomes clear if we compare the field of view of the WFPC2 detector with the tidal radii of the clusters. The tidal radii are always larger than 30 pc (\citealt{McLaughlin05}, assuming a \citet{King66} profile) whereas the field of view of the WFPC2 is $\sim$36 pc in diameter (assuming a distance of 50 kpc to the LMC). However, an over-subtraction is not a serious issue as it affects mostly the well populated regions in the CMDs and keeps the overall structure unchanged. 

\begin{figure*}
  \centering
  
  \subfloat[NGC 2249]{\label{fig:ngc2249_cmd}\includegraphics[width=47mm]{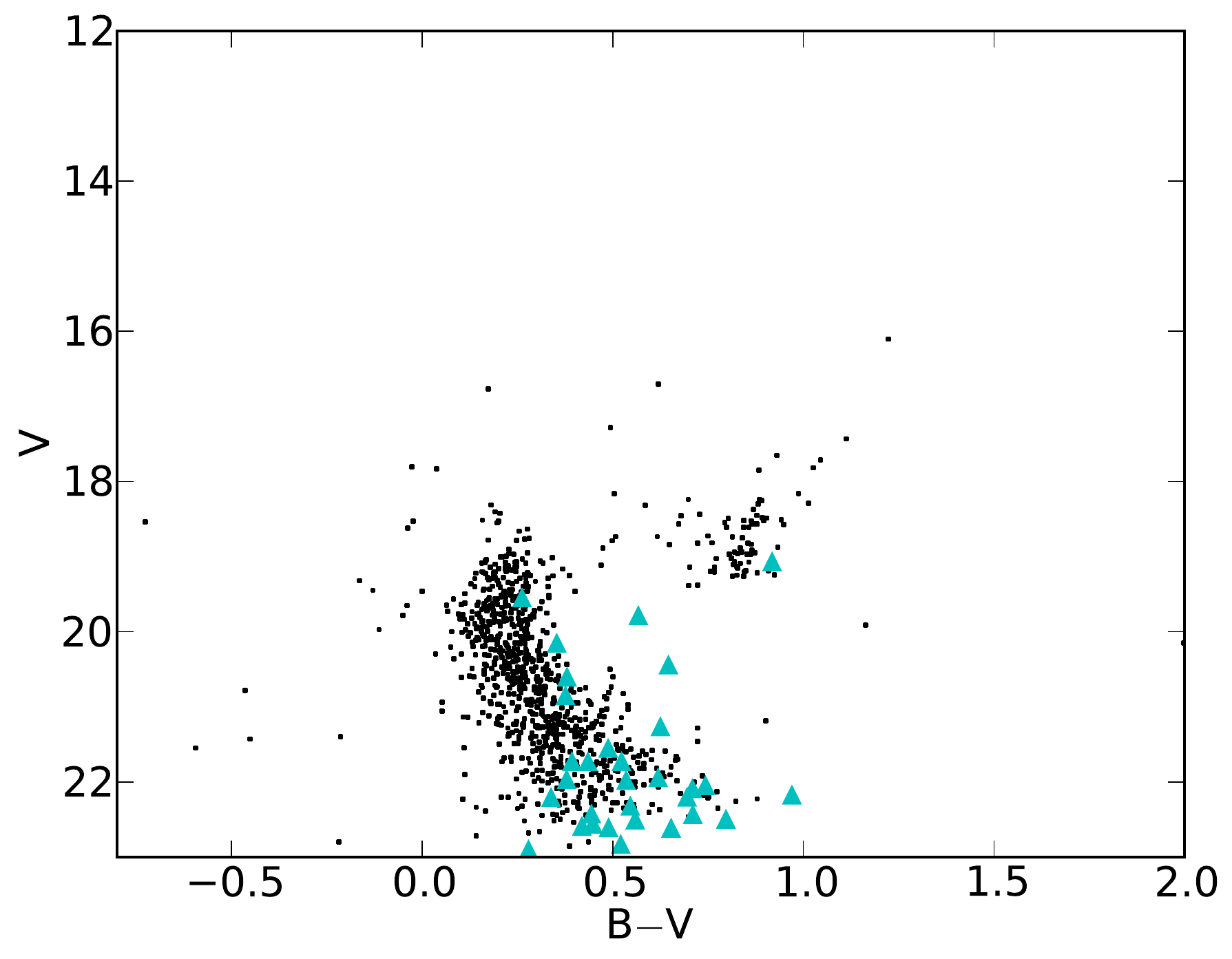}}
  \subfloat[NGC 1831]{\label{fig:ngc1831_cmd}\includegraphics[width=47mm]{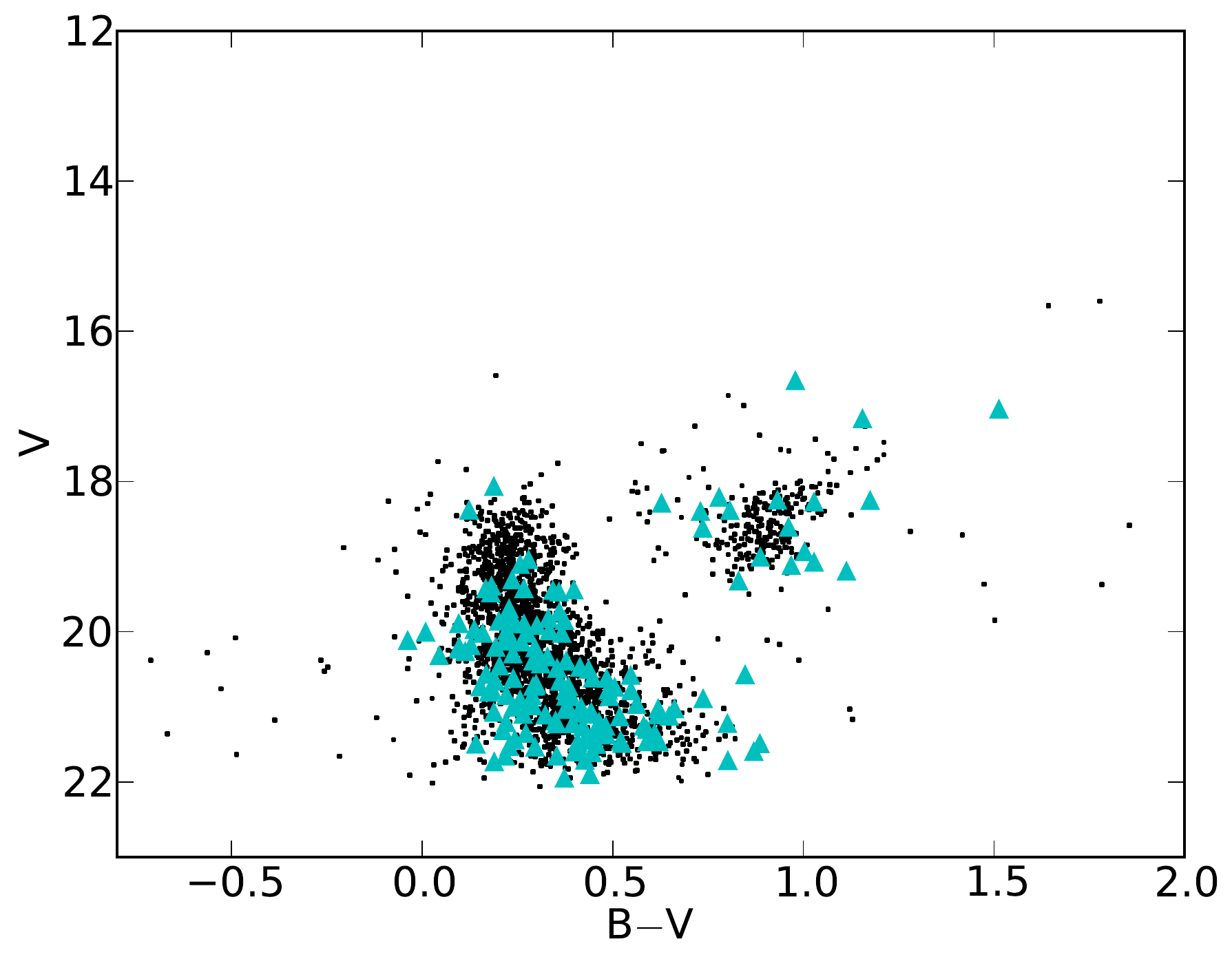}}
  \subfloat[NGC 2136]{\label{fig:ngc2136_cmd}\includegraphics[width=47mm]{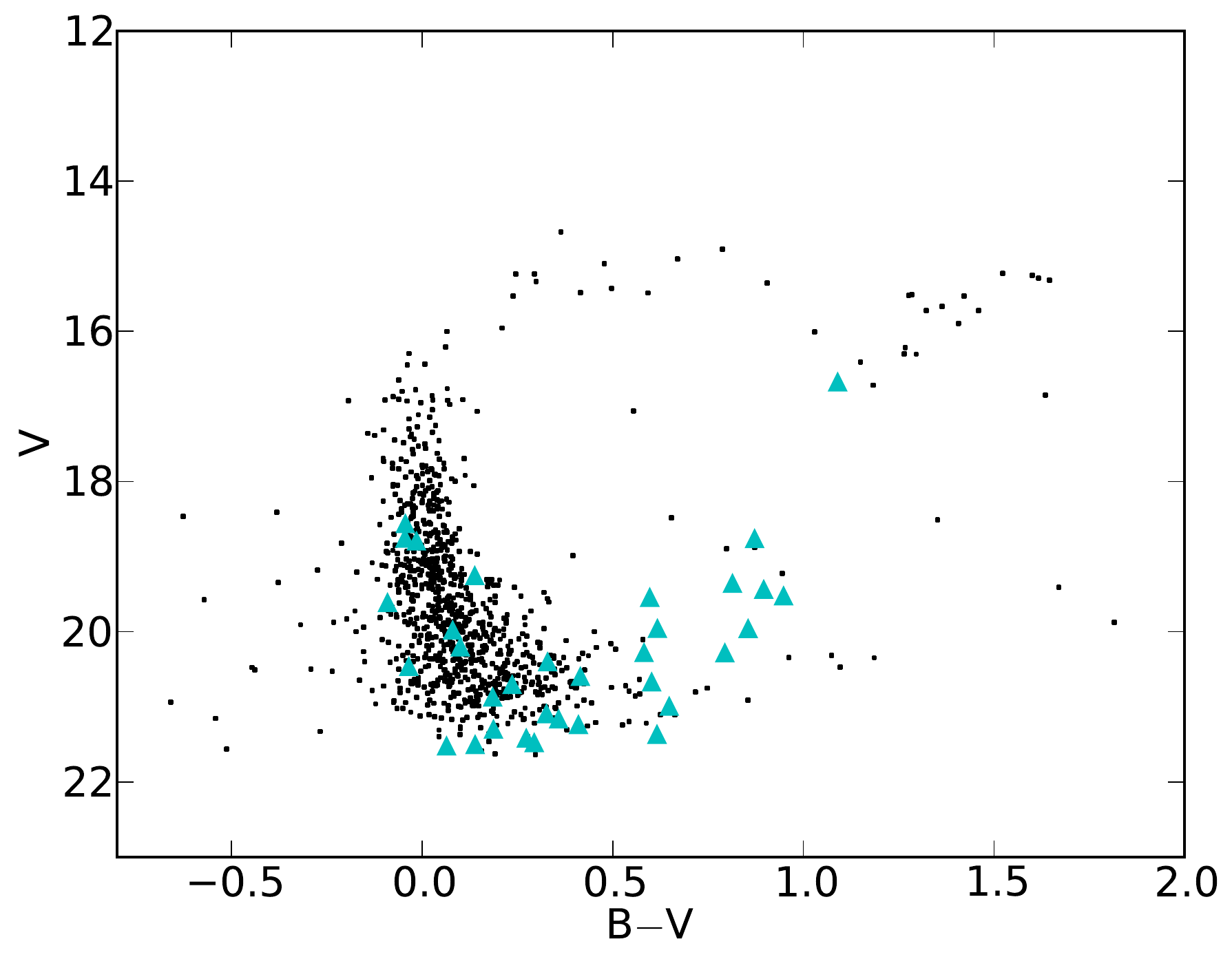}} 
  \subfloat[NGC 2157]{\label{fig:ngc2157_cmd}\includegraphics[width=47mm]{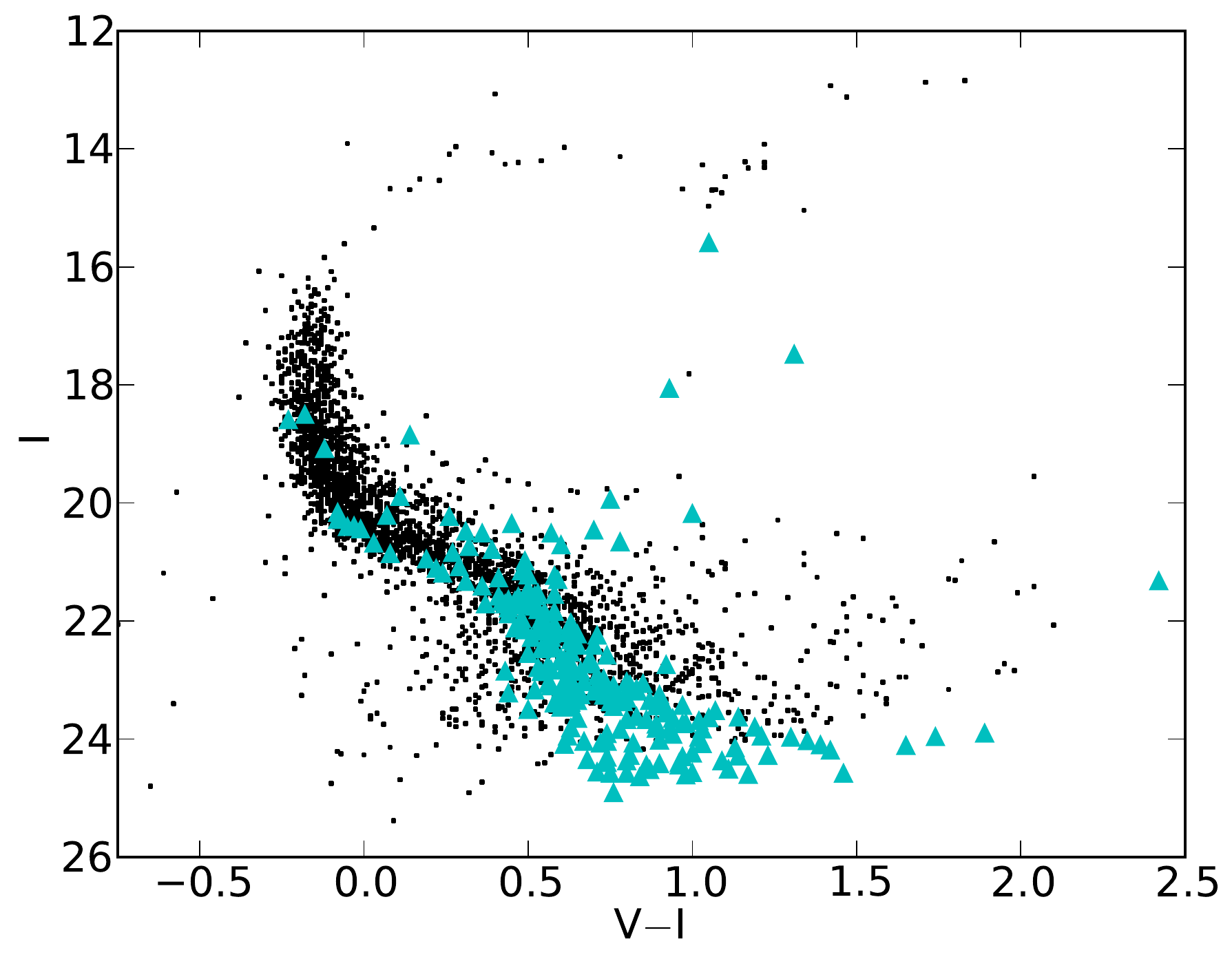}}
  \\
  \subfloat[NGC 1850]{\label{fig:ngc1850_cmd}\includegraphics[width=47mm]{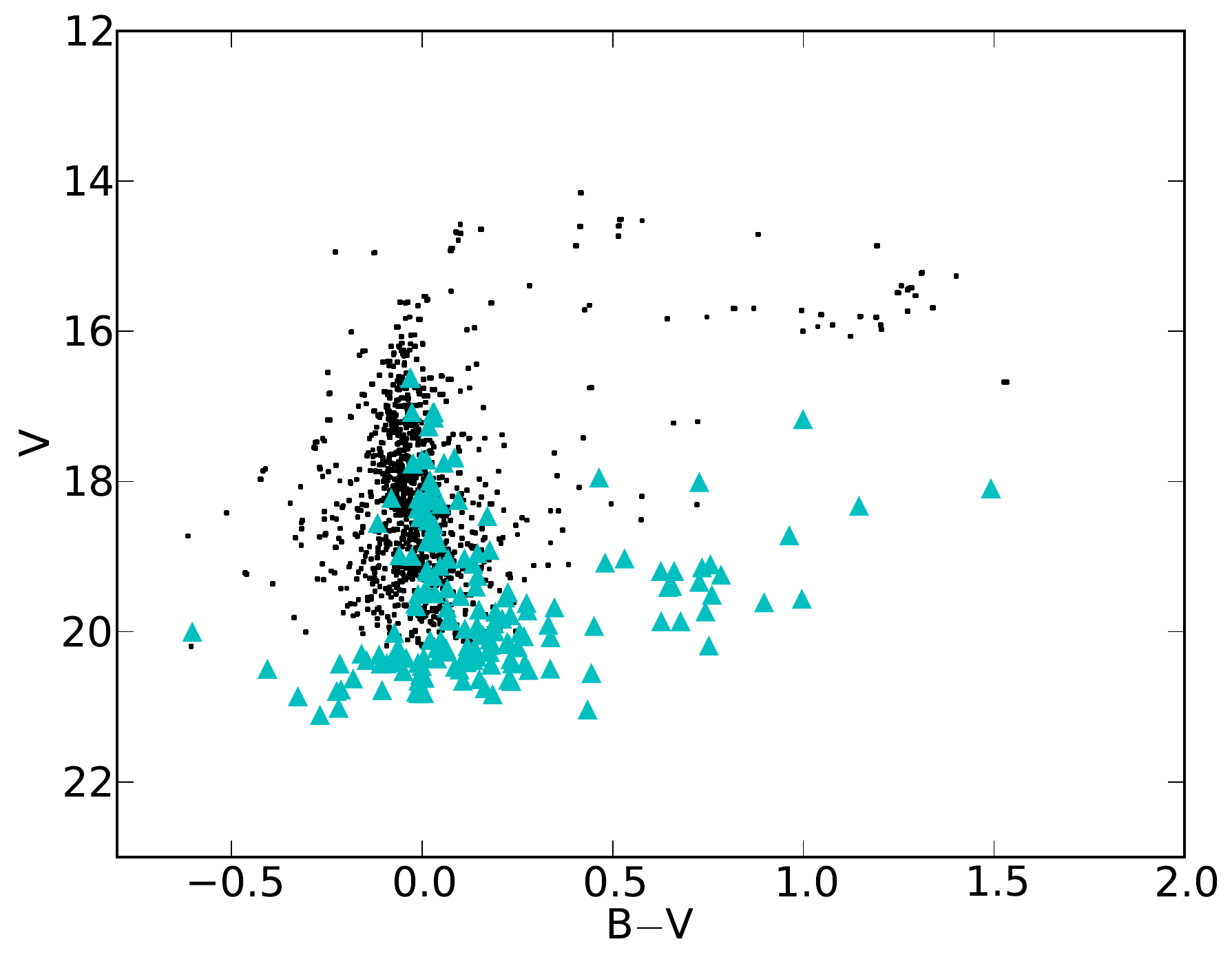}}
  \subfloat[NGC 1847]{\label{fig:ngc1847_cmd}\includegraphics[width=47mm]{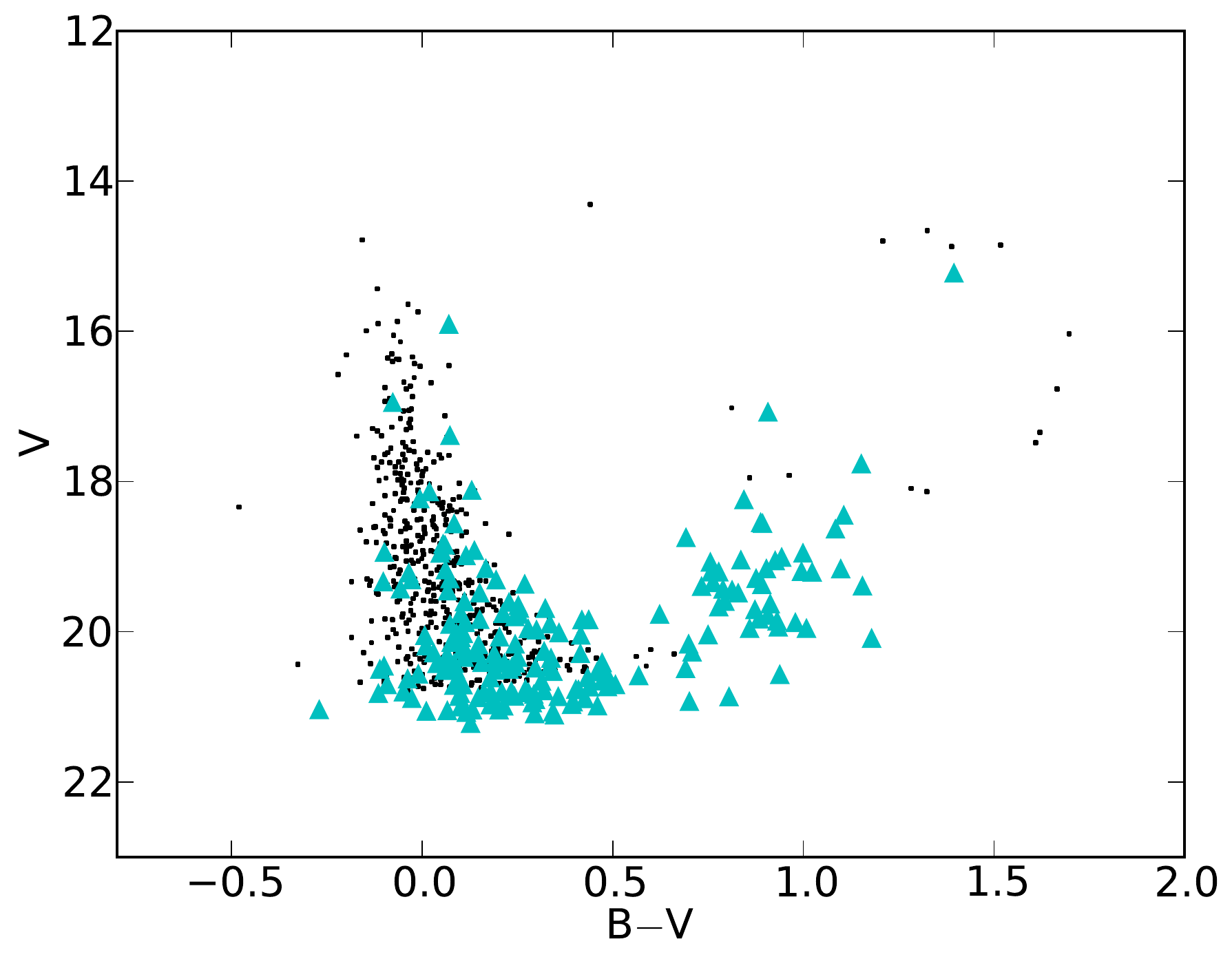}}
  \subfloat[NGC 2004]{\label{fig:ngc2004_cmd}\includegraphics[width=47mm]{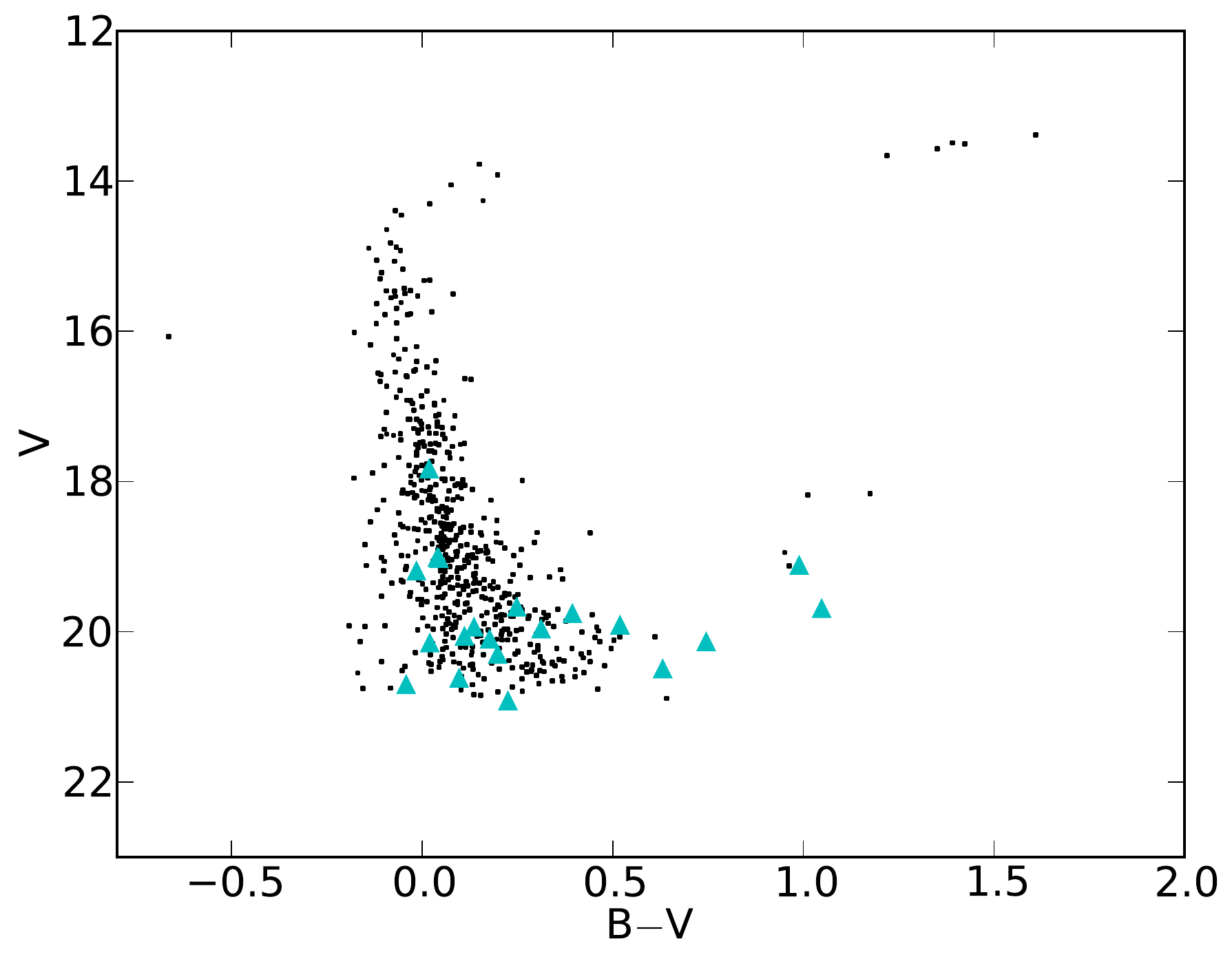}}
  \subfloat[NGC 2100]{\label{fig:ngc2100_cmd}\includegraphics[width=47mm]{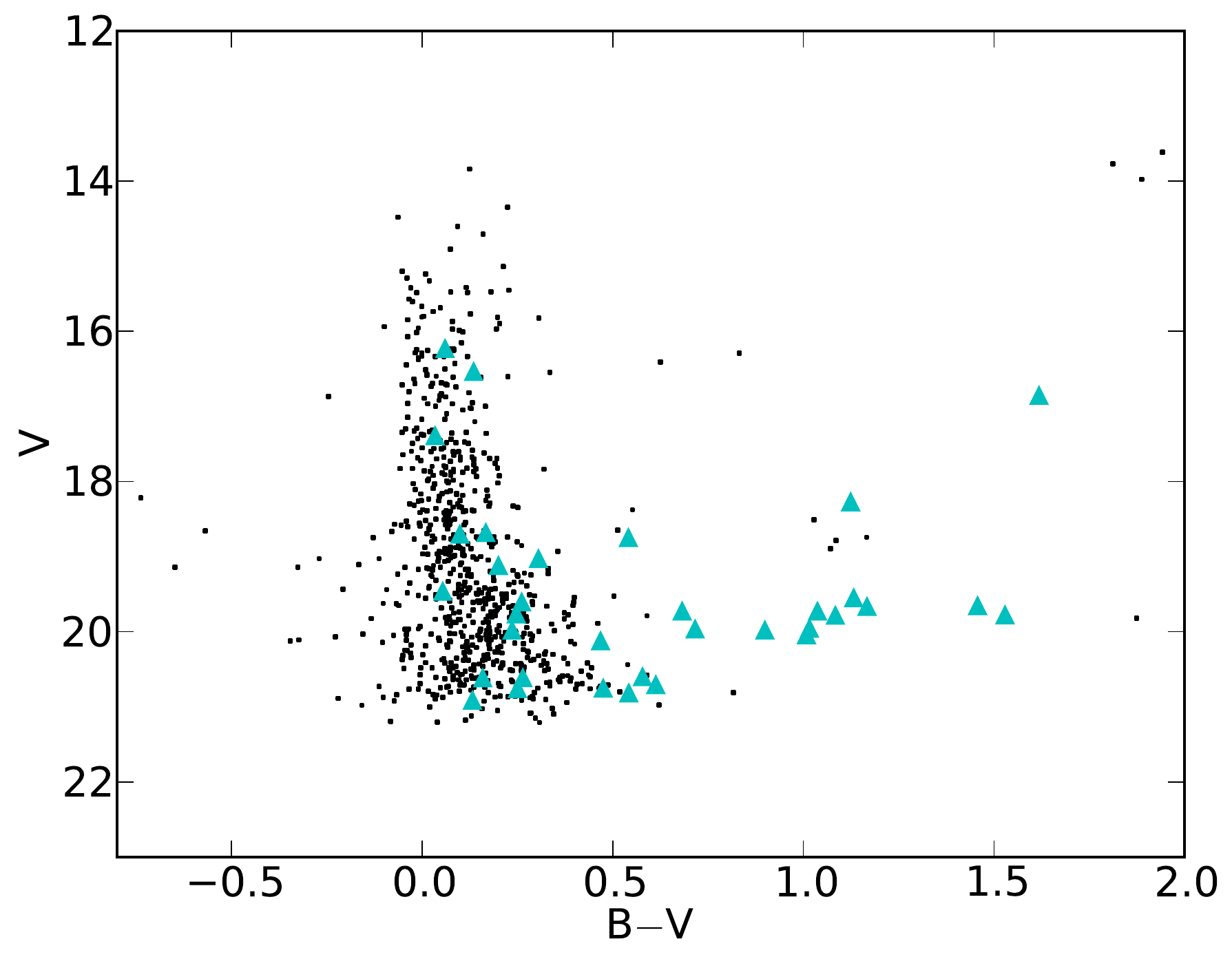}}
  \caption{CMDs of the clusters in our sample. The black dots are the stars that were used for the further analysis, whereas the (cyan) triangles indicate stars that were subtracted as field stars. Obviously, also some cluster stars were excluded in our SFH fits, but this does not affect our results. All CMDs contain only the stars in the inner regions of the clusters (two times the core radius). Note that the CMD of NGC 2157 is in the ($V-I$) vs $I$ space.}
  \label{fig:Cluster_cmds}
\end{figure*}

As already mentioned we are provided with an image of field stars near NGC 2157. Therefore, in this case, we made CMDs of the stars in the field and the cluster region of every chip and subtracted the field stellar population of every chip separately the same way as we did for the previous clusters. Figure \ref{fig:Cluster_cmds} shows the CMDs of all clusters. The black dots indicate all stars that were used in our analysis and the triangles (cyan) are the subtracted 'field' stars.

\citet{Brocato01} and \citet{Fischer98} performed artificial star tests to infer the completeness curves of their clusters. 
We adopt limiting $V$-band magnitudes that are at or brighter than the 90 \% completeness limit (cf. their Table 3). \citet{Fischer98} determined for NGC 2157 a 90\% completeness limit in the $I$ band of $\sim$20.5 mag (cf. their Figure 3). For NGC 1850, however, we do not have any measure of the completeness. To be on the safe side, we choose a limit of 18.5 mag in the $V$ band for the further analysis of this cluster. This limit is more than 1 magnitude brighter than the completeness limits of the cluster in the \citet{Brocato01} data set with comparable exposure times.

\section{Tests with Artificial Star Clusters\label{sec:art_cluster_test}}

\begin{figure*}[htp]
  \centering
  
  \subfloat[20 Myr ($B,V$ photometry)]{\label{fig:cluster730bv_cmd}\includegraphics[width=47mm]{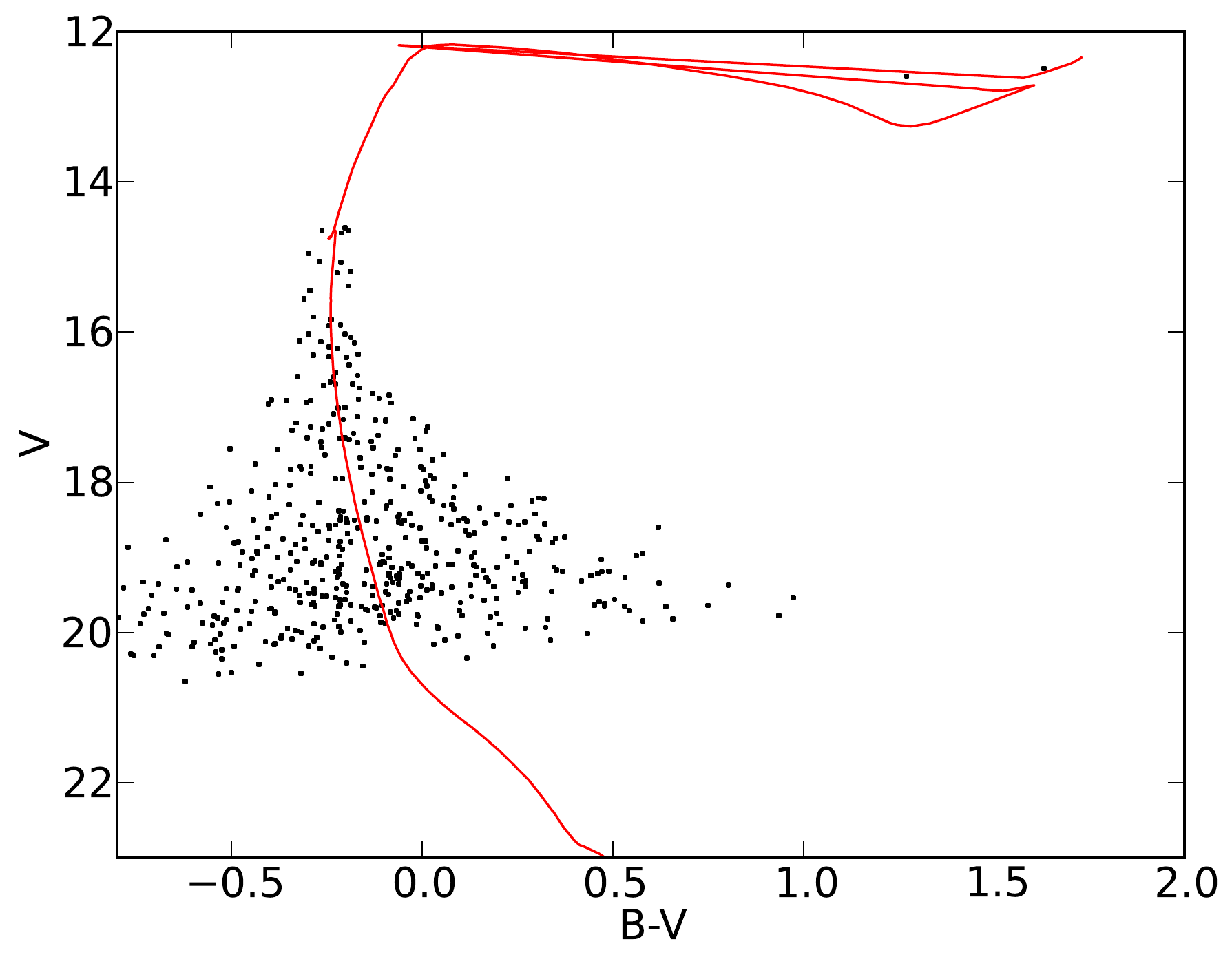}}
  \subfloat[100 Myr ($V,I$ photometry)]{\label{fig:cluster800vi_cmd}\includegraphics[width=47mm]{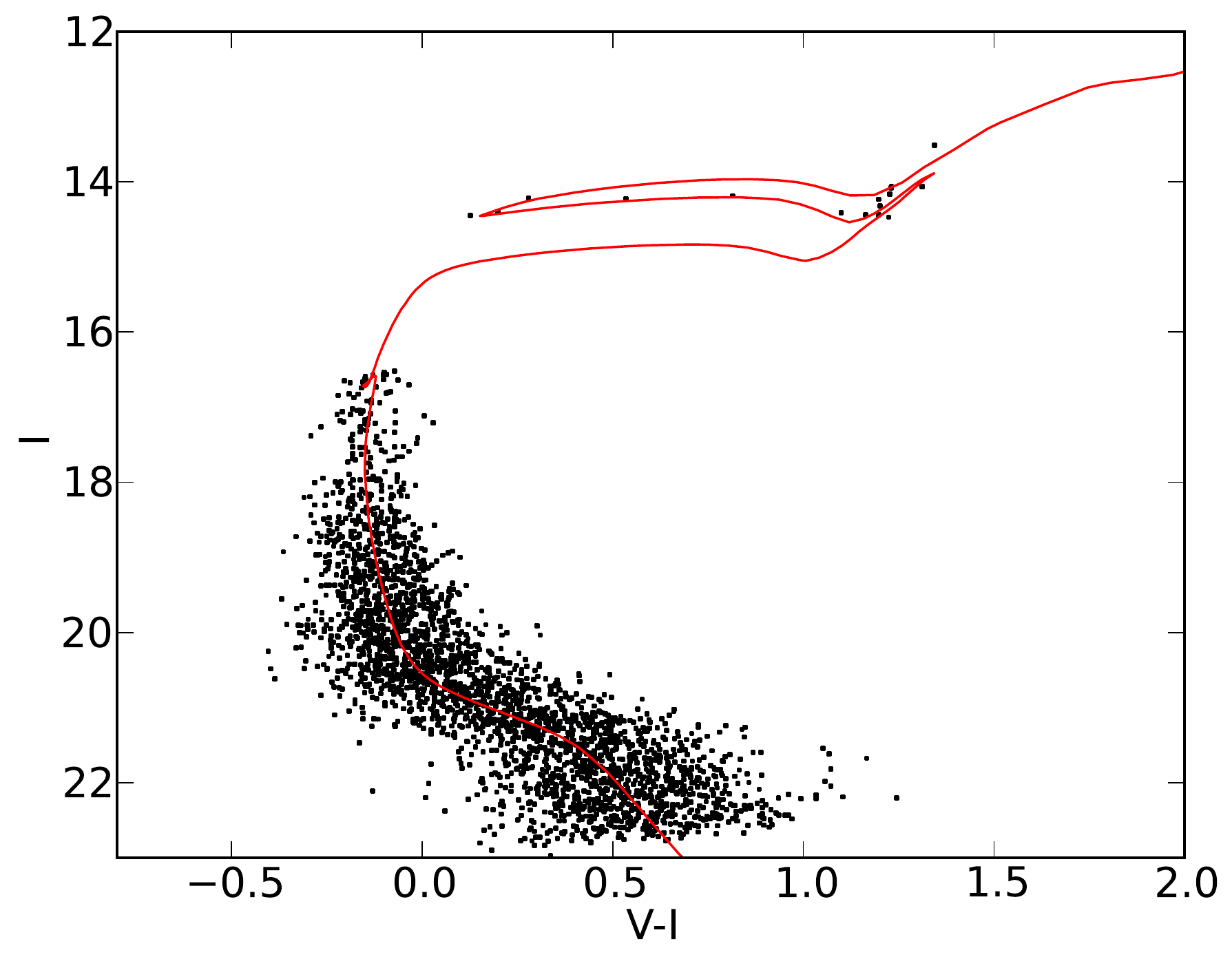}}
    \subfloat[125 Myr ($B,V$ photometry)]{\label{fig:cluster800bv_cmd}\includegraphics[width=47mm]{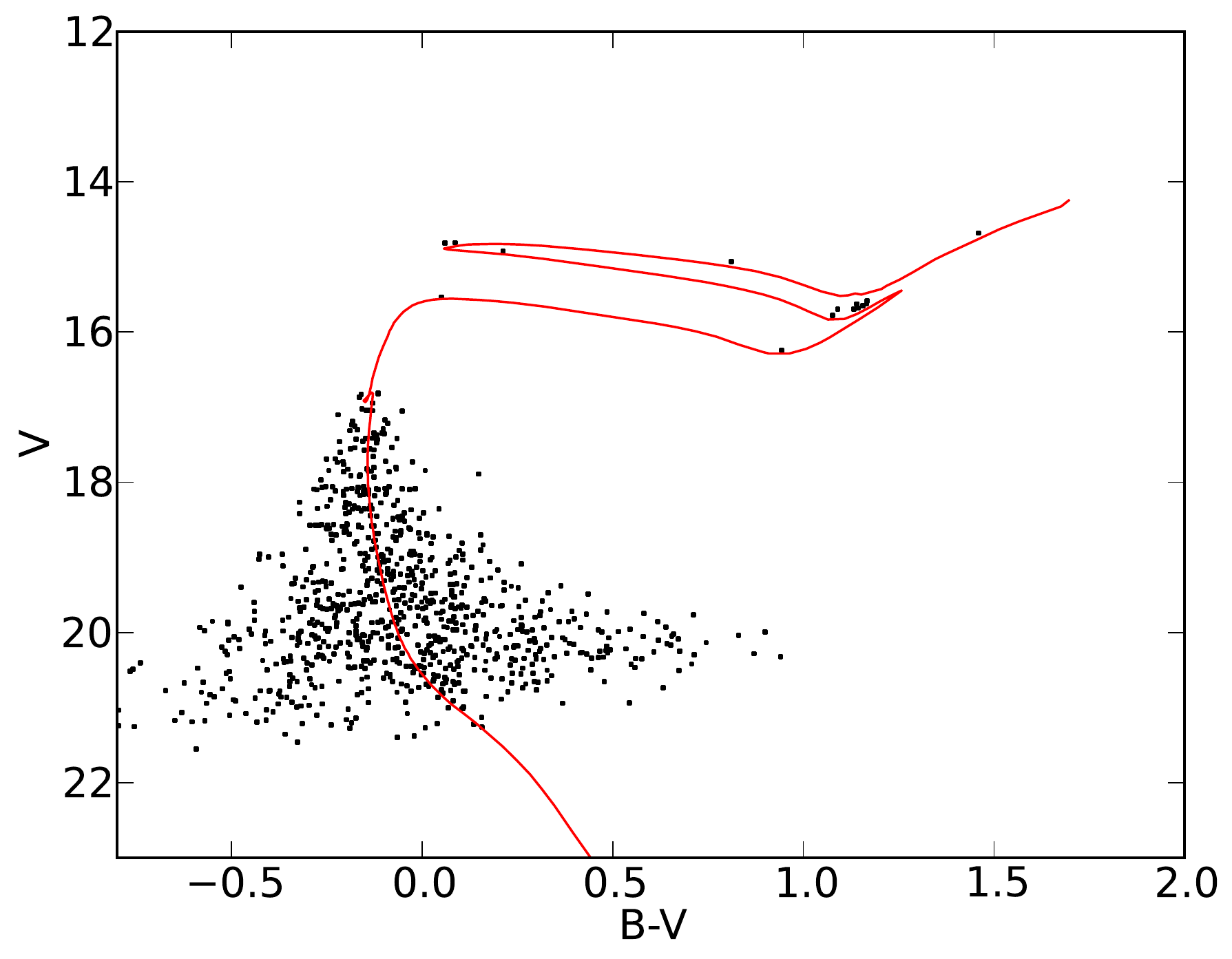}}
  \subfloat[1.1 Gyr ($B,V$ photometry)]{\label{fig:cluster904bv_cmd}\includegraphics[width=47mm]{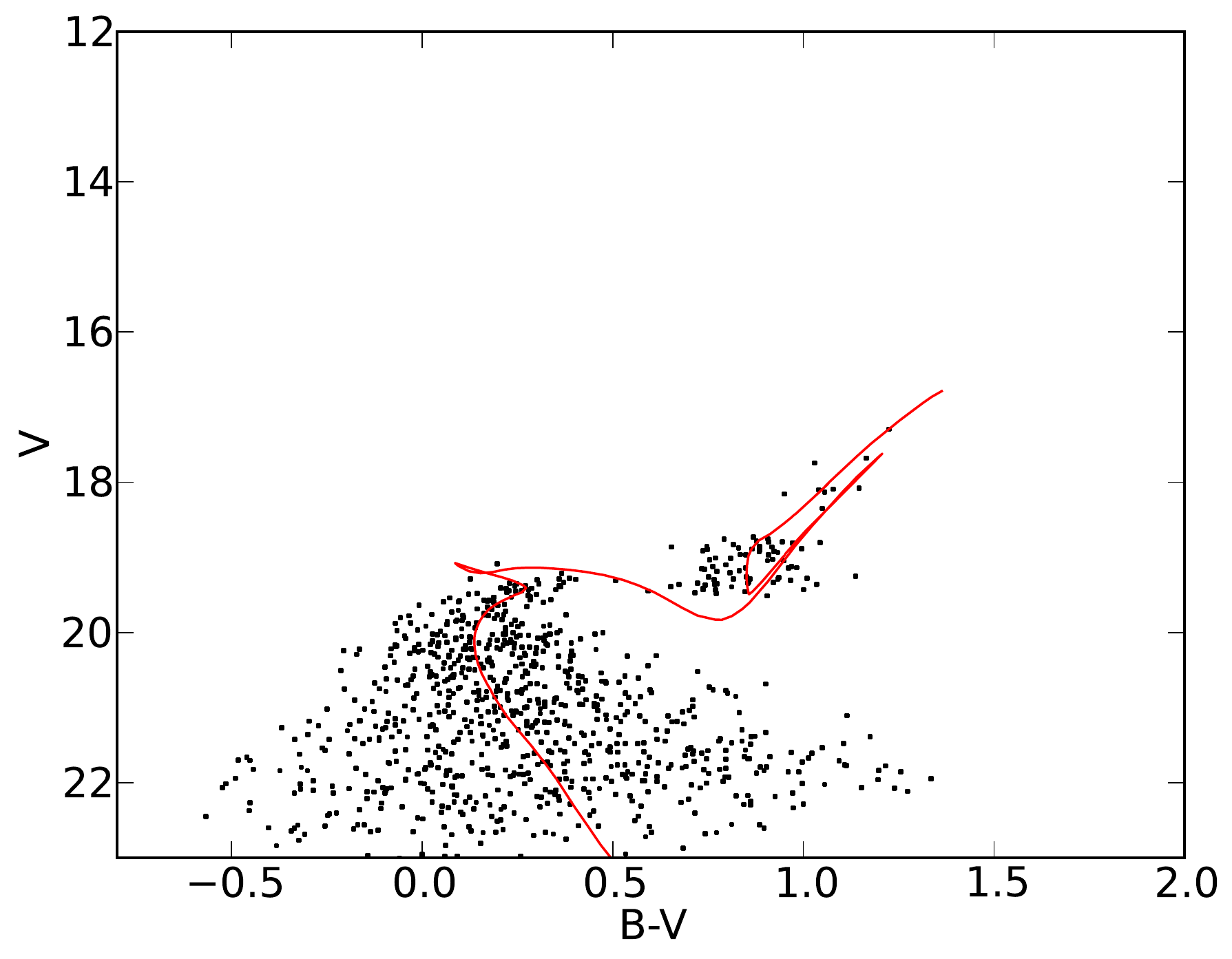}}
  \\
  \subfloat[20 Myr ($B,V$ photometry)]{\label{fig:cluster730bv_sfh_fit}\includegraphics[width=48.5mm]{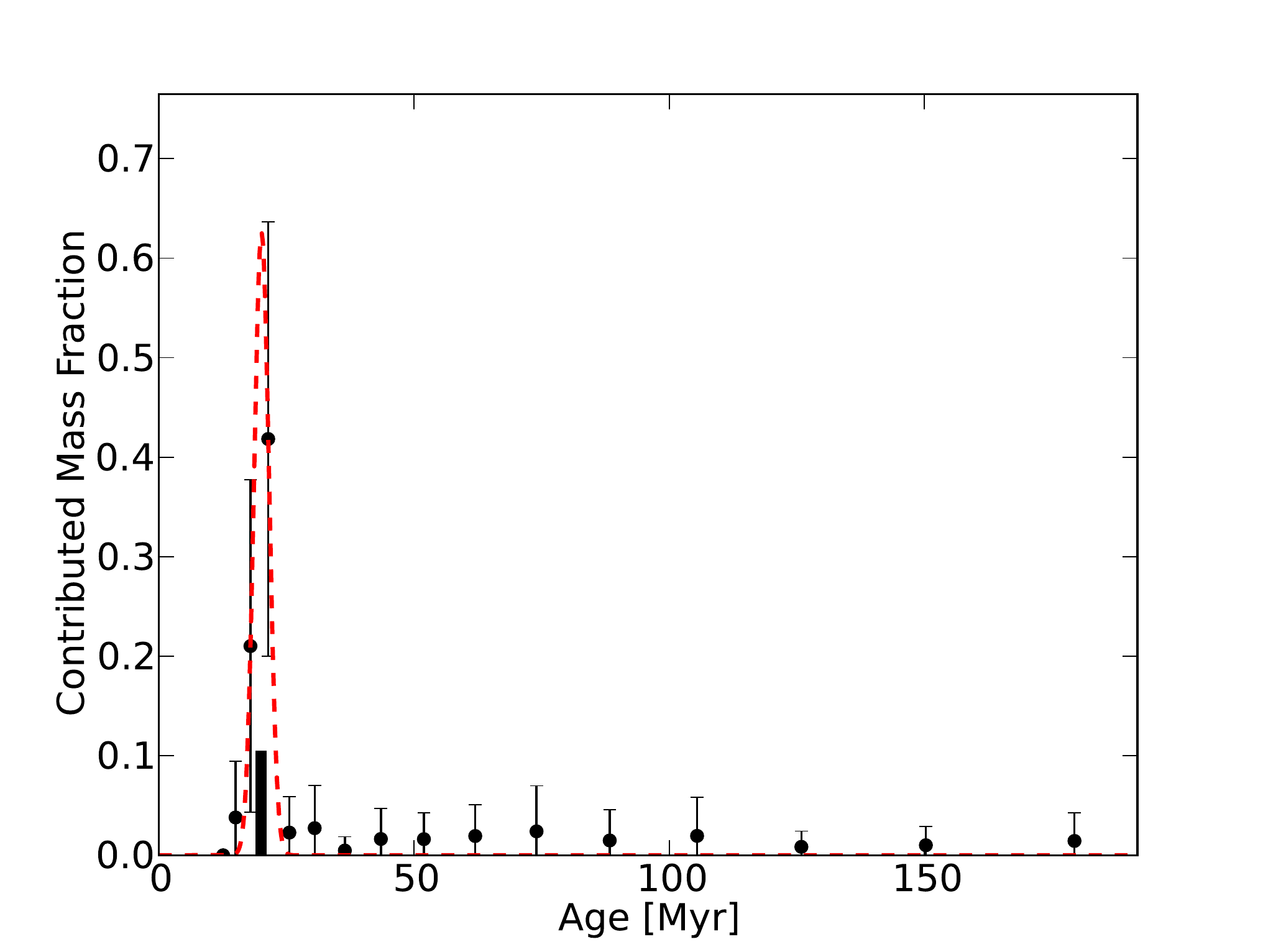}}
    \subfloat[100 Myr ($V,I$ photometry)]{\label{fig:cluster800vi_sfh_fit}\includegraphics[width=48.5mm]{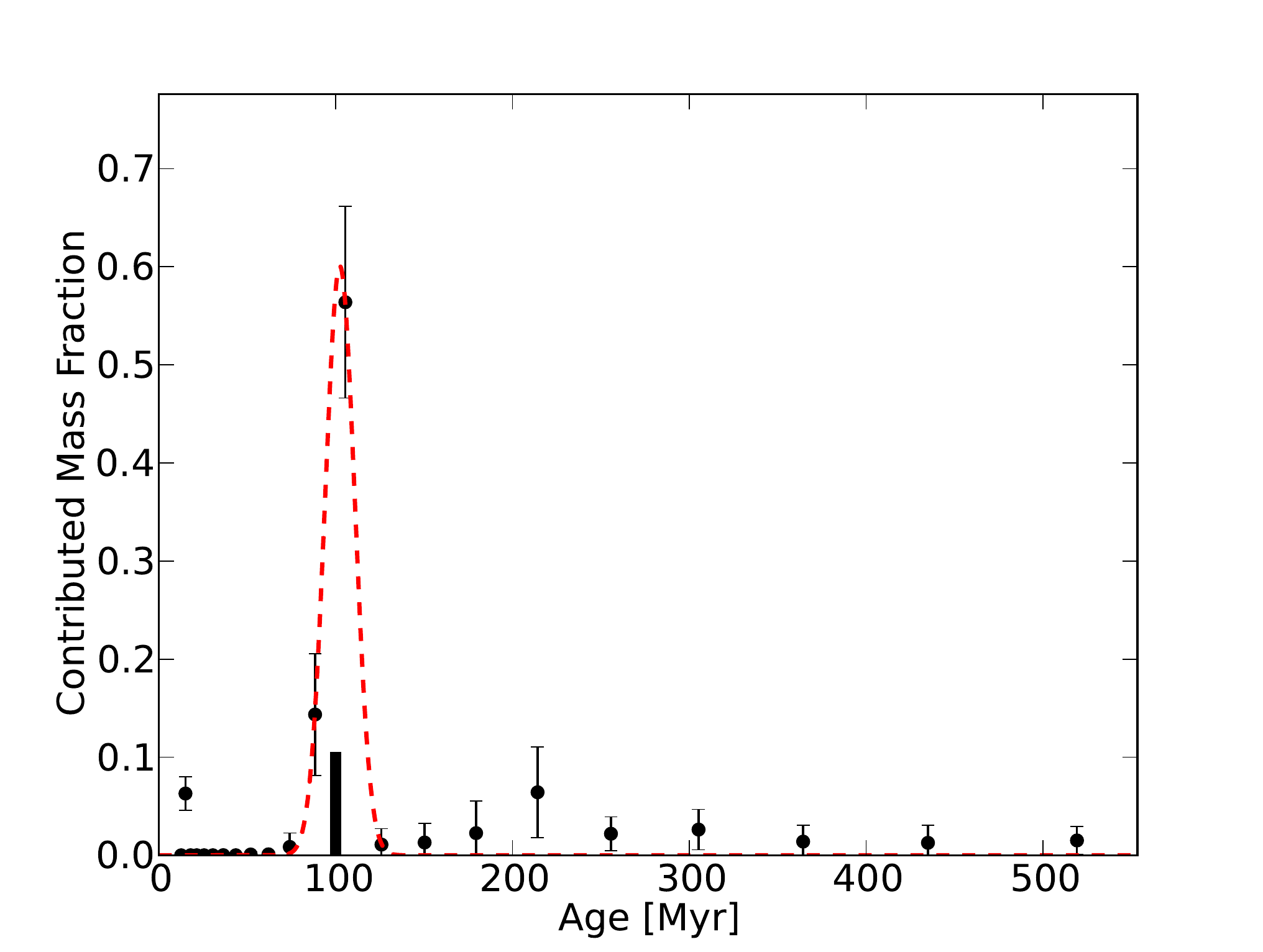}}
  \subfloat[125 Myr ($B,V$ photometry)]{\label{fig:cluster800bv_sfh_fit}\includegraphics[width=48.5mm]{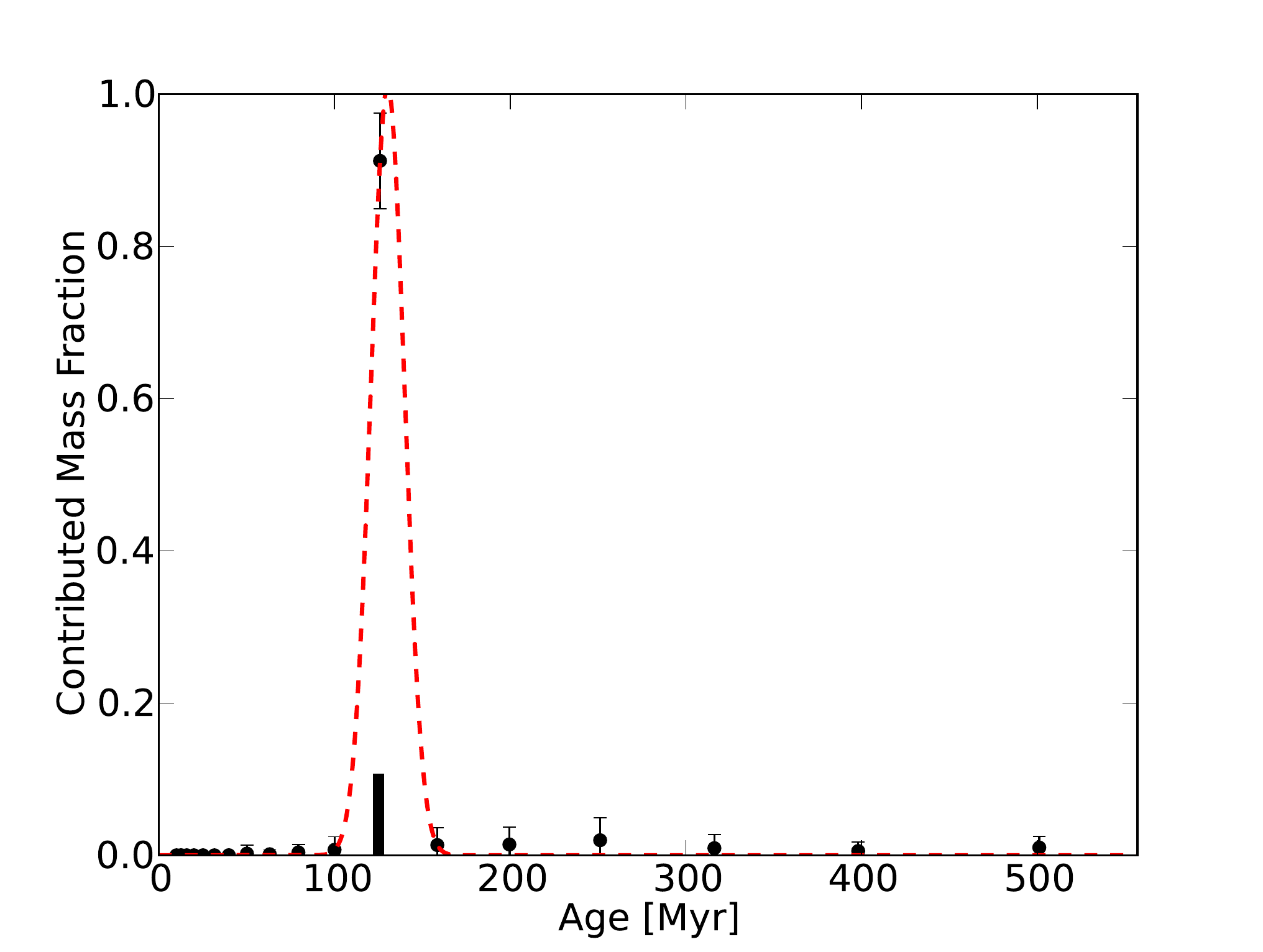}}
  \subfloat[1.1 Gyr ($B,V$ photometry)]{\label{fig:cluster905bv_sfh_fit}\includegraphics[width=48.5mm]{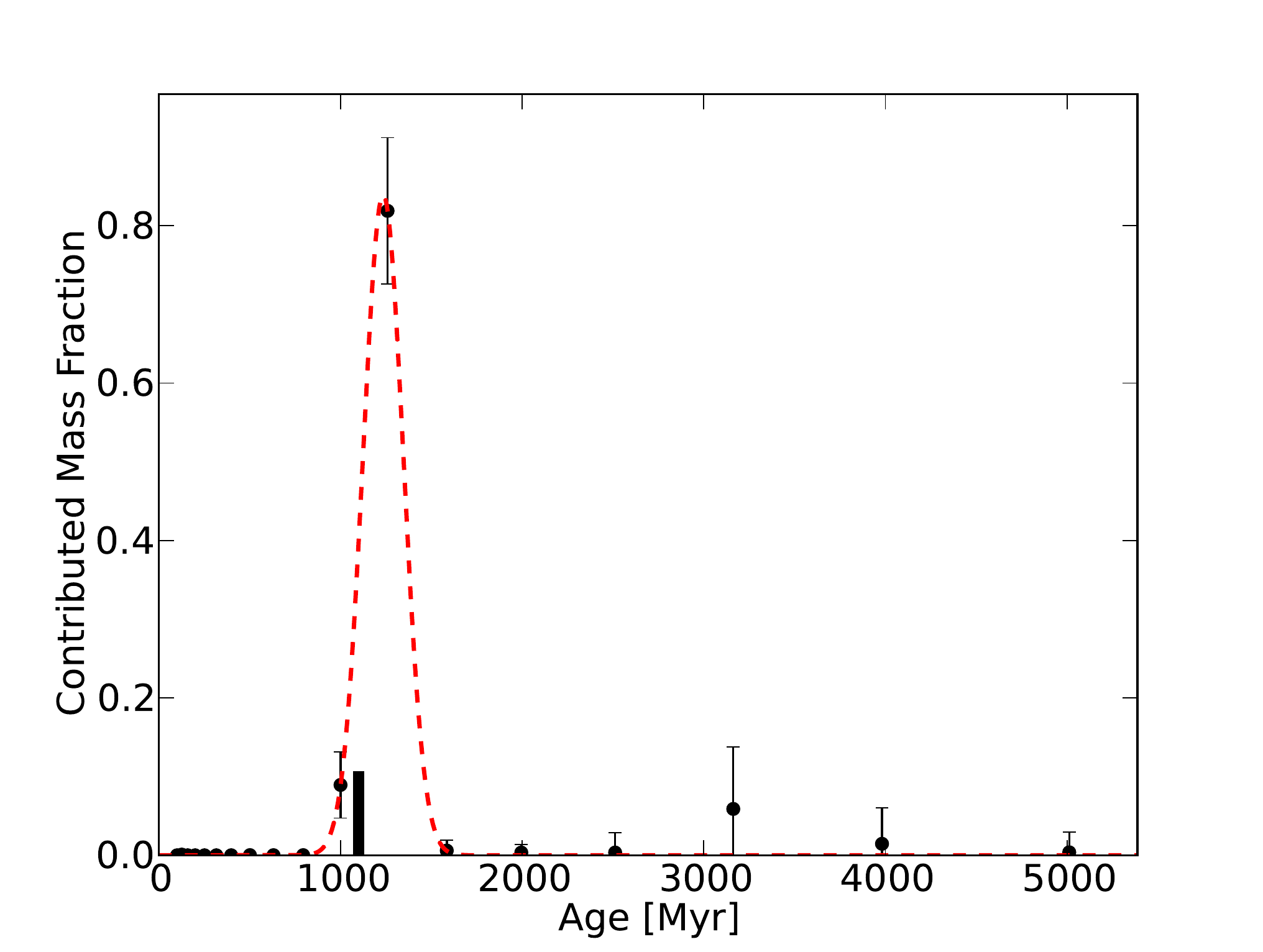}}  
  \caption{\textit{Upper panels}: CMDs of four of the artificial clusters that we created. We simulated these clusters to estimate the resulting age spread of a single-age population given by our SFH fitting code caused only by photometric errors. All stars in one cluster have the same age and the photometric errors are estimated from the ones of the real clusters at the respective age. The clusters shown here have ages of 20, 100 Myr, 125 Myr and 1.1 Gyr, spanning the range of ages of the real clusters in our sample. The (red) line is the theoretical Parsec 1.1 Padova isochrone at the respective age of the cluster. \newline
  \textit{Lower panels}: Results of the SFH fits of the four artificial star clusters shown in the upper panel. The dots represent the mean contributed mass fraction at individual ages of the 100 realizations whereas the errorbars are the standard deviation. The dashed \textbf{(red)} line is the best-fit Gaussian to the highest peak of the points taking into account the errorbars. Indicated as a vertical think line at the x-axis is the actual age of the respective cluster. The fit to the 20 Myr old cluster gives a peak at 20.1 Myr with a dispersion of 1.5 Myr. We got an age of 102 Myr for the 100 Myr old cluster with an standard deviation of 8.4 Myr and an age of 130 Myr with a dispersion of 9.6 Myr for the 125 Myr old cluster. The 1.1 Gyr old cluster has a fitted age of 1.2 Gyr. The standard deviation is 111 Myr.}
  \label{fig:artificial_cluster}
\end{figure*}

Before we fit the SFH of our sample of LMC clusters we made tests with artificial coeval star clusters to assess the magnitude of apparent age spreads that are purely induced by photometric errors. For each cluster in our sample we modeled a corresponding artificial cluster with the same metallicity and age that we found for the real cluster. We assigned every cluster a random number of stars that is drawn from a Gaussian distribution that peaks at the number of stars present in the data table of the respective observed cluster. The masses of the cluster stars were drawn stochastically from a stellar initial mass function (IMF) with an index $\alpha$ of $-$2.35 \citep{Salpeter55}. The lower limits of the stellar masses were chosen such that the CMDs of the artificial clusters cover the same magnitude range as the real clusters. We create the photometry for each star by linearly interpolating the isochrone grid (Parsec 1.1 from the Padova set) at the respective ages of the clusters and adding photometric uncertainties to the synthetic photometry that follow the same behavior as the observed errors. For all observed clusters we fitted an exponential curve to the photometric standard deviations as a function of the magnitude in each band. We then assigned every star an error that is drawn from a Gaussian distribution with a width that corresponds to the value of the exponential function at the respective magnitude plus a small random scatter comparable to the scatter of the real errors around the fitted curve.  
The upper panels of Figure \ref{fig:artificial_cluster} show as an example the CMDs of four of the modeled clusters together with the theoretical isochrones that were used to create the photometry.

When comparing the artificial clusters with the observed ones (cf. Figure \ref{fig:Cluster_cmds} and \ref{fig:artificial_cluster}) we note that the MS of the synthetic clusters is much broader at fainter magnitudes than the MS of the real clusters. This is due to the fact that we modeled the clusters using the photometric errors of the corresponding real clusters. If we look at Figure \ref{fig:ngc2136_error_vs_mag} we see that the MS at fainter magnitudes is narrower than would be expected from the photometric errors. However, this does not affect our analysis as we do not use these regions in the CMD for the fitting of the SFH (see Section \ref{sec:res}).

We carried out 100 Monte Carlo realizations of each cluster and fitted their SFH the same way as we did for the observed clusters (see Section \ref{sec:res}). The results of the clusters that are shown as an example are presented in the lower panels of Figure \ref{fig:artificial_cluster}. The dots represent the mean contributed mass fraction that results from the 100 Monte Carlo realizations, at individual ages. The errorbars are the standard deviations. The (red) dashed line is the best Gaussian fit to the highest peak in the distribution, taking into account the errors. The thick vertical line at the x-axis marks the input age of the cluster. All fits reproduce the input age of the clusters very well within the errors. We note that the fits to the clusters show some additional low amplitudes of star formation at higher ages as it is observed in the real clusters (see Section \ref{sec:res}). This is a first sign that these features are due to the fitting process and not intrinsic to the cluster itself. The fitted ages and standard deviations of all modeled clusters are summarized in Table \ref{tab:Fitting_results} where we compare them with the results from the observed clusters. The dispersion of ages is lower than the spreads we found for the highest peak of most of the real clusters which suggest that the spread in the fitted SFH of the observed clusters is not only due to photometric errors. We will discuss this in Section~\ref{sec:disc}.

\section{Results\label{sec:res}}

\begin{figure*}
\centering
\begin{tabular}{cc}
\includegraphics[width=8cm]{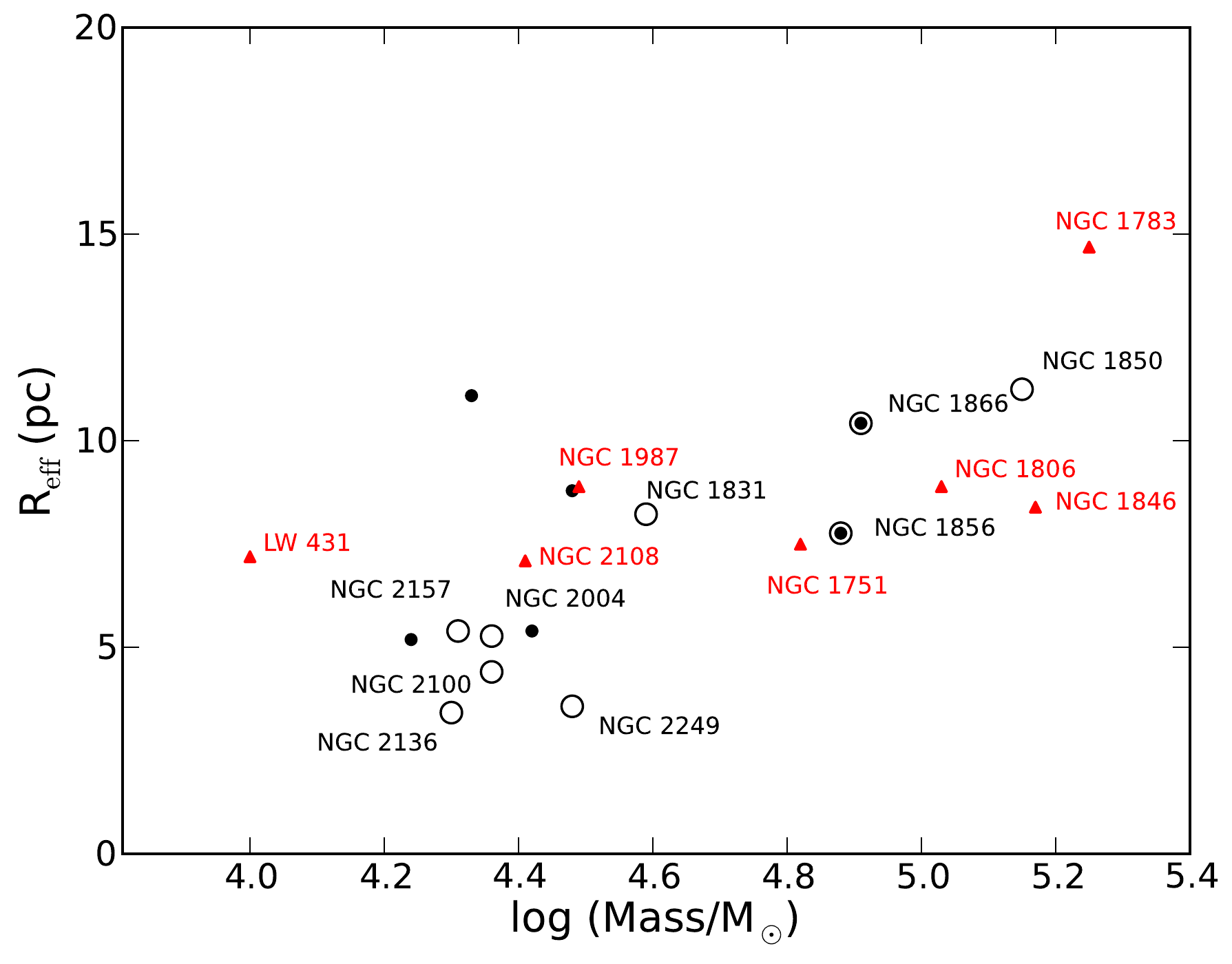} & \includegraphics[width=8cm]{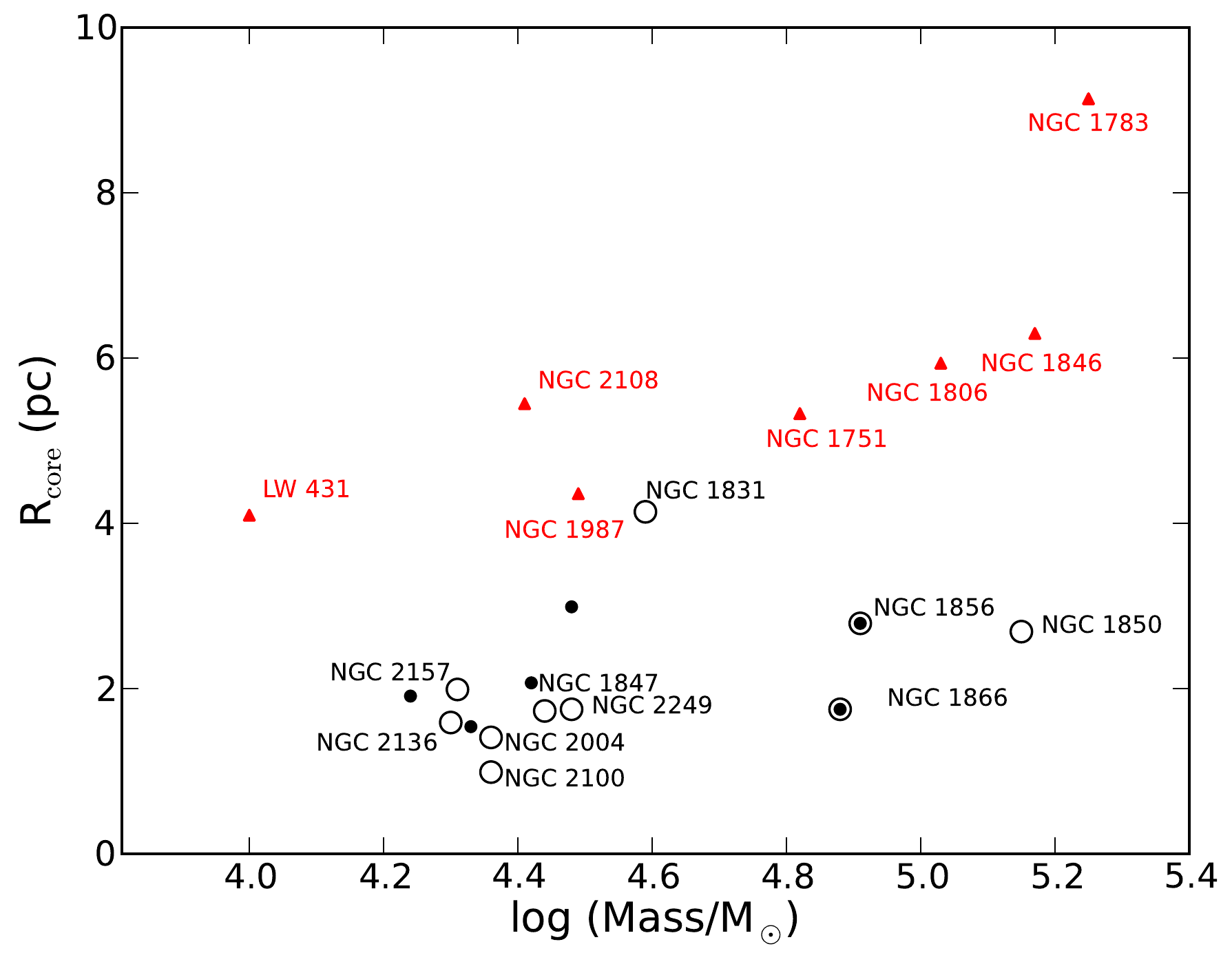}\\
\end{tabular}
\caption{\textit{Left panel}: Plot of the mass versus the effective radius for our sample of YMCs (black circles) and some additional clusters with similar properties (black dots), taken from the catalog of \citet{McLaughlin05}. The two clusters that are marked with a black dot surrounded by a black circle are the two clusters analyzed by \citet{BastianSilva13}. The (red) triangles are clusters of intermediate ages that display extended MSTOs, from \citet{Goudfrooij09, Goudfrooij11a}. In this plot NGC 1847 is not shown as its effective radius has a large uncertainty due to its shallow profile. We increased the mass of NGC 1850 by a factor of 2 with respect to the literature value to account for the new age that we found (see Section \ref{sec:ngc1850}). \newline
\textit{Right panel}: Same as the left panel but for the core radius. It is shown, e.g. in \citet{Keller11}, that the core radii of clusters have larger spreads the older the clusters are. We note here that the core radii of all clusters from the \citet{Goudfrooij09, Goudfrooij11a} sample are systematically higher than the ones from our sample. One reason for this could be that the core radii given in \citet{Goudfrooij09, Goudfrooij11a} are obtained from the surface number density, whereas \citet{McLaughlin05} used the surface brightness profiles. Both methods do not yield the same value (see text).}
\label{fig:Mass_vs_Reff}
\end{figure*}

\begin{table*} 
\caption{Parameters of the LMC clusters\label{tab:Cluster_Param}}
\centering
\begin{tabular}{l c c c c c c c c c c} 
\hline\hline
\noalign{\smallskip}
Cluster & \multicolumn{2}{c}{Age (Myr)} & log Mass/M$_{\sun}$ & \multicolumn{2}{c}{Z (Z$_{\sun}$=0.0152)}  & R$_{\mathrm{core}}$ (pc) & V$_{\mathrm{esc}}$ (km/s)& V$_{\mathrm{esc}}$ (km/s) & \multicolumn{2}{c}{$E(B-V)$}\\
& lit.& this work &  & lit. & this work & & & at 10 Myr & lit. & this work \\
\noalign{\smallskip}
\hline
\noalign{\smallskip}
NGC 1831 & 700\tablefootmark{a} & 926 & 4.59\tablefootmark{b} & 0.016\tablefootmark{a} & 0.016 & 4.24\tablefootmark{c} / 4.13\tablefootmark{b} & 6.6\tablefootmark{b} & 9.3 & 0.01\tablefootmark{a} & 0.0\\

NGC 1847 & 26\tablefootmark{d} & 57 & 4.44\tablefootmark{b} & 0.006\tablefootmark{e} & 0.006 & 3.35\tablefootmark{c} / 1.73\tablefootmark{b} & 4.2\tablefootmark{b} & 4.5 & 0.1\tablefootmark{f} & 0.16\\

NGC 1850 & 30\tablefootmark{g} & 93 & 4.86\tablefootmark{b} / 5.15 & 0.008\tablefootmark{h} & 0.006 & $--$ / 2.69\tablefootmark{b}  & 8.77\tablefootmark{b} / 12.3 & 15.5 & 0.17\tablefootmark{h} & 0.1\\

NGC 2004 & 20\tablefootmark{i} & 20 & 4.36\tablefootmark{b} & 0.004\tablefootmark{j} & 0.004 & 2.18\tablefootmark{c} / 1.41\tablefootmark{b} & 7.0\tablefootmark{b} & 7.5 & 0.08\tablefootmark{k} & 0.23\\

NGC 2100 & 16\tablefootmark{i} & 21 & 4.36\tablefootmark{b} & 0.007\tablefootmark{j} & 0.007 & 3.03\tablefootmark{c} / 0.99\tablefootmark{b} & 7.9\tablefootmark{b} & 8.5 & 0.24\tablefootmark{k} & 0.17\\

NGC 2136 & 100\tablefootmark{l} & 124 & 4.30\tablefootmark{b} & 0.004\tablefootmark{l} & 0.005 & 2.91\tablefootmark{c} / 1.59\tablefootmark{b} & 7.4\tablefootmark{b} & 9.3 & 0.1\tablefootmark{l} & 0.13\\

NGC 2157 & 100\tablefootmark{m} & 99 & 4.31\tablefootmark{b} & 0.008\tablefootmark{m} & 0.008 & $--$ / 2.00\tablefootmark{b}  & 6.2\tablefootmark{b} & 7.8 & 0.1\tablefootmark{m} & 0.1\\

NGC 2249 & 1000\tablefootmark{n} & 1110 & 4.48\tablefootmark{n} & 0.007\tablefootmark{a} & 0.008 & 2.79\tablefootmark{c} / 1.75\tablefootmark{b} & 9.4\tablefootmark{n} & 12.1 & 0.01\tablefootmark{a} & 0.02\\
\hline
\noalign{\smallskip}
NGC 1856 & \multicolumn{2}{c}{281\tablefootmark{o}} & 4.88\tablefootmark{b} & \multicolumn{2}{c}{0.008\tablefootmark{o}} & $--$ / 1.75\tablefootmark{b} & 11.0\tablefootmark{b} & 14.7 & \multicolumn{2}{c}{0.26\tablefootmark{o}}\\

NGC 1866 & \multicolumn{2}{c}{177\tablefootmark{o}} & 4.91\tablefootmark{b} & \multicolumn{2}{c}{0.008\tablefootmark{o}} & $--$ / 2.79\tablefootmark{b} & 9.5\tablefootmark{b} & 12.5 & \multicolumn{2}{c}{0.05\tablefootmark{o}}\\
\noalign{\smallskip}
\hline
\end{tabular}\\

\tablefoot{In the columns that give the age, the metallicity Z and the extinction $E(B-V)$, the first value is from the literature and the second one is determined in this work. Additionally, we also found new values for the mass and the escape velocity of NGC 1850 (second values in the respective columns). The core radius of the clusters can be determined in two ways: Using the surface number density (first value) or the surface brightness density (second value).\\ The last two clusters (NGC 1856 and NGC 1866) are not studied in this work. They were already analyzed by \citet{BastianSilva13}. \\ \textbf{References:}}
\tablefoottext{a}{\citet{Kerber07}};
\tablefoottext{b}{\citet{McLaughlin05} assuming \citet{King66} profiles};
\tablefoottext{c}{\citet{Brocato01}};
\tablefoottext{d}{\citet{ElsonFall88}};
\tablefoottext{e}{\citet{Mackey03}};
\tablefoottext{f}{\citet{Nelson83}};
\tablefoottext{g}{\citet{Baumgardt13}};
\tablefoottext{h}{\citet{Fischer93}};
\tablefoottext{i}{\citet{Elson91}};
\tablefoottext{j}{\citet{Jasniewicz94}};
\tablefoottext{k}{\citet{Keller2000}};
\tablefoottext{l}{\citet{Dirsch2000}};
\tablefoottext{m}{\citet{Fischer98}};
\tablefoottext{n}{\citet{Correnti14}};
\tablefoottext{o}{\citet{BastianSilva13}}\\

\end{table*}

The aim of this work is to search for potential age spreads in a sample of eight young ($<$1.1 Gyr) massive ($>10^4\mathrm{M_{\sun}}$) star clusters in the LMC. Our sample of clusters covers an age range from 20 Myr to about 1 Gyr. The clusters analyzed in this work and in \citet{BastianSilva13} have similar properties as the intermediate age (1-2 Gyr) LMC clusters that show extended or double MSTOs. The left panel of Figure \ref{fig:Mass_vs_Reff} shows the masses of the clusters of our sample as a function of the effective radius $R_{\rm eff}$ along with the clusters presented in \citet{Goudfrooij09, Goudfrooij11a}. Both, young and intermediate age clusters follow the same trend of increasing $R_{\rm eff}$ with increasing mass. NGC 1847 is not included in this plot as it has a large uncertainty in its effective radius due to its shallow profile. The right panel of Figure \ref{fig:Mass_vs_Reff} shows the core radius $R_{\rm core}$ as a function of logarithmic cluster mass for the same clusters.  It is expected that, due to dynamical evolution, the spread in $R_{\rm core}$ increases with the age of the clusters (e.g. \citealt{Keller11}). We note that all the YMCs have systematically smaller core radii than the intermediate age clusters. One reason for this could be that the core radii given by \citet{Goudfrooij09, Goudfrooij11a} are constructed from the surface number density profiles, whereas \citet{McLaughlin05} used surface brightness profiles. Both methods do not yield necessarily the same value for $R_{\rm core}$. Due to dynamical mass segregation inside clusters the surface brightness profile is more concentrated towards the center and therefore results in a smaller value for the core radius. 

We fitted theoretical isochrones from the Parsec 1.1 isochrone set \citep{Bressan12} to the observed CMDs to estimate the clusters' metallicity, distance modulus and the reddening towards the clusters. This set of isochrones uses a value of 0.0152 for the solar metallicity $\mathrm{Z_{\sun}}$. Table \ref{tab:Cluster_Param} lists the basic properties of the LMC clusters that are the subject of this work. For the further analysis we dereddened the magnitudes of all stars in each cluster by the same value of the derived reddening (we assume no differential reddening, except for NGC2100). The data of NGC 2157 is already corrected for a reddening value of $E(B-V)$ = 0.1 by \citet{Fischer98}.

As a second step we fitted the SFH of each cluster using the code $FITSFH$ \citep{Silva-Villa10}. For the fitting we used the previously determined metallicity and distance and assumed a \citet{Salpeter55} IMF. 

We performed the fitting in two regions (a "blue" and a "red" fitting box) of the CMD for each cluster. The blue box contains the MS of the cluster whereas the red box covers the regions of the evolved stars. The limits of all the boxes and the total number of stars in those boxes are summarized in Table \ref{tab:Fitting_boxes}. The faint limits of the boxes containing the MS are chosen such that they are at least 0.5 mag brighter than the 90\% completeness limits. 
We did several fits with different choices of the box limits. Thereby we noticed that the overall result does not depend on the exact choice of the boxes. 
To assess the statistical errors of the fittings that result from the stochastic IMF population of stars we performed additional bootstrapping tests. We created for each cluster 50 bootstrap samples and ran the SFH for each of these samples. The figures of the SFH fits presented in the next subsections show the mean values at individual ages (black dots) and the one sigma errorbars that follow from the bootstrapping procedure.  

The results of the SFH fitting will give us upper limits of potential age spreads as we do not take into account binaries and differential reddening. 

In the following we present the age and SFH fitting results of each cluster in order of decreasing age.
\subsection{NGC 2249}

\begin{figure*}
\includegraphics[width=19cm]{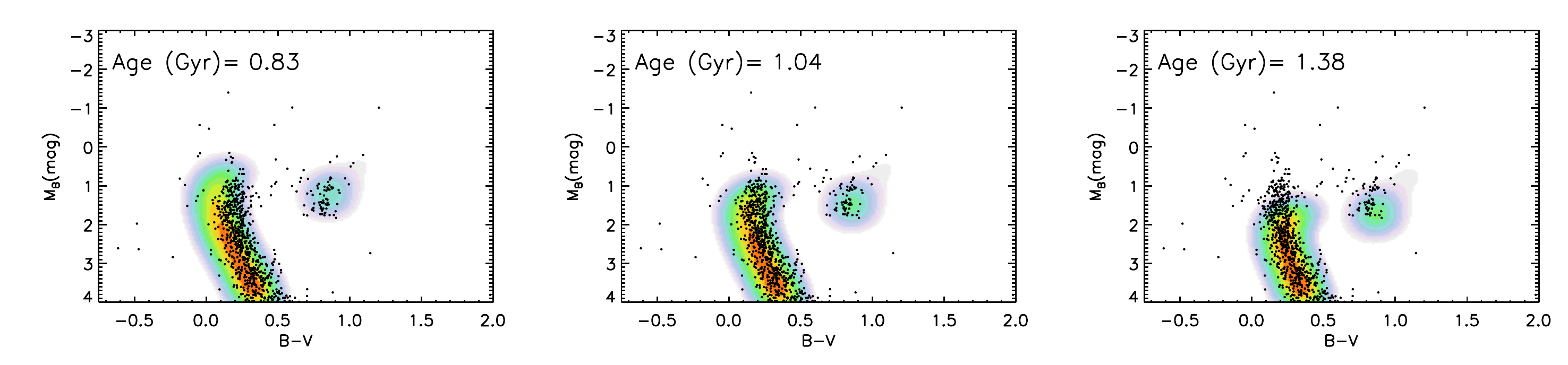}\\
\caption{CMD of NGC 2249 with overlaid model Hess diagrams at three different ages. The black dots are the individual observed stars and the filled (colored) contours are theoretical Hess diagrams. Their respective ages are indicated in the upper left corner of each panel.}
\label{fig:ngc2249_hess}
\end{figure*}

NGC 2249 is the oldest cluster in our sample. The literature age of the cluster is between 660 Myr \citep{Baumgardt13} and about 1 Gyr (e.g. \citealt{Kerber07}, \citealt{Correnti14}). It is therefore not a YMC any more but rather belongs to the category of intermediate age LMC clusters. Adopting an age of 1 Gyr, \citet{Correnti14} found a mass of 3.0$\cdot10^4\mathrm{M_{\sun}}$ for this cluster. We matched theoretical isochrones to NGC 2249 to determine its basic parameters for the further analysis. We found a reddening $E(B-V)$ of 0.02, a metallicity of Z=0.008 and a DM $(m-M)_0$ of 18.3 mag. These values are in good agreement with previous studies. \citet{Kerber07}, who also used the \citet{Brocato01} data set, modeled the CMDs of various clusters using Padova isochrones from \citet{Girardi02}. For NGC 2249 they find a value of 0.01 for the reddening, a metallicity Z of 0.007 and a DM of 18.27 mag. The recent study by \citet{Correnti14} that used deep HST photometry finds similar values fitting \citet{Marigo08} isochrones to the CMDs ($E(B-V)$ = 0.02, $(m-M)_0$ = 18.2 mag, Z = 0.006).

Figure \ref{fig:ngc2249_hess} shows the CMD of NGC 2249. The black dots indicate individual observed stars. The CMD shows two evident features: The MS of the cluster which extends up to a $B$ band magnitude of $\sim$0.8 and the red clump at $B \sim$1.3 mag and $B-V \sim$0.8. Overplotted are model Hess diagrams at three different ages from 830 Myr to 1.4 Gyr (filled contours). The position of the red clump in the CMD and the MSTO is best reproduced by an age of about 1 Gyr, in agreement with the findings by \citet{Kerber07} and \citet{Correnti14}. At an age of 830 Myr the MS extends to brighter magnitudes than it is observed and at 1.4 Gyr the models predict a MSTO that is fainter than observed.

\begin{figure}
\centering
\resizebox{\hsize}{!}
{\includegraphics{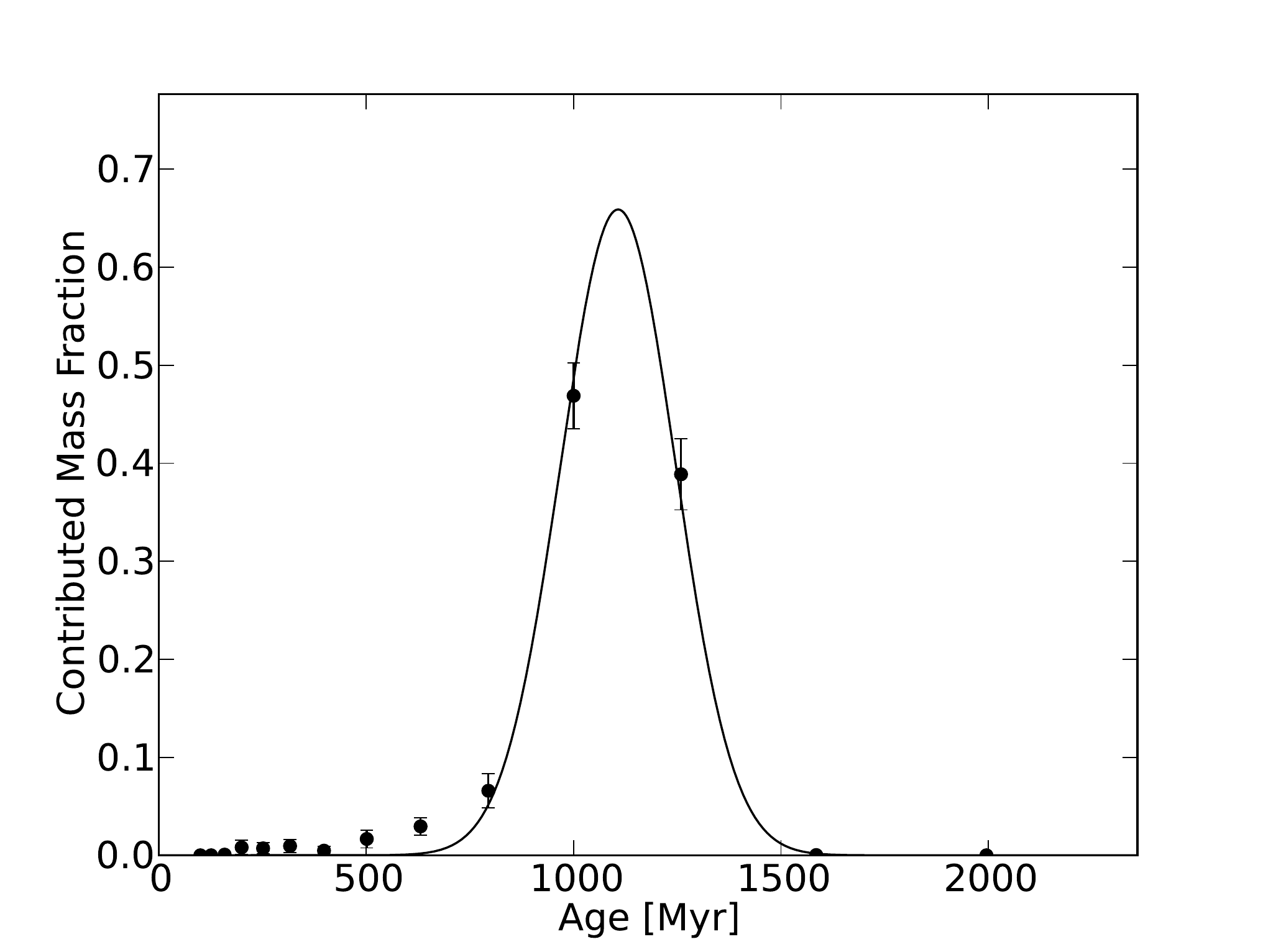}}
\caption{Results of the SFH fit of NGC 2249. The dots represent the contributed mass fraction at individual ages and the solid line shows the best Gaussian fit to the points with a peak at 1.11 Gyr and a standard deviation of 139 Myr.}
\label{fig:ngc2249_sfh_fit}
\end{figure}

For a more quantitative analysis of the cluster age we fit the SFH of the cluster using the \textit{FITSFH} code \citep{Silva-Villa10} providing the observed magnitudes and photometric errors of the stars as input.  We did the fit of the SFH in two regions in the CMD. The limits of the chosen boxes can be found in Table \ref{tab:Fitting_boxes}. The blue fitting box contains the MS down to a magnitude of 2.0  which is about half a magnitude above the 90\% completeness limit given by \citet{Brocato01}. The red fitting box covers the red clump stars. 
Figure \ref{fig:ngc2249_sfh_fit} shows the results of the SFH fit. The points are the mean contributed mass fraction of the fit at individual ages with one sigma errorbars that result from the bootstrapping. The solid line is the best fit Gaussian taking into account the errorbars. The curve peaks at an age of $\sim$ 1.11 Gyr. The dispersion of 139 Myr is a measure of the maximum age spread that is present in the cluster.

\begin{table*} 
\caption{Regions in the CMDs of the clusters where the SFH was fitted\label{tab:Fitting_boxes}}
\centering
\begin{tabular}{l c c c c c} 
\hline\hline
\noalign{\smallskip}
Cluster & Color limits of the  & Magnitude limits of the  & Color limits of the  & Magnitude limits of the  & Number of stars\\
&blue fitting box&blue fitting box&red fitting box&red fitting box & inside the boxes\\
\noalign{\smallskip}
\hline
\noalign{\smallskip}
NGC 1831 & $-0.1\leq (B-V)\leq 0.5$ &$-6.0\leq V\leq 2.0$ & $0.5\leq (B-V)\leq 1.4$& $-6.0\leq V\leq 1.5$ & 1015\\
NGC 1847 & $-0.4\leq (B-V)\leq 0.4$& $-6.0\leq V\leq 0.5$& $0.4\leq (B-V)\leq 1.5$& $-6.0\leq V\leq -2.5$ & 280\\
NGC 1850 & $-0.3\leq (B-V)\leq 0.1$& $-6.0\leq V\leq 0.0$& $0.1\leq (B-V)\leq 1.8$& $-6.0\leq V\leq -2.5$ & 995\\
NGC 2004 & $-0.4\leq (B-V)\leq 0.3$ & $-6.0\leq V\leq 0.5$ & $0.3\leq (B-V)\leq 1.7$ & $-6.0\leq V\leq -1.5$ & 410\\
NGC 2100 & $-0.4\leq (B-V)\leq 0.2$ & $-6.0\leq V\leq 0.0$ & $0.2\leq (B-V)\leq 1.9$ & $-6.0\leq V\leq -1.0$ & 375\\
NGC 2136 & $-0.3\leq (B-V)\leq 0.2$ & $-6.0\leq V\leq 0.5$ & $0.2\leq (B-V)\leq 1.7$ & $-6.0\leq V\leq -0.5$ & 360\\
NGC 2157 & $-0.3\leq (V-I)\leq 0.2$ & $-6.0\leq I\leq 1.5$ & $0.2\leq (V-I)\leq 1.9$ & $-6.0\leq I\leq -1.0$ & 1000\\
NGC 2249 & $0.0\leq (B-V)\leq 0.5$ & $-6.0\leq V\leq 2.0$ & $0.5\leq (B-V)\leq 1.7$ & $-6.0\leq V\leq 1.5$ & 390\\

\noalign{\smallskip}
\hline
\end{tabular}\\

\end{table*}

\subsection{NGC 1831}

\begin{figure*}
\centering
\includegraphics[width=19cm]{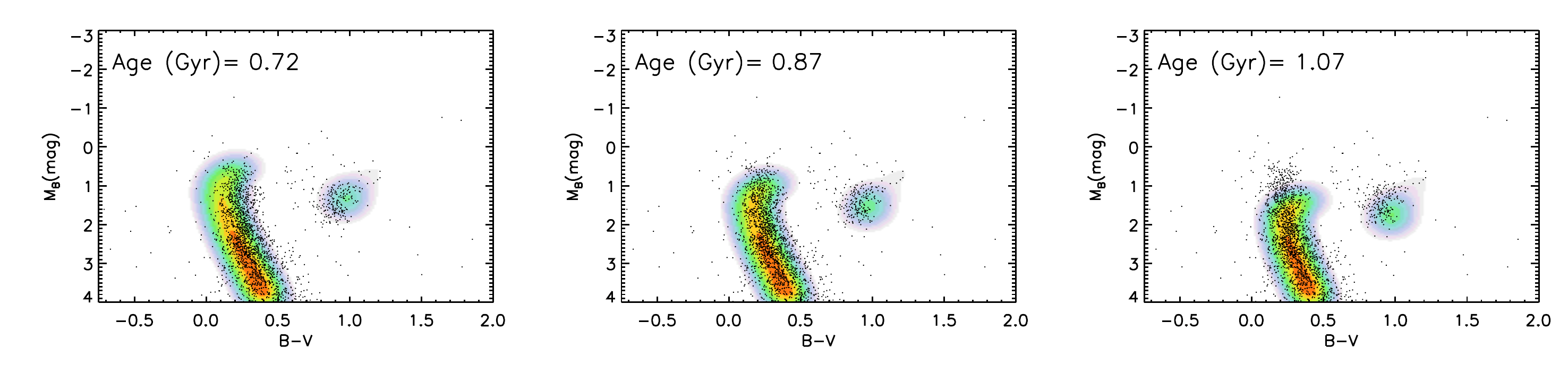}\\
\caption{CMD of NGC 1831 with overlaid model Hess diagrams at three different ages from 750 Myr to 1.04 Gyr.}
\label{fig:ngc1831_hess}
\end{figure*}

NGC 1831 is the second oldest cluster in our sample. \citet{Kerber07} found an age of 700 Myr for this cluster, a metallicity Z of 0.016, an extinction $E(B-V)$ of 0.01 and a DM of 18.23. \citet{Li14} analyzed the CMD of NGC 1831 and found that the MSTO is broader than expected. They conclude that this feature is best explained with stars of different rotation velocities and a dispersion of ages between 550 and 650 Myr. For their analysis they used the following parameters: Z=0.012, $E(B-V)$=0.03 and $(m-M)_0$=18.4 mag. We determined an extinction $E(B-V)$ of 0.0, a metallicity Z of 0.016 and a distance modulus of 18.1 mag that we used for the analysis of the SFH of NGC 1831. The CMD of the cluster with superimposed model Hess diagrams is shown in Figure \ref{fig:ngc1831_hess}. It is very similar to the CMD of NGC 2249 suggesting that both clusters have comparable ages. The MS of NGC 1831 extends up to $M_B \sim$0.5 mag. The position of the red clump is at a $B$ magnitude of about 1.2 and a $B-V$ color of $\sim$0.9. Looking at the Hess diagrams we see that the model at 870 Myr fits best the data.

\begin{figure}
\centering
\resizebox{\hsize}{!}
{\includegraphics{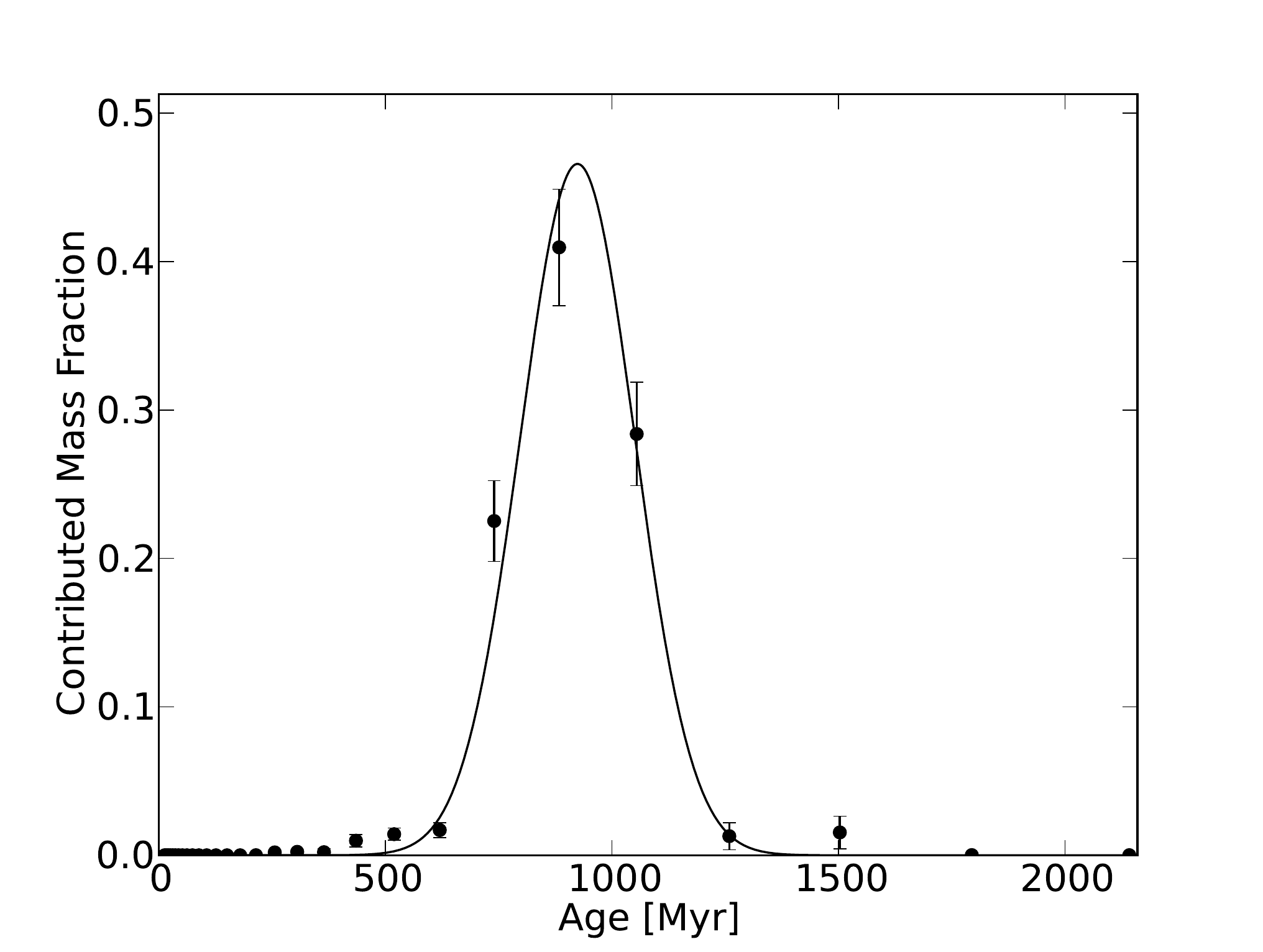}}
\caption{Results for the fitting of the SFH of NGC 1831. The dots represent the results at individual ages and the solid line shows the best Gaussian fit to the points with a peak at 924 Myr and a standard deviation of 126 Myr.}
\label{fig:ngc1831_sfh_fit}
\end{figure}

We fitted the SFH of the cluster in two regions in the CMD. The limits of the fitting boxes are given in Table \ref{tab:Fitting_boxes}. The boxes contain the MS as well as the He-burning stars. The results of the fitting are shown in Figure \ref{fig:ngc1831_sfh_fit}. A Gaussian fit to the individual points that represent the contributed mass fraction at single ages gives a peak at an age of 924 Myr. The standard deviation of the Gaussian (126 Myr) is the upper limit of the age spread as we do not consider other effects like binaries, rotating stars or differential extinction that might effect the CMD.

\subsection{NGC 2136\label{sec:ngc2136}}
\begin{figure*}
\centering
\includegraphics[width=19cm]{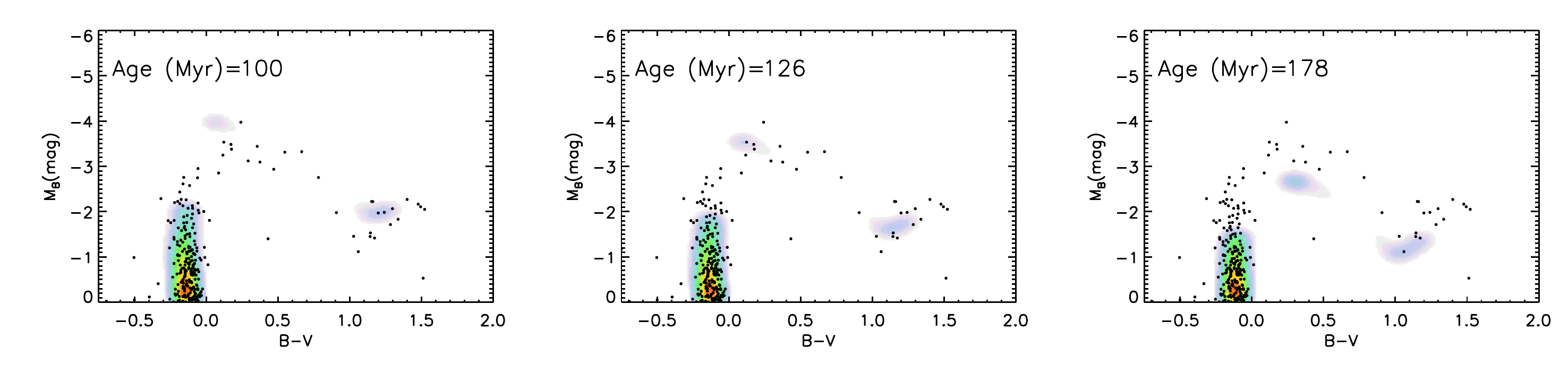}\\
\caption{CMD of NGC 2136 with overplotted theoretical Hess diagrams at 100, 126 and 178 Myr.}
\label{fig:ngc2136_hess}
\end{figure*}

NGC 2136 is younger than the previous two clusters. Its age from the literature is about 100 Myr (e.g \citealt{Dirsch2000}). By overplotting isochrones onto the cluster's CMD we find a reddening $E(B-V)$ of 0.13, which is in agreement with the value of 0.1$\pm$0.03 found by \citet{Dirsch2000}. Furthermore, we determined the metallicity Z to be 0.005 which is somewhat higher than the value of 0.004 by \citet{Dirsch2000} and a distance modulus of 18.5 mag that is consistent with the mean distance to the LMC \citep{deGrijs14}.
Figure \ref{fig:ngc2136_hess} shows the CMD of NGC 2136 overlaid with Hess diagrams at ages of 100, 125 and 177 Myr. The top of the MS in this cluster is at $B\sim -2$ followed by an almost continuous sequence of already evolved stars which reaches a top brightness of $M_{B}\sim -4$ and extends to a $B-V$ color of $\sim$1.5. We note that there are stars in the so called "Blue-Hertzsprung-Gap" between the MS and the subgiant branch where no stars are predicted by the models. These stars could be stars in the dense cluster center that are affected by crowding, fast rotating stars, interactive binaries or higher mass stars that formed out of the merging or collision of two low mass stars ("blue stragglers"). The blue loop region in the CMD that is populated by He-burning stars is well sampled in this cluster which makes it easier to constrain a possible age spread just by looking at the theoretical Hess diagrams. We see that an age of 125 Myr fits best the position of the evolved stars, especially the red side of the blue loop (Figure \ref{fig:ngc2136_hess} middle). Also an age of 100 Myr is compatible with the data, although the model seems to be a bit too bright (Figure \ref{fig:ngc2136_hess} left). On the other hand, an age of $\sim$180 Myr is too old for NGC 2136 (Figure \ref{fig:ngc2136_hess}). The theoretical Hess diagram fails to fit the blue loop stars as well as the position of the MS turn off. From this we can already rule out a possible age spread of more than $\pm$50 Myr.

\begin{figure}
\centering
\resizebox{\hsize}{!}
{\includegraphics{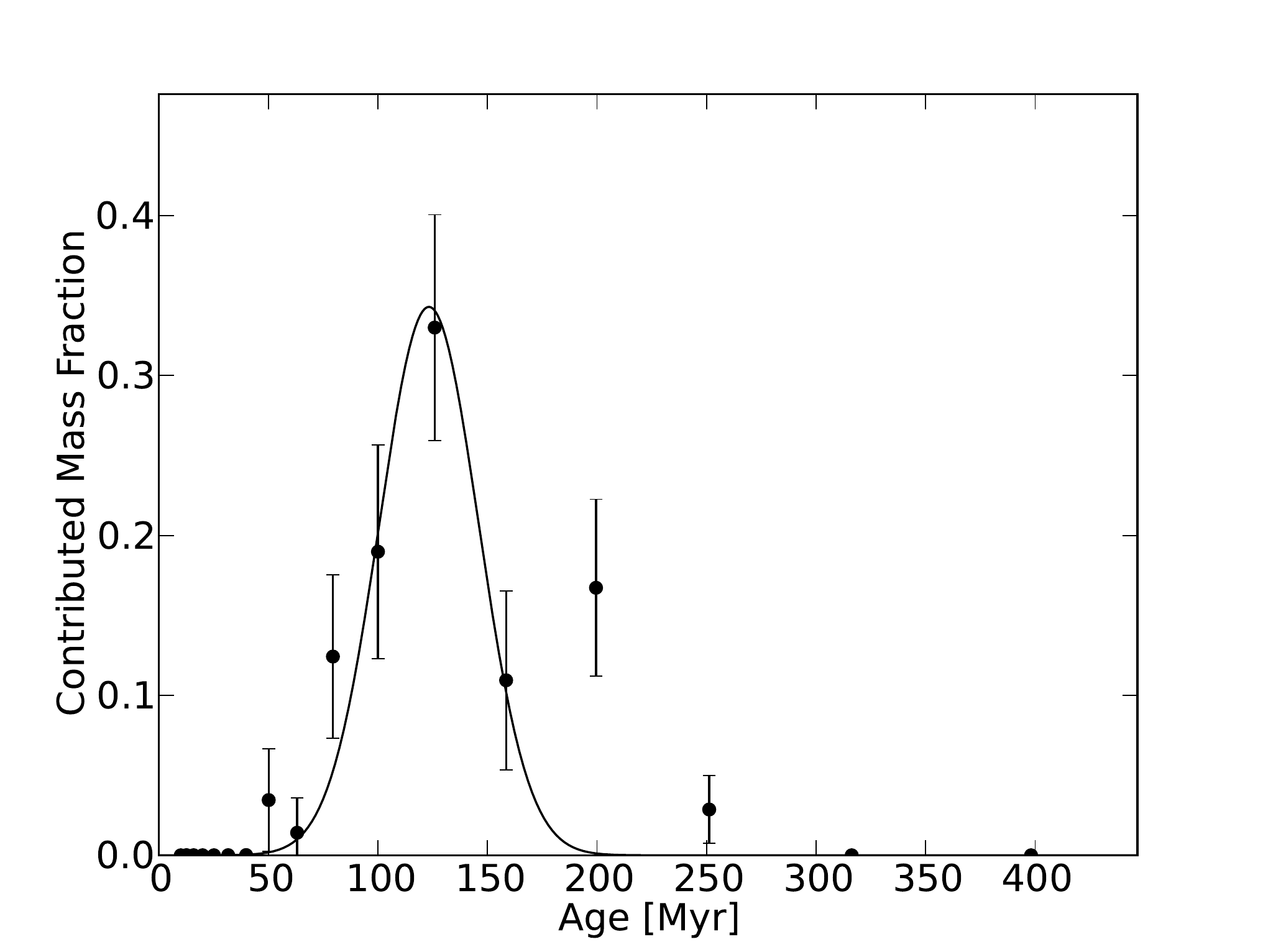}}
\caption{Results of the SFH fit of NGC 2136. The dots represent the results at individual ages and the solid line shows the best Gaussian fit to the points with a peak at 123.3 Myr and a standard deviation of 22.6 Myr.}
\label{fig:ngc2136_sfh_fit}
\end{figure}

\begin{figure}
\centering
\includegraphics[width=6.5cm]{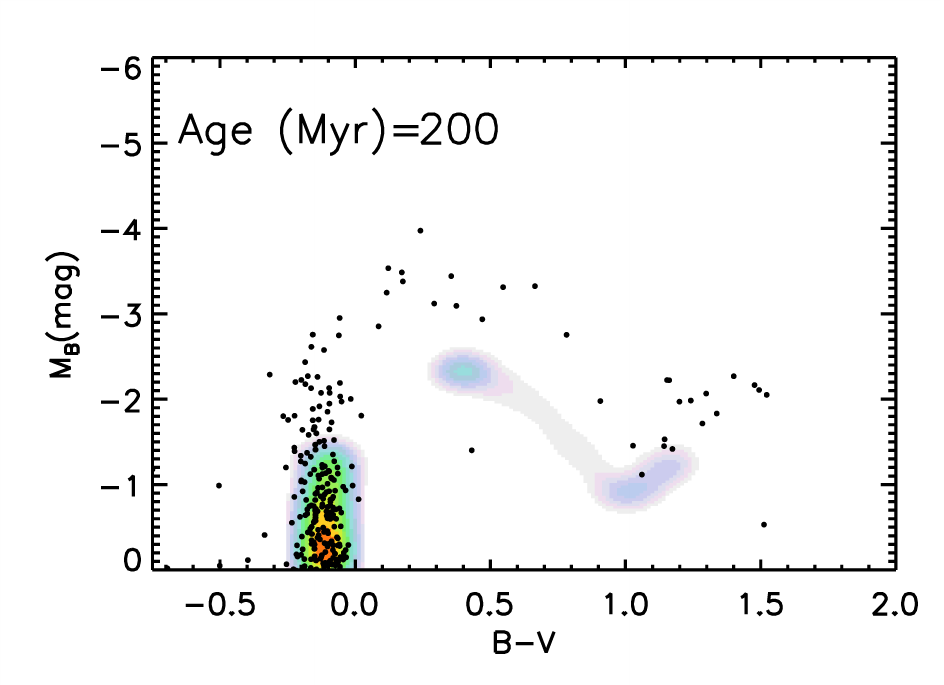}
\caption{CMD of NGC 2136 with an overplotted theoretical Hess diagram at 200 Myr.}
\label{fig:ngc2136_200myr}
\end{figure}

To put our initial estimate on a more quantitative basis we fitted the SFH using \textit{FITSFH}, as we did in the previous clusters. We adopted two fitting boxes that were chosen such that they include the main features of the cluster CMD but exclude the leftover contamination of faint red field stars (compare Figure \ref{fig:Cluster_cmds}). The faint limit of the blue fitting box is one magnitude above the 90\% completion limit given by \citet{Brocato01}. The fitting of the SFH of NGC 2136 gives a period of star formation which peaks at an age of $\sim$123 Myr with an upper limit of $\pm$23 Myr for its duration (see Figure \ref{fig:ngc2136_sfh_fit}), comparable with our first estimate. 
We also note, that there is an additional significant peak at 200 Myr. Figure \ref{fig:ngc2136_200myr} shows a Hess diagram at 200 Myr superimposed over the CMD of NGC 2136. The fitted star formation at 200 Myr is due to the faintest stars at the red end of the blue loop at $B-V \sim$1.0. But at this age we would also expect a higher density of blue loop stars at a $B-V$ color of about 0.4 and a $B$ magnitude of about $-$2.5 where no stars are observed. Therefore, we can rule out the peak at 200 Myr.

\subsection{NGC 2157}

\begin{figure*}
\centering
\includegraphics[width=19cm]{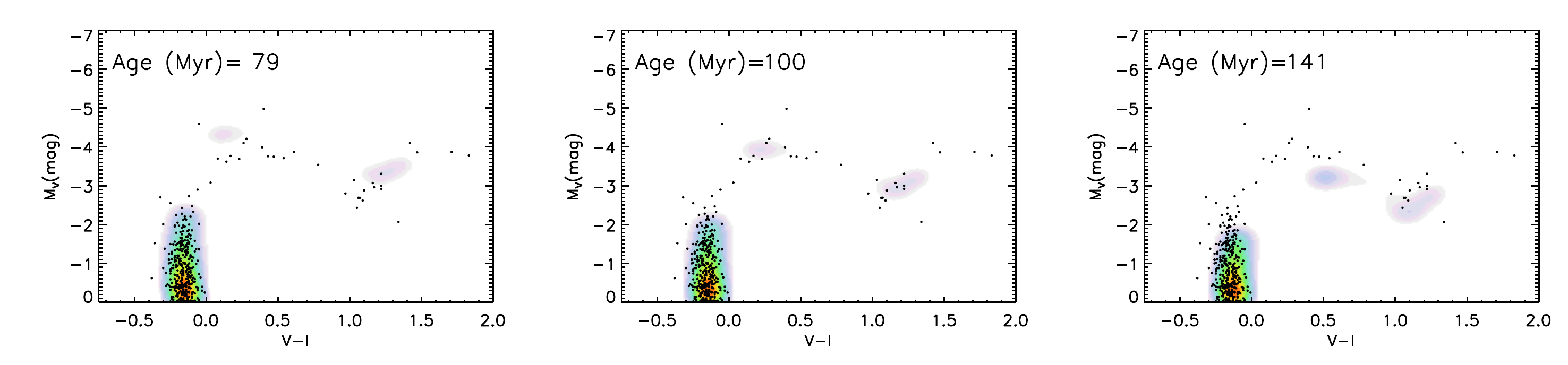}\\
\caption{CMD of NGC 2157 with overlaid Hess diagrams at the ages 80 Myr, 100 Myr and 140 Myr. Note that this CMD, in contrast to the other cluster CMDs, is in the $(V-I)$ vs. $M_V$ space.}
\label{fig:ngc2157_hess}
\end{figure*}

NGC 2157 is not part of the \citet{Brocato01} data set and its data is taken from \citet{Fischer98} who provides photometry in the $V$ and $I$ filters. The photometry is already corrected for extinction by \citet{Fischer98} who adopted a reddening value $E(B-V)$ of 0.1. We derived the same value and therefore did not change it for our analysis. By plotting theoretical isochrones over the observed CMD of the cluster they determined an age of 100 Myr for the cluster. We estimated a metallicity Z of 0.008 and a distance modulus of 18.5 mag for the cluster, consistent with the values assumed for the cluster by \citet{Fischer98}. Figure \ref{fig:ngc2157_hess} shows the CMD of the cluster together with theoretical Hess diagrams overplotted. As NGC 2136, which has a similar age, the CMD of NGC 2157 shows a clear MS turn off at about $M_V=-3$ and a well populated blue loop with a slight over-density of stars at $V-I \sim 1.2$ that marks the red envelope of the loop. In agreement with \citet{Fischer98} we note that an age of 100 Myr fits best the observed stellar distribution in the CMD (Figure \ref{fig:ngc2157_hess} middle). 
From the CMD we can constrain an age spread to be less than about $\pm$30 Myr.

\begin{figure}
\centering
\resizebox{\hsize}{!}
{\includegraphics{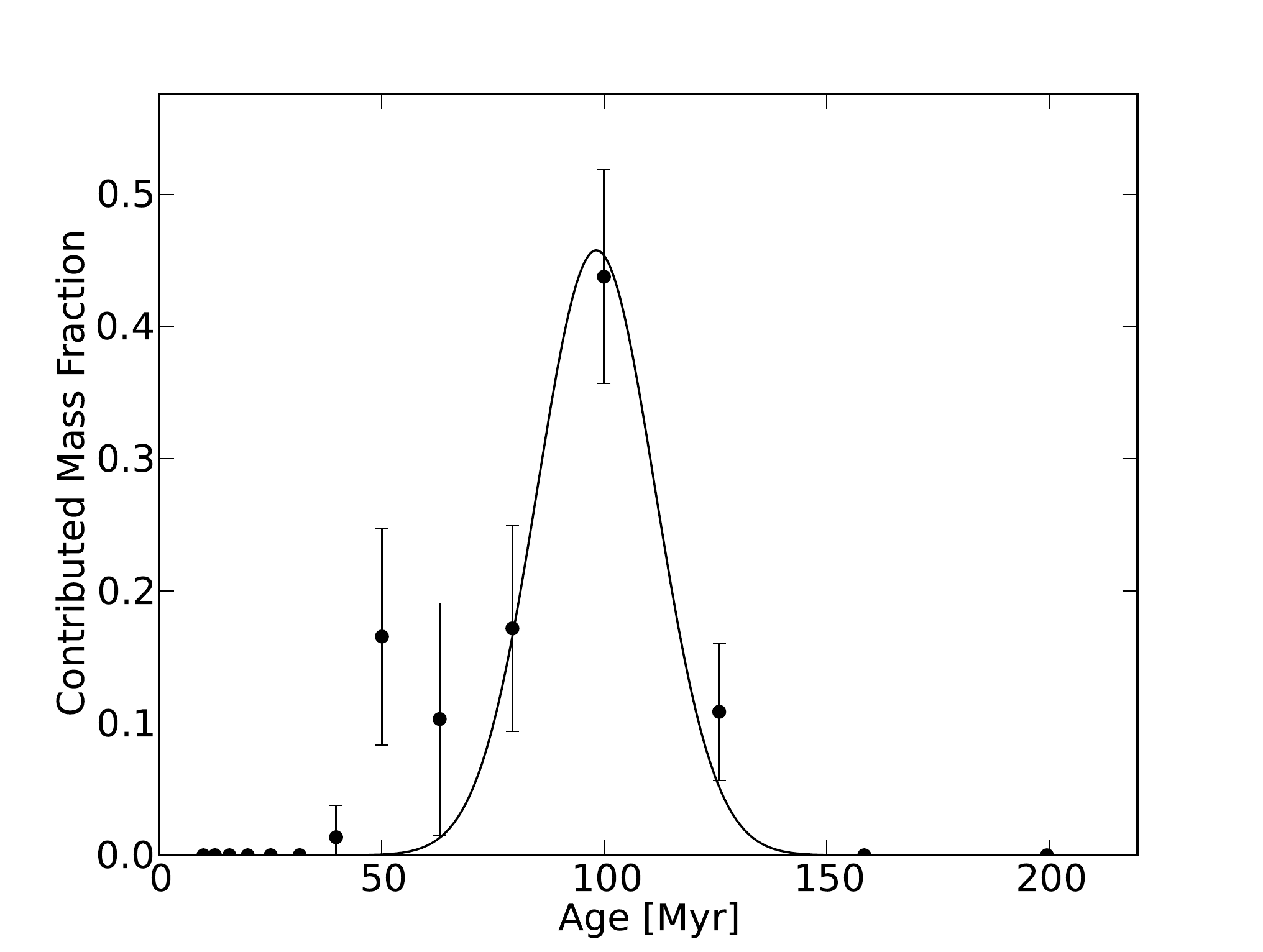}}
\caption{Results of the SFH fit of NGC 2157. The dots represent the results at individual ages and the solid line shows the best Gaussian fit to the points with a peak at 98.3 Myr and a standard deviation of 13.2 Myr.}
\label{fig:ngc2157_sfh_fit}
\end{figure}

We chose the regions in the CMD that we used to fit the SFH of the cluster such that they contain the main features of the cluster plus some additional space below and above the sequence of evolved stars to better constrain younger and older ages (see Table \ref{tab:Fitting_boxes} for the exact limits). The results of the fitting procedure are displayed in Figure \ref{fig:ngc2157_sfh_fit}. As expected, the star formation has a clear maximum at about 100 Myr with a upper limit of an age spread of $\pm$13.2 Myr. Additionally, the fit yields a high value of star formation at 50 Myr with a 1.5$\sigma$ significance. This age corresponds to the stars that are at the top of the MS at $M_V=-3$. However, the position of the evolved stars at 50 Myr do not agree with the observations. Therefore this age can be excluded.

\subsection{NGC 1850\label{sec:ngc1850}}

\begin{figure*}
\centering
\begin{tabular}{l}
\includegraphics[width=19cm]{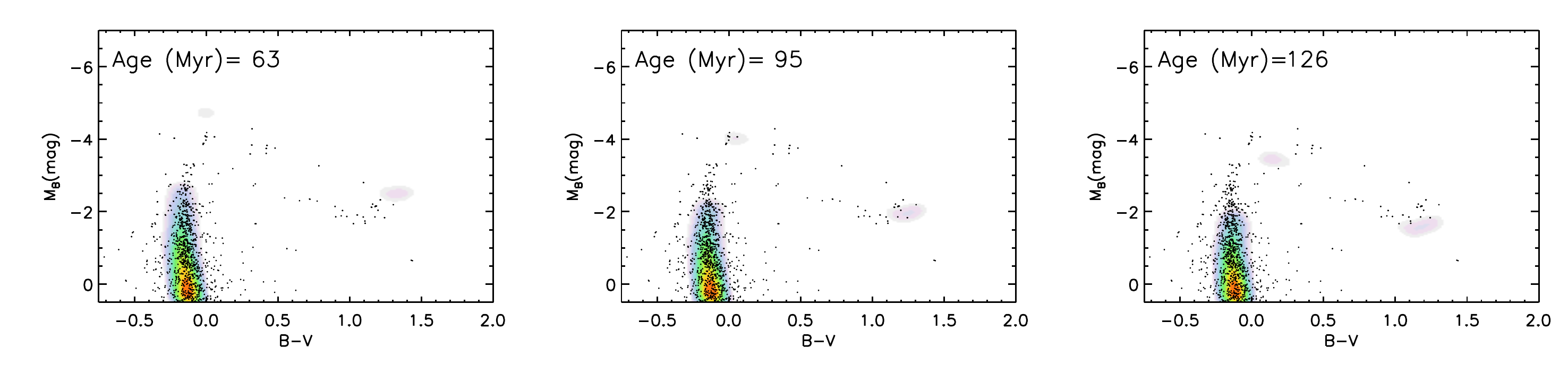}\\
\includegraphics[width=19cm]{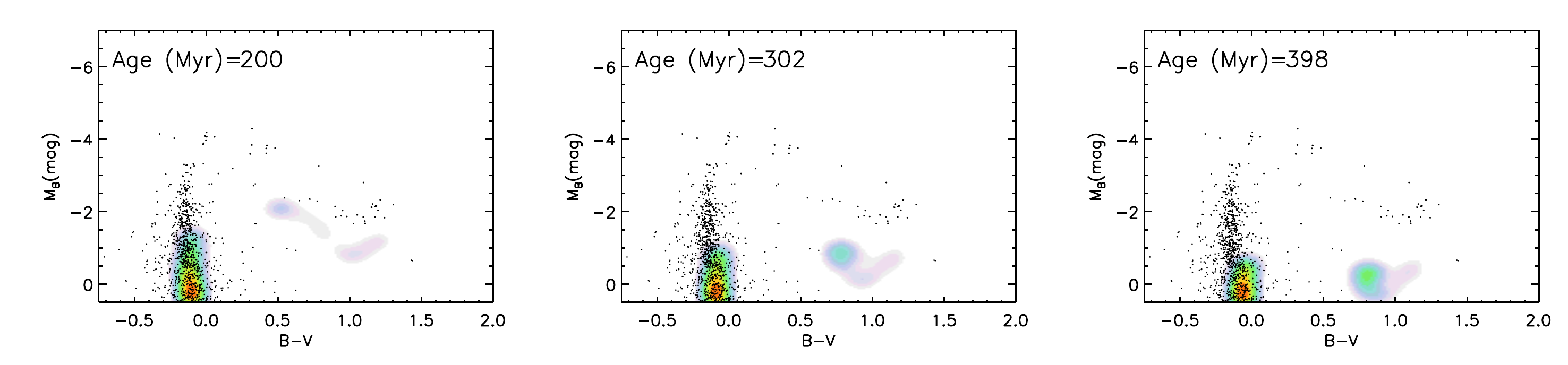}
\end{tabular}
\caption{CMD of NGC 1850 with superimposed model Hess diagrams at ages from 60 to 400 Myr.}
\label{fig:ngc1850_hess}
\end{figure*}

NGC 1850 is a binary cluster system consisting of a main cluster (NGC 1850A) and a small and dense secondary cluster (NGC 1850B) composed mainly of young O and B stars. This cluster is the only one for which we obtained the data from the HST Legacy Archive. The data catalog consists of the spatial positions of the stars on the detector chip, the RA and Dec coordinates and the AB magnitudes in the F450W and F555W filter, however without photometric errors. We converted the magnitudes to Vega magnitudes and afterwards to the Johnson $BV$ system and created artificial photometric errors that follow the same behavior as the observed ones in the \citet{Brocato01} data set of the other clusters. In our study we consider only the central parts of the main cluster NGC 1850A (all stars within two times the core radius). The young stars of the secondary cluster are outside this area and therefore they will not affect our analysis. We determined a reddening $E(B-V)$ of 0.1 which is smaller than the value of 0.17 that is assumed by \citet{Fischer93}. Additionally, a metallicity of Z=0.006 and a distance modulus of 18.5 mag reproduces best the position and extent of the blue loop. The distance modulus is a bit smaller than 18.6 mag that was determined by \citet{Sebo95} by measurements of Cepheids.
The resulting CMD of NGC 1850A is shown in Figure \ref{fig:ngc1850_hess}. It shows a MS that extends up to a $B$ magnitude of about -3.5 and a track of evolved blue loop stars that goes to a $B-V$ color of about 1.3. 
Comparing the observed CMD with the model we see that for an age of about 90 Myr the best accordance is found. The magnitude and the morphology of the evolved stars matches the observations best at this age. We can exclude ages older than 125~Myr and younger than about 70~Myr. We note that also in this cluster the MS extends to brighter magnitudes as would be expected from the models.

\begin{figure}  
   \resizebox{\hsize}{!}
{\includegraphics{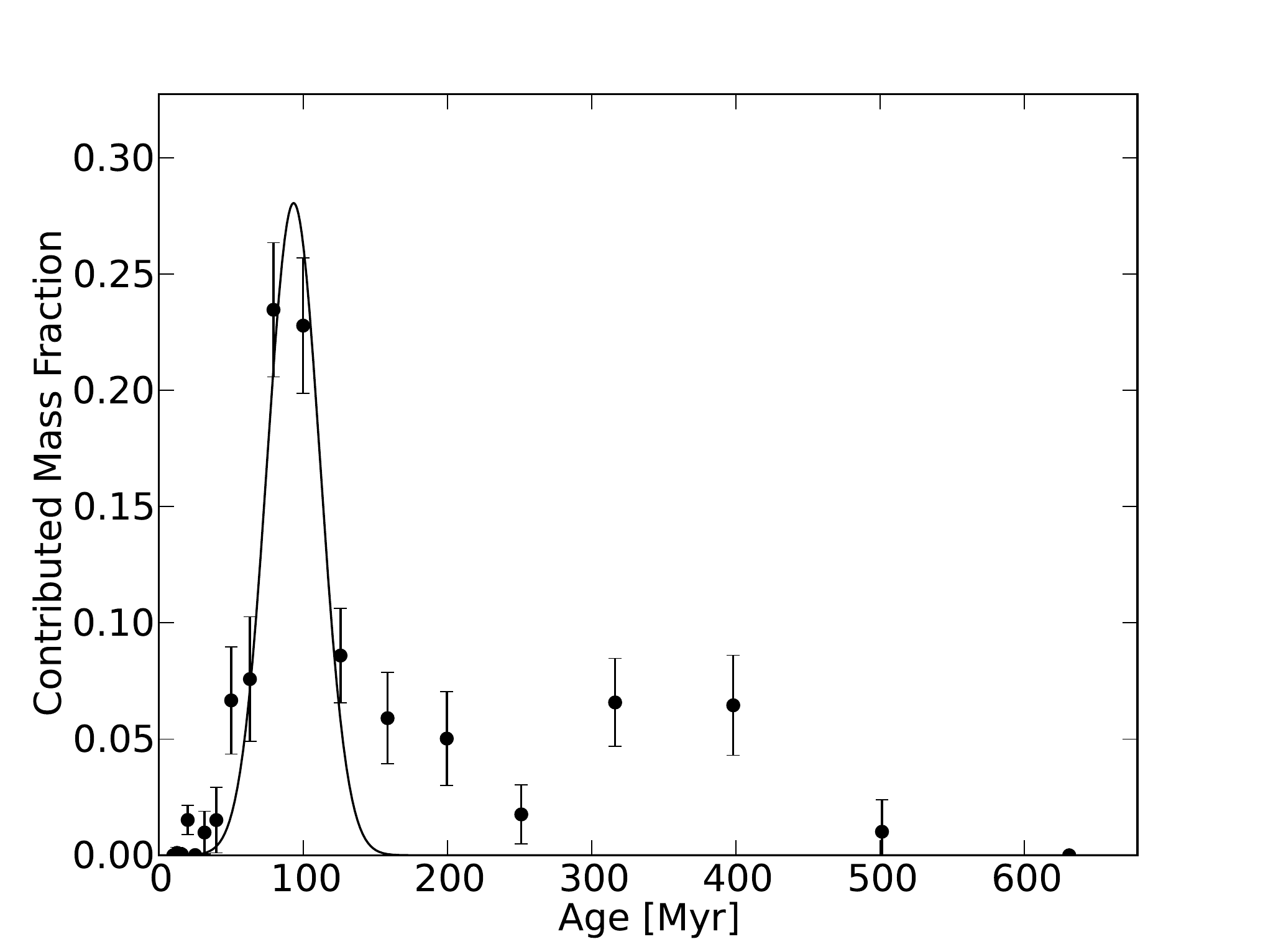}}
   \caption{Results for the fitting of the SFH of NGC 1850. The dots represent the results at individual ages and the solid line shows the best Gaussian fit to the points with a peak at 93.4 Myr and a standard deviation of 18.3 Myr.}
              \label{fig:ngc1850_sfh_fit}
    \end{figure}

\begin{figure}  
   \resizebox{\hsize}{!}{\includegraphics{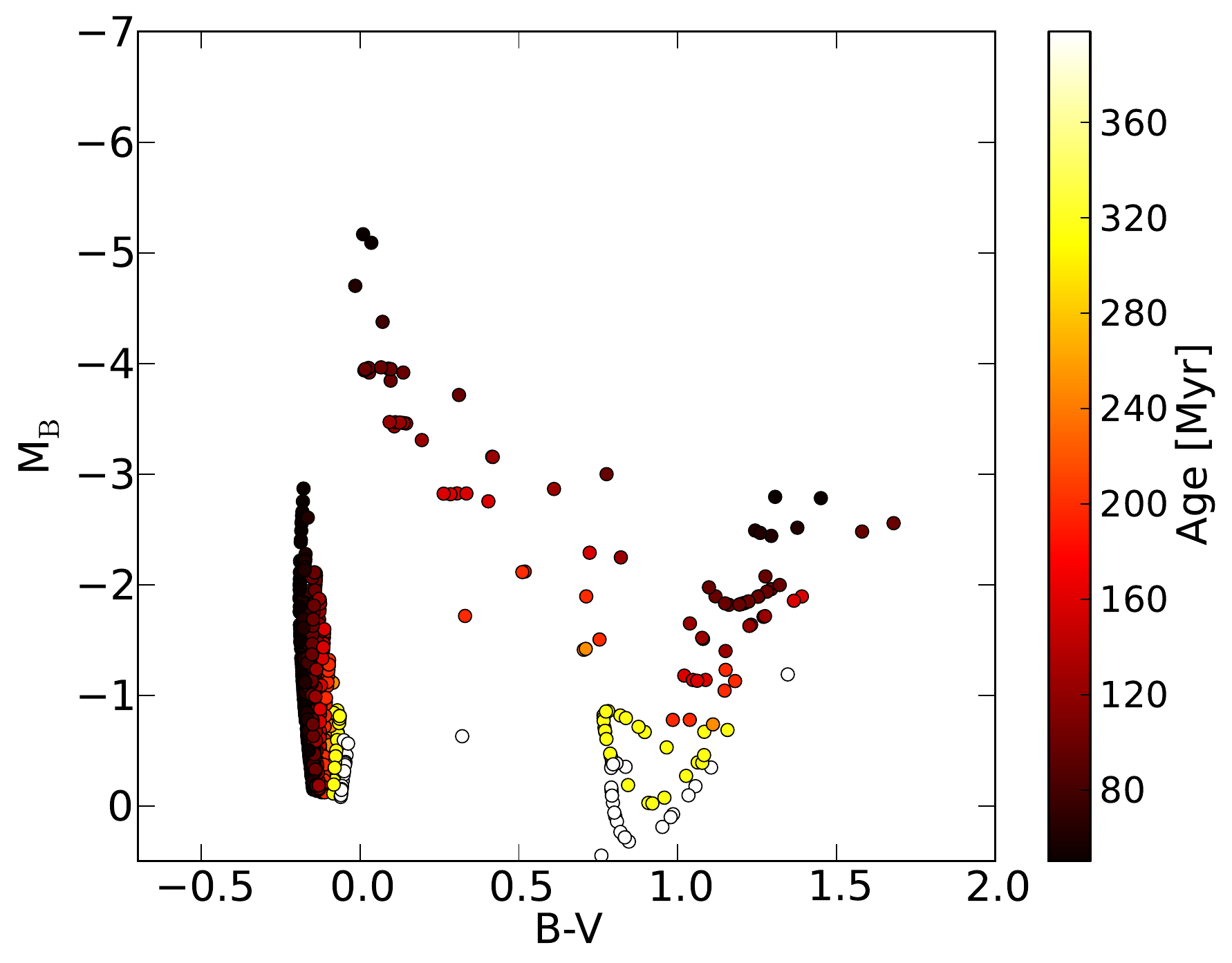}}
   \caption{CMD of an artificial cluster created using the fitted SFH of NGC 1850 normalized to the number of stars present in the used data set of NGC 1850 that are brighter than 0.0 mag in the $V$ band. The stars are color-coded by their ages going from black (the youngest) to white (the oldest stars).}
              \label{fig:art_sfh_ngc1850}
    \end{figure}

We fitted the SFH in two regions in the CMD that contain the MS and the post-MS track (see Table \ref{tab:Fitting_boxes}). As already mentioned at the end of Section \ref{sec:obs}, we made a conservative estimate of the completeness limit (18.5 mag in the $V$ band) of NGC 1850 and chose the limits of the fitting boxes according to that. The results of the fitting is displayed in Figure \ref{fig:ngc1850_sfh_fit}. We find a distribution that peaks at an age of about 93 Myr with a maximum spread of 18 Myr. This is in agreement with the result by \citet{Fischer93} who found an age of 90$\pm$30 Myr. However, besides the main peak at 93 Myr the fitting of the SFH gives a considerable amount of stars that formed at ages between 200 and 400 Myr. The lower three panels of Figure \ref{fig:ngc1850_hess} show the CMD of NGC 1850 with superimposed Hess diagrams at ages of 200, 300 and 400 Myr. At these older ages there are no evolved stars observed where the models would predict them. As an additional test to rule out the fitted ages of a few hundreds of Myrs, we created an artificial cluster with the fitted SFH of NGC 1850, 
normalized to the total number of stars in NGC 1850 brighter than an absolute magnitude of 0.0 in the $V$ band. We chose this limit to be sufficiently bright in order to avoid any effects of incompleteness.
The result is shown in Figure \ref{fig:art_sfh_ngc1850}. In this figure, the stars are color-coded by their ages, whereas the youngest stars are black and the oldest are white. From the synthetic cluster we expect stars to be between 0.7 and 1.2 in $B-V$ color and between $-2$ and 0.5 in $B$ magnitude. However, there are no stars found in this region of the CMD of NGC 1850. We can therefore rule out stars that have ages older than 200 Myr. 
An alternative explanation of these older ages could be contributions from field stars as NGC 1850 has a high level of field star contamination (see Figure \ref{fig:Cluster_cmds}).

The age that we found for NGC 1850A is significantly higher than the widely used value of 30 Myr in the literature (e.g. \citealt{Elson91, McLaughlin05, Baumgardt13}). Adopting this new age of about 100 Myr results in a mass that is twice as large as the one that is estimated from an age of 30 Myr. So, the new mass of NGC 1850 is 1.4$\cdot 10^5\mathrm{M_{\sun}}$ instead of 7.2$\cdot 10^4\mathrm{M_{\sun}}$. 
This new mass is consistent with the expected mass-to-light ratio for a ~100 Myr population \citep{McLaughlin05}, which further supports our result.

\subsection{NGC 1847}

NGC 1847 is one of the younger clusters in our sample. \citet{ElsonFall88} dated this cluster to an age of 26 Myr. We find a reddening $E(B-V)$ of 0.16 which is somewhat higher than the value by \citet{Nelson83} who adopted  $E(B-V)=$ 0.1 for their studies and a metallicity Z of 0.006, in agreement with \citet{Mackey03}.

According to \citet{McLaughlin05} NGC 1847 has an effective radius of more than 30 pc (assuming \citealt{King66} profiles). This is more than 4 times as large as it would be expected from its mass assuming it follows the same $R_{\rm eff}$-log(M) relation as the other young and intermediate age LMC clusters (see Figure \ref{fig:Mass_vs_Reff}). However, NGC 1847 has a very shallow density profile (e.g. \citealt{Mackey03}) and therefore the value of $R_{\rm eff}$ is afflicted with a large uncertainty.

Figure \ref{fig:ngc1847_cmd} shows the CMD of the central part of NGC 1847. The black dots are the cluster stars used for the analysis and the red crosses are the stars subtracted as field stars. Beneath the cluster population there is a second numerous population of stars at older ages (about 800 Myr, corresponding to the position of the red clump) that is not present in such an extent in the other CMDs. This population is most likely due to the background LMC stars, as NGC 1847 is close to the LMC bar. 

\begin{figure*}
\centering
\begin{tabular}{l}
\includegraphics[width=19cm]{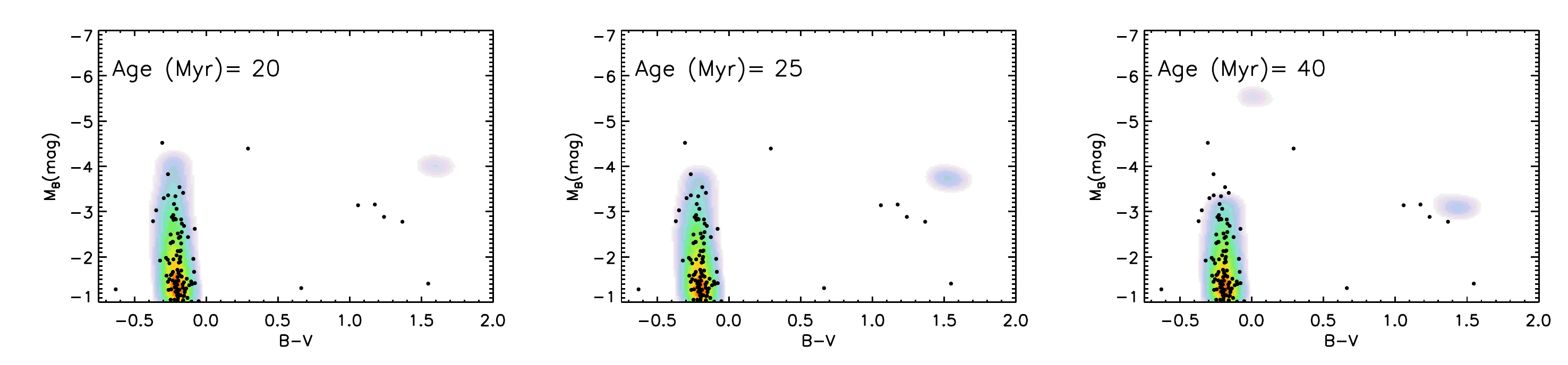}\\
\includegraphics[width=19cm]{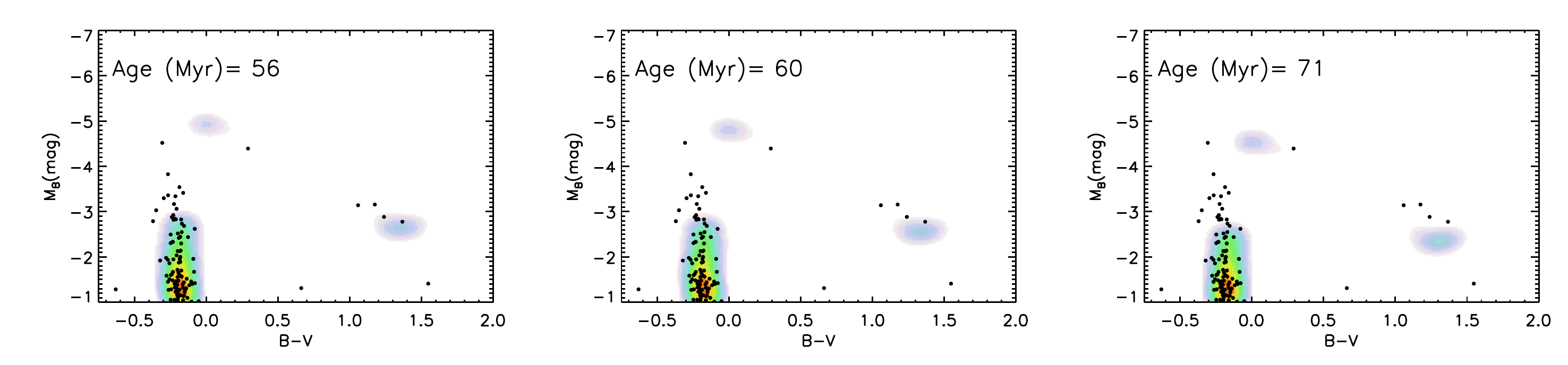}
\end{tabular}
\caption{CMD of NGC 1847 with overlaid model Hess diagrams at six different ages ranging from 20 to 70 Myr.}
\label{fig:ngc1847_hess}
\end{figure*}

Figure \ref{fig:ngc1847_hess} shows model Hess diagrams at different ages overlaid over the CMD of NGC 1847. An age of about 56 Myr reproduces best the position of the He-burning stars. On the one hand, there are no evolved stars that would be suggestive of ages older than 70 Myr. On the other hand, at 56 Myr the observed MS extends beyond the model by more than one magnitude. Only ages of about 20 to 25 Myr can account for these stars. By just comparing the CMD with the models we can restrict the ages to be between 20 and 70 Myr.

\begin{figure}
\centering
\resizebox{\hsize}{!}
{\includegraphics{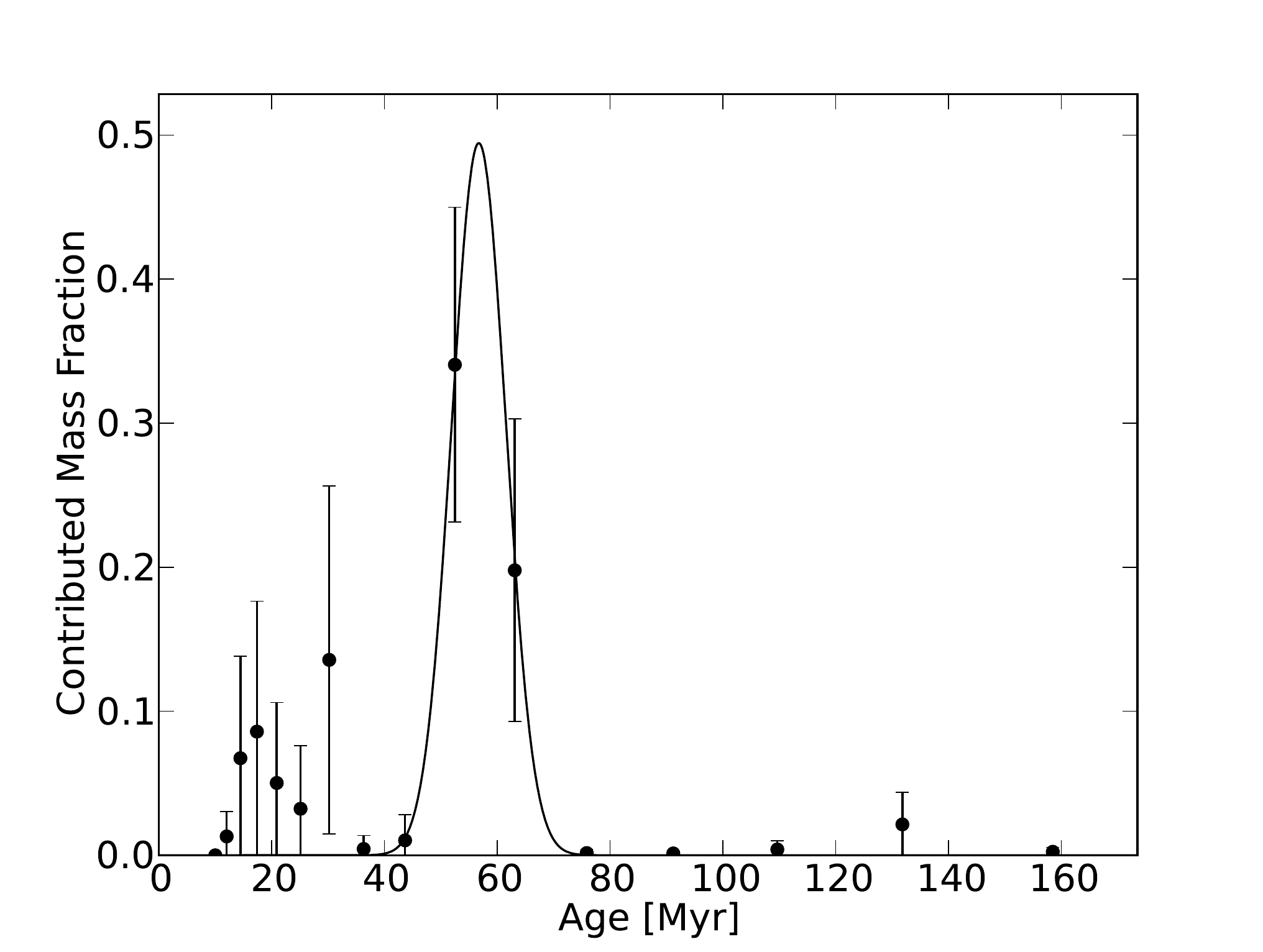}}
\caption{Results for the fitting of the SFH of NGC 1847. The dots represent the results at individual ages. The solid line show the best Gaussian fits to the points with a peak 56.7 Myr and a standard deviation of 4.8 Myr. The irregular star formation rate between ages of 14 and 30 Myr is due to the MS stars that are brighter than would be expected for a 57 Myr old population.}
\label{fig:ngc1847_sfh_fit}
\end{figure}

\begin{figure}
\centering
\resizebox{\hsize}{!}{\includegraphics{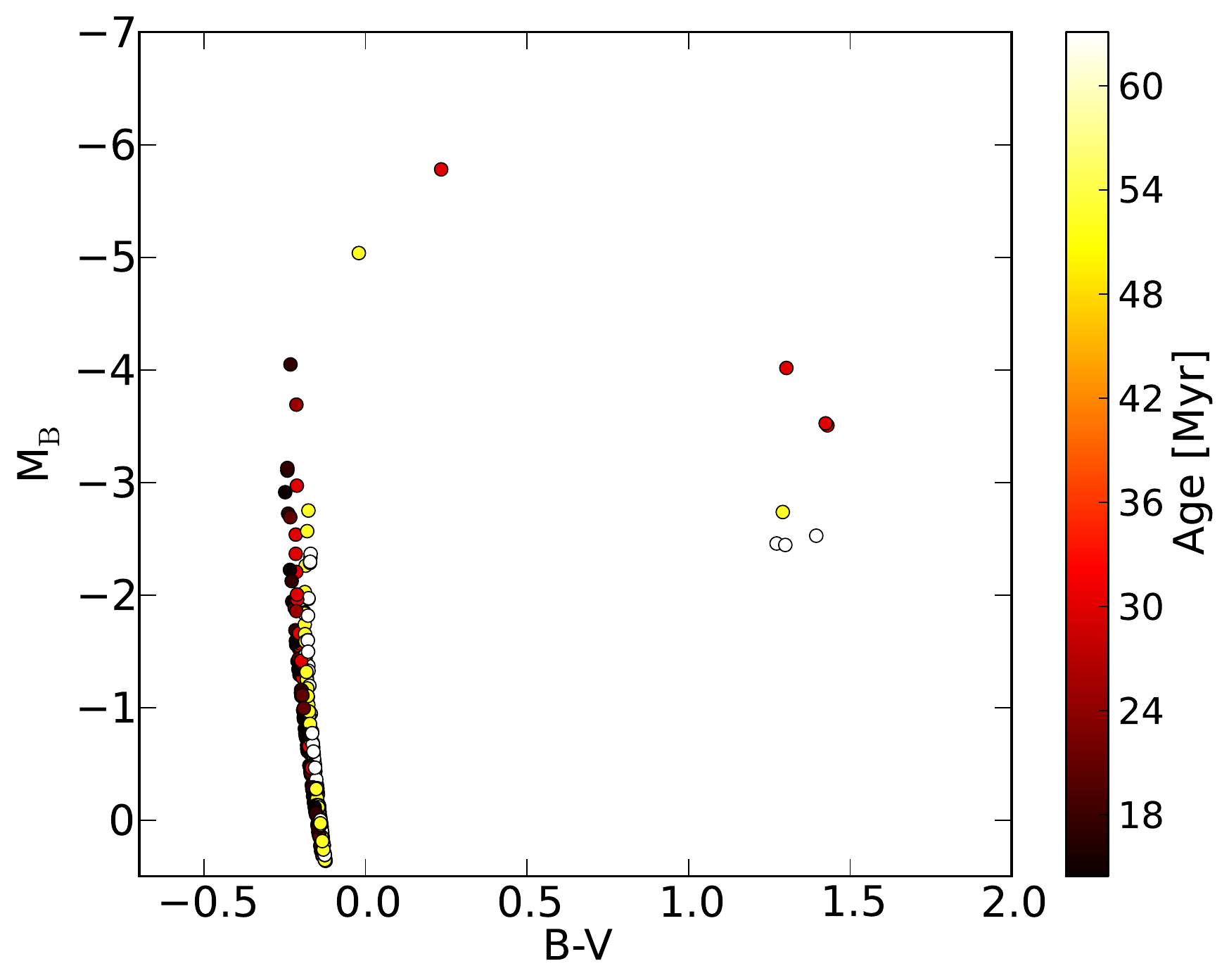}}
\caption{CMD of an artificial cluster created using the fitted SFH of NGC 1847 normalized to the number of stars in NGC 1847 that have a $V$ band magnitude brighter than 0.5 mag. All the stars are color-coded by their age.}
\label{fig:ngc1847_art_cluster_sfh}
\end{figure}

\begin{figure}
\centering
\begin{tabular}{ll}
\includegraphics[width=4.5cm]{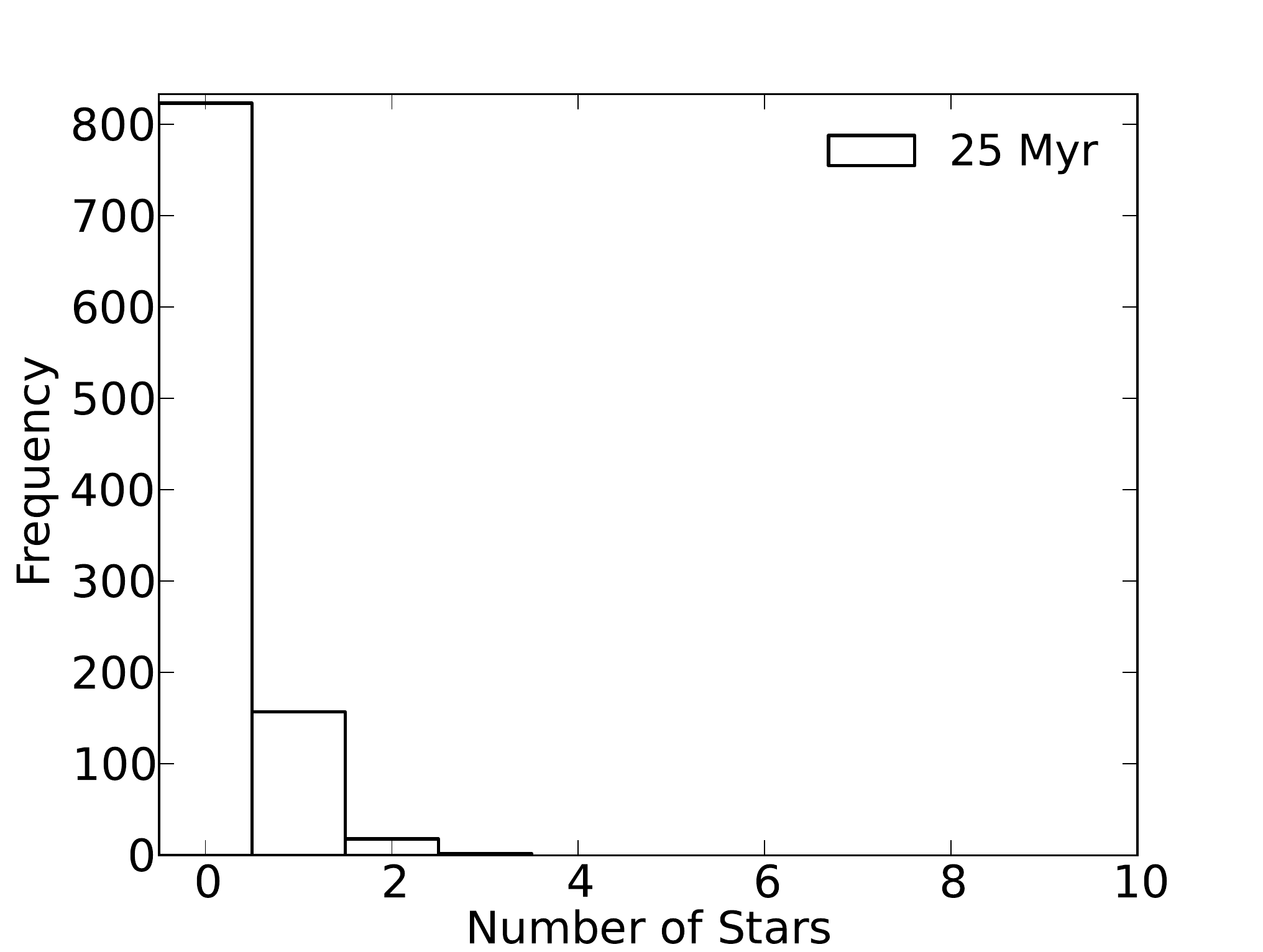}
\includegraphics[width=4.5cm]{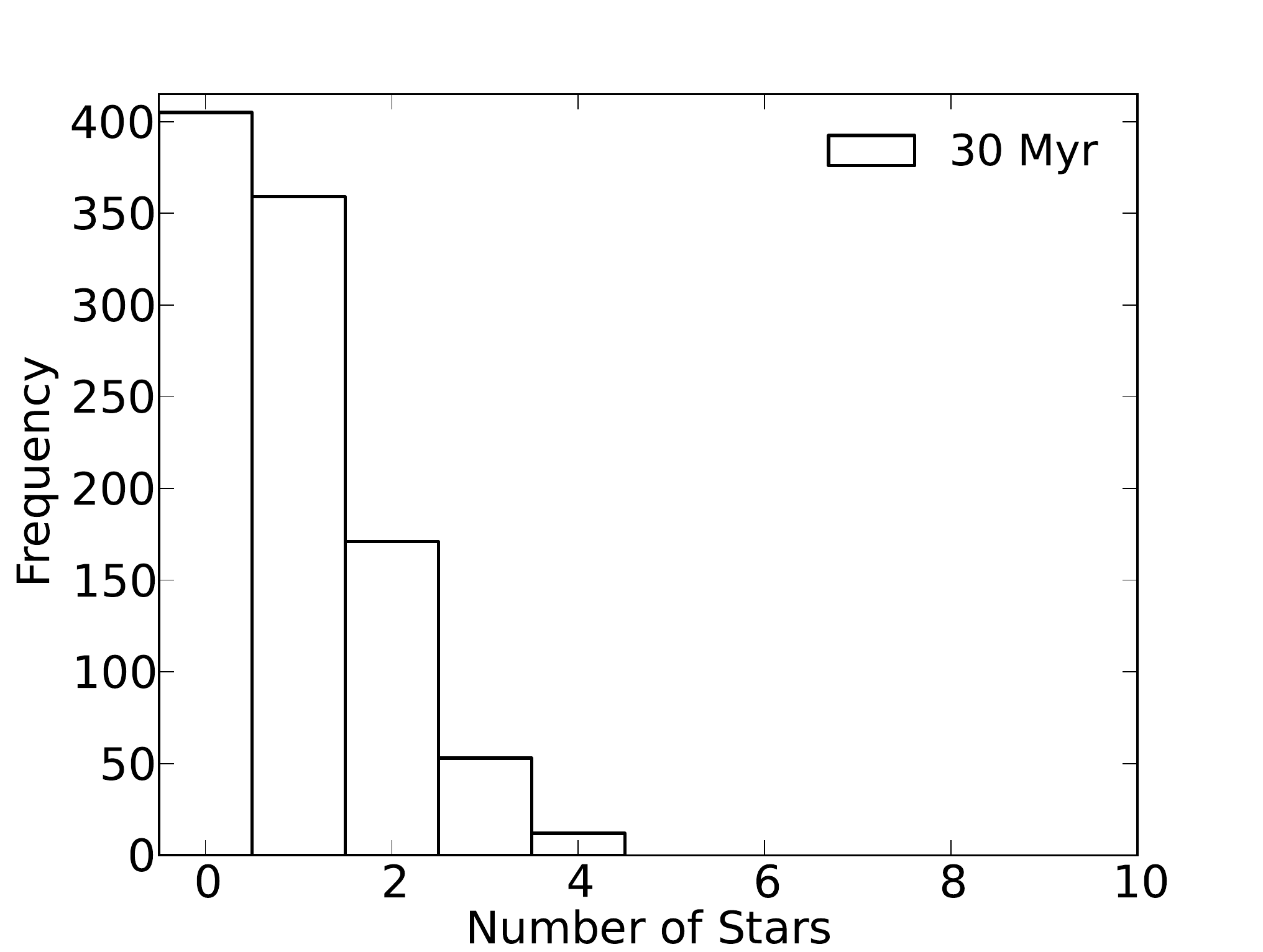}\\
\includegraphics[width=4.5cm]{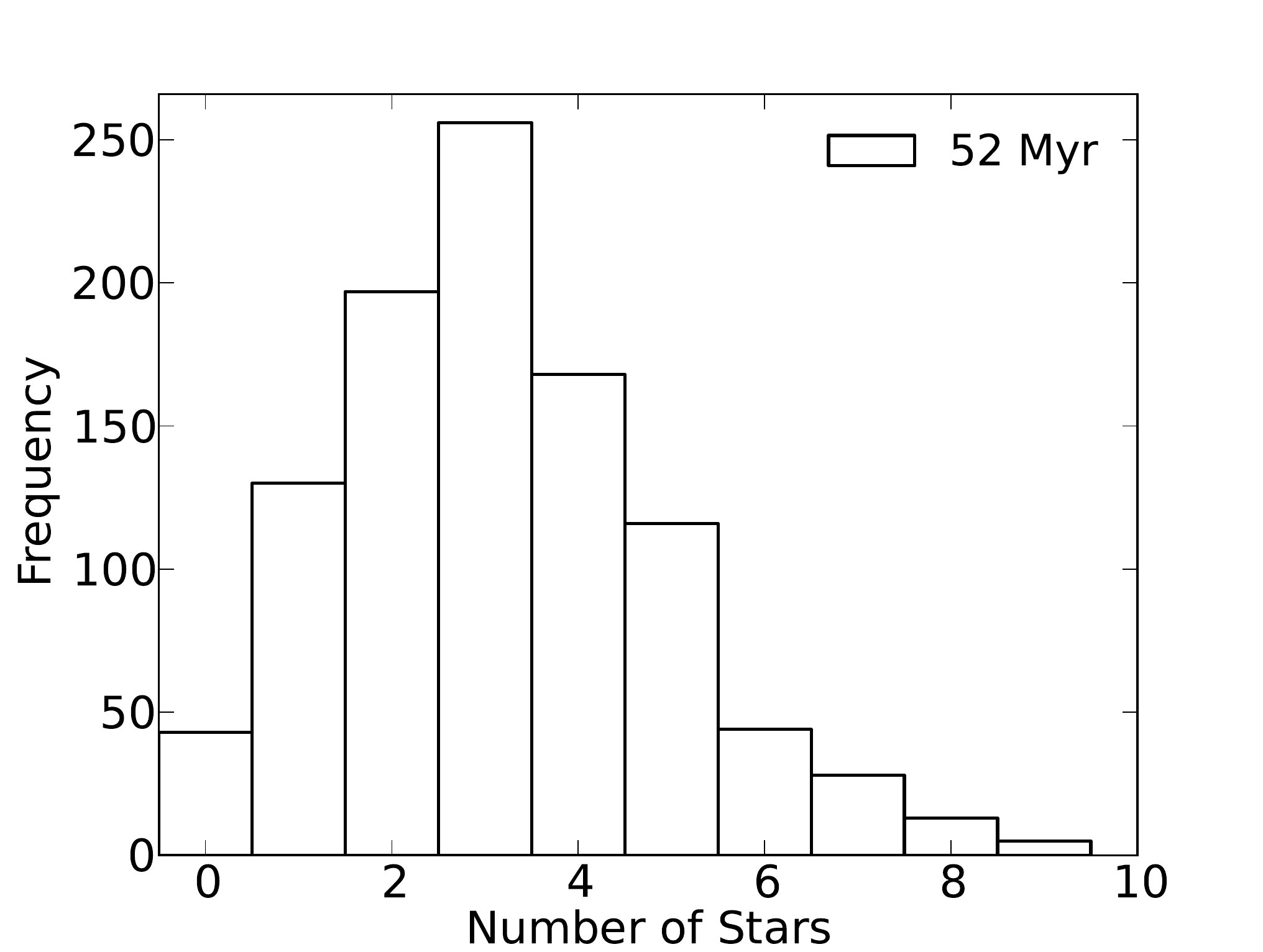}
\includegraphics[width=4.5cm]{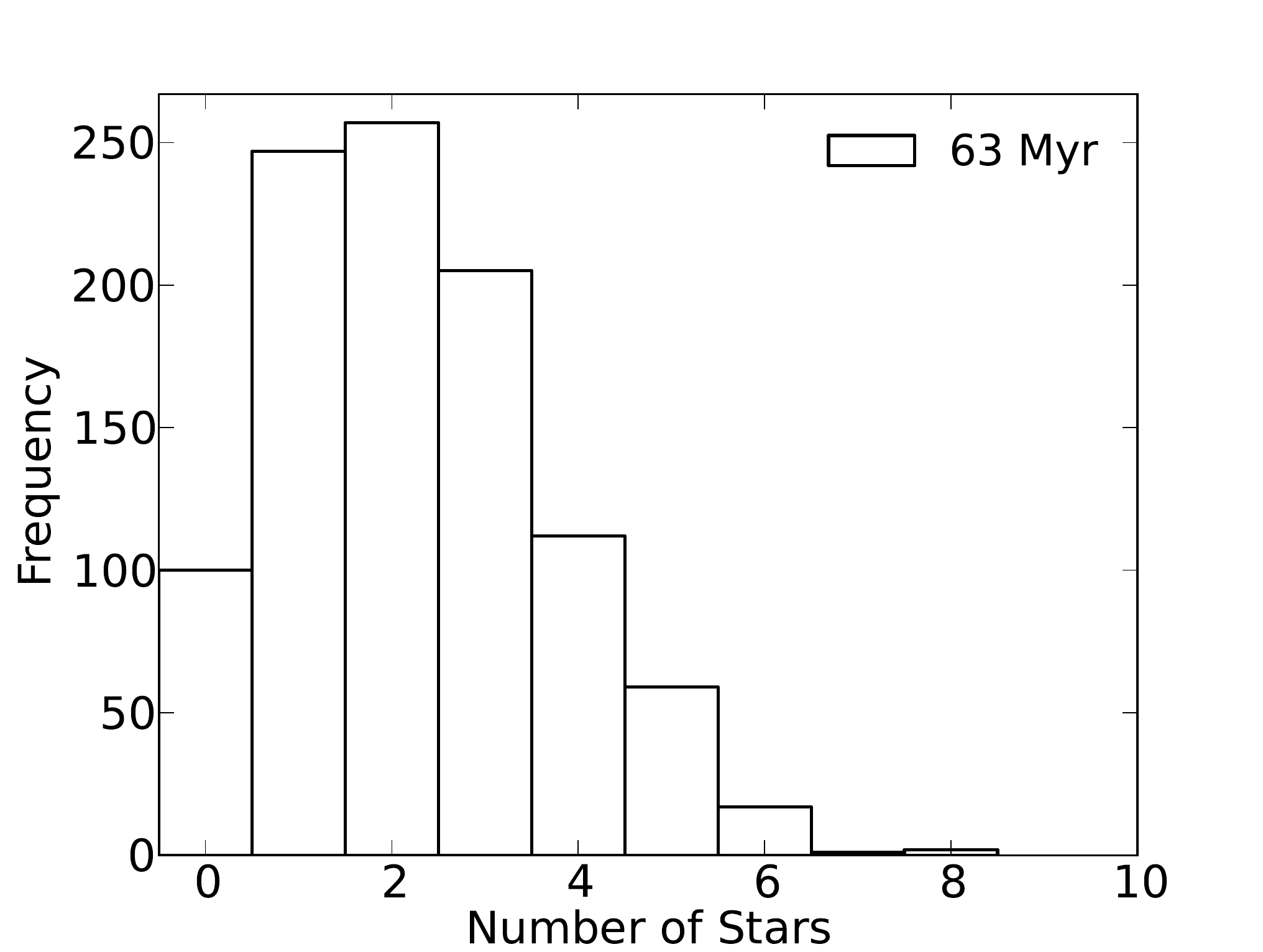}\\
\end{tabular}
\caption{Occurrence rate of post-MS stars in a simulation of 1000 artificial clusters constructed from the fitted SFH of NGC 1847. The four panels show four different ages (25 Myr, 30 Myr, 52 Myr and 63 Myr, going from top left to bottom right).}
\label{fig:ngc1847_stars_hist}
\end{figure}

\begin{figure}
\centering
\resizebox{\hsize}{!}{\includegraphics{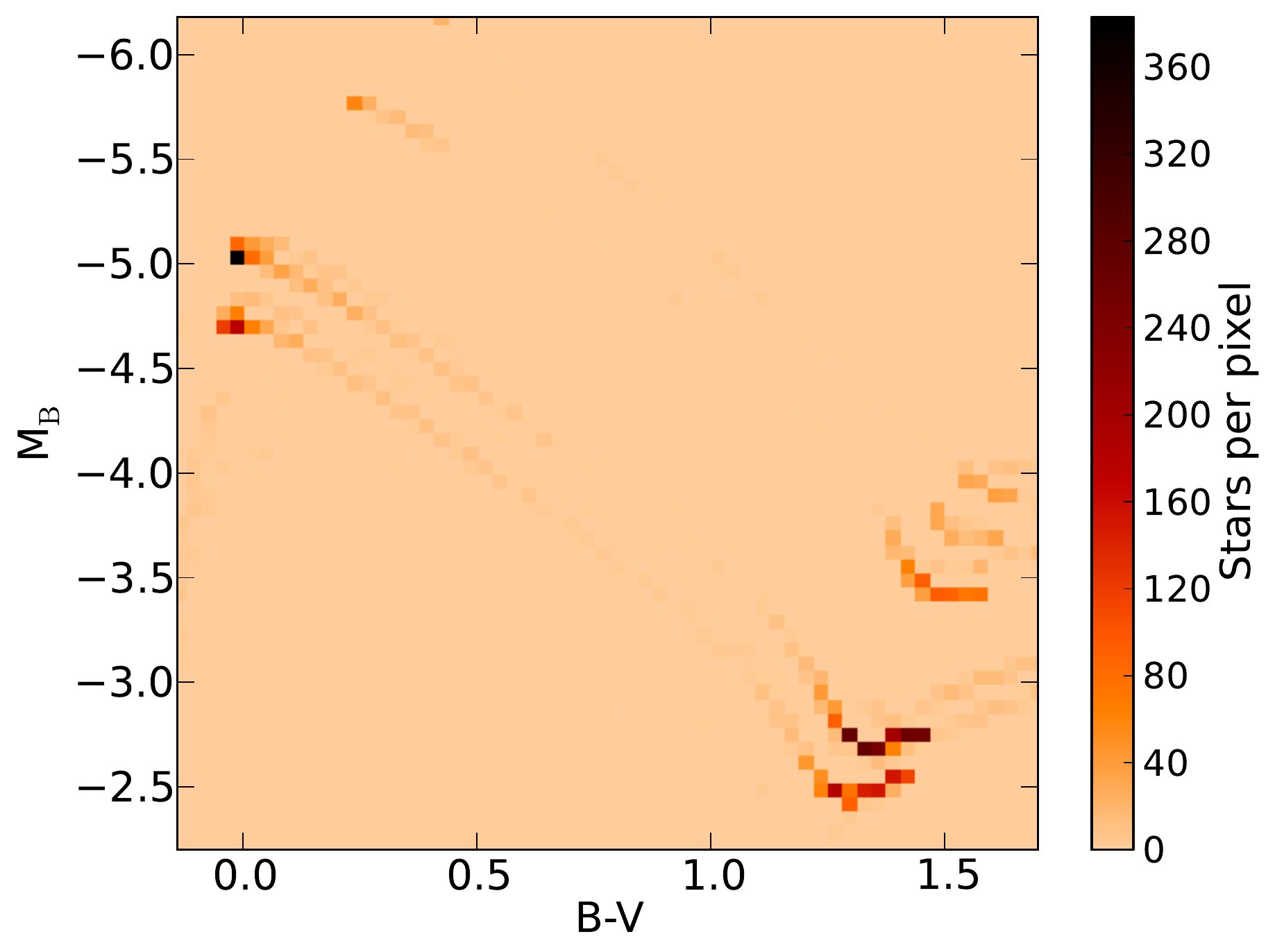}}
\caption{Stellar density plot of the post-MS tracks of all 1000 synthetic clusters with the fitted SFH of NGC 1847 stacked together. The density is a measure of the probability to find a star in the respective part of the CMD.}
\label{fig:ngc1847_art_cluster_sfh_density_plot}
\end{figure}

In the fitting of the SFH we include the regions in the CMD that contain the MS down to a $V$ magnitude of 0.5 and the red He-burning stars. The results of the fitting (Figure \ref{fig:ngc1847_sfh_fit}) show 
a large peak at 57 Myr and an irregularly shaped lower star formation at ages between 14 and 30 Myr. The result at younger ages is due to the MS extending to brighter magnitudes than would be expected for a 57 Myr old population. There is no sign for this younger population in the post-MS part of the CMD. This might be due to the random sampling of the IMF. Therefore,
to evaluate the results of the SFH fit statistically, we create 1000 synthetic clusters with the same SFH as we got for NGC 1847 normalized to the number of stars present in the CMD of NGC 1847 
brighter than a $V$ band magnitude of 0.5 to avoid effects resulting from incompleteness.
Figure \ref{fig:ngc1847_art_cluster_sfh} shows the CMD of one of the clusters where the individual stars are color-coded by their age. 
We analyzed the occurrence of post-MS stars of all individual ages to get a measure of how many stars we should expect at a given age (see Figure \ref{fig:ngc1847_stars_hist}). A density plot of the post-MS tracks of all 1000 clusters stacked together is shown in Figure \ref{fig:ngc1847_art_cluster_sfh_density_plot} to visualize where in the CMD the stars are expected to be. We see from Figure \ref{fig:ngc1847_stars_hist} that in a considerable fraction of cases (82\% at an age of 25 Myr and 40\% at an age of 30 Myr) there are no post-MS stars present at ages where the young peak is. So from a statistical point of view we cannot rule out the second peak at 14-30 Myr.
If the star formation at the younger ages is real, this would result in an overall total age spread of $<$ 45 Myr which is inconsistent with the models proposed by \citet{Goudfrooij09,Goudfrooij11a,Goudfrooij11b} and \citet{Rubele13}.

However, there might be other causes for the bright MS stars that are suggestive of a 26 Myr old population: They could be binaries, blue straggler stars or fast rotating stars. 
In an upcoming paper [Niederhofer et al., in prep.] we will investigate the role of rotation and test the last hypothesis.

\subsection{NGC 2004\label{sec:ngc2004}}

\begin{figure*}
\centering
\includegraphics[width=19cm]{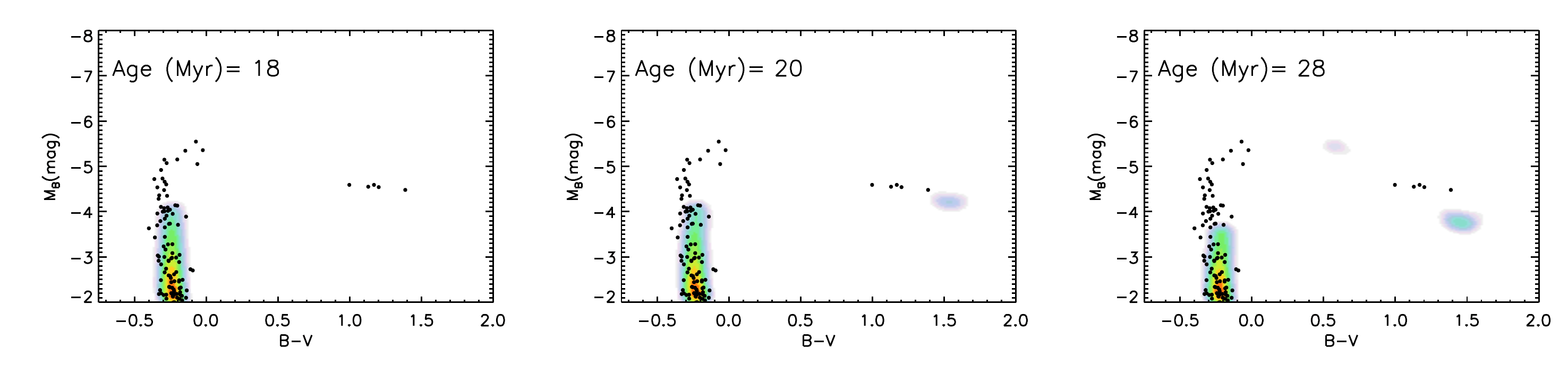}\\
\caption{CMD of NGC 2004 together with theoretical isochrones at three ages from 18 to 28 Myr.}
\label{fig:ngc2004_hess}
\end{figure*}

\begin{figure*}
\centering
\includegraphics[width=19cm]{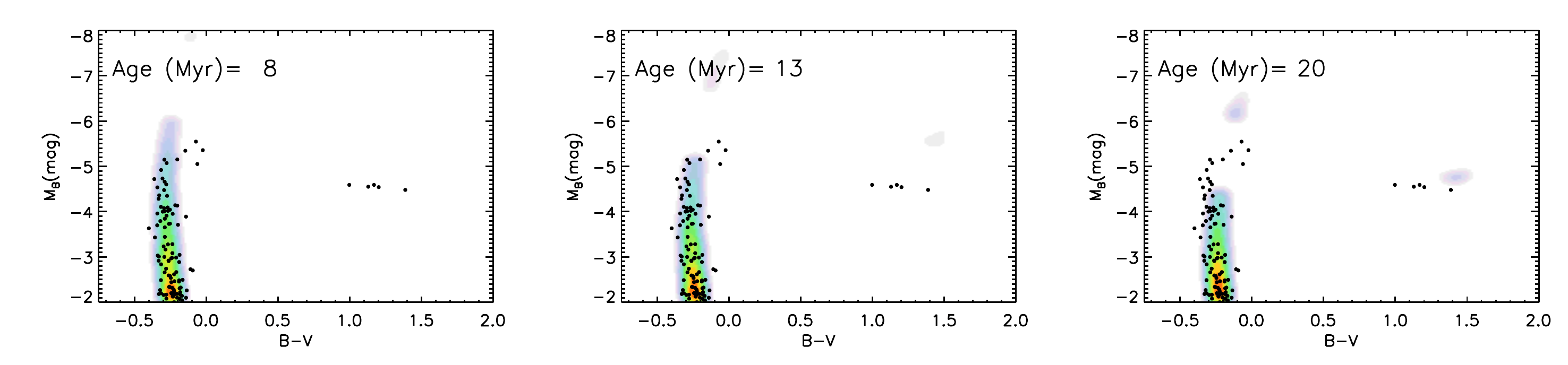}\\
\caption{Same as Figure \ref{fig:ngc2004_hess} but using \citet{Marigo08} isochrones from 8 to 20 Myr.}
\label{fig:ngc2004_hess_marigo}
\end{figure*}

NGC 2004 also belongs to the youngest group of clusters in our YMC sample. \citet{Elson91} found an age of 20 Myr for this cluster. It has a total mass of $\sim$2.3$\cdot$10$^4$M$_{\sun}$ \citep{McLaughlin05}. By matching theoretical isochrones to the MS of the cluster we find a reddening value of $E(B-V)=0.23$ which is considerably higher than the value of 0.08 found by \citet{Keller2000}. Figure \ref{fig:ngc2004_hess} shows the CMD of this cluster together with Hess diagrams at three different ages. The observed CMD consists of two main features: The MS that extends to magnitudes of about $-$5 in the $B$ band and a handful of red He-burning stars at $(B-V) \sim$1.2. The position of this He-burning stars is best reproduced by an age of 20 Myr (Figure \ref{fig:ngc2004_hess} middle panel). Ages that are older than about 30 Myr can be excluded because there are no He-burning stars at the position in the CMD predicted by the models. Younger models do not show any evolved stars. 
We note that the observed MS for all ages extends to brighter magnitudes than is predicted by the theoretical model. This might be caused by the used set of isochrones as the Parsec 1.1 isochrones only include masses up to 12 M$_{\sun}$. At an age of $\sim$18 Myr, the turn-off mass is about 10.4 M$_{\sun}$ so the left panel of Figure \ref{fig:ngc2004_hess} shows only the MS and no stars that already evolved off the MS. For a comparison we show in Figure \ref{fig:ngc2004_hess_marigo} the CMD of NGC 2004 with Hess diagrams constructed using Padova isochrones from the \citet{Marigo08} set which contain models for stars with masses up to 100 M$_{\sun}$. The age which matches best the helium burning stars is about 22 Myr and therefore a bit older than the one from the Parsec 1.1 isochrones. But also this model cannot explain the bright stars that extend over the MS at this age. In order to explain the bright MS stars an age of 12.5 Myr for the cluster is needed (see Figure \ref{fig:ngc2004_hess_marigo}). However, the position of the He-burning stars at this age does not correspond to the observations. If this discrepancy is due to an age spread, we can limit it to $\pm$8 Myr as on the one hand there is no sign of recent star formation younger than 12 Myr and on the other hand the CMD positions of the post-MS stars older than 30 Myr are inconsistent with the observations.
As we already mentioned in Section \ref{sec:ngc2136}, this stars in the Blue-Hertzsprung-Gap can also be crowded stars, fast rotating stars, interacting binaries or merged stars.

\begin{figure}  
\resizebox{\hsize}{!}
{\includegraphics{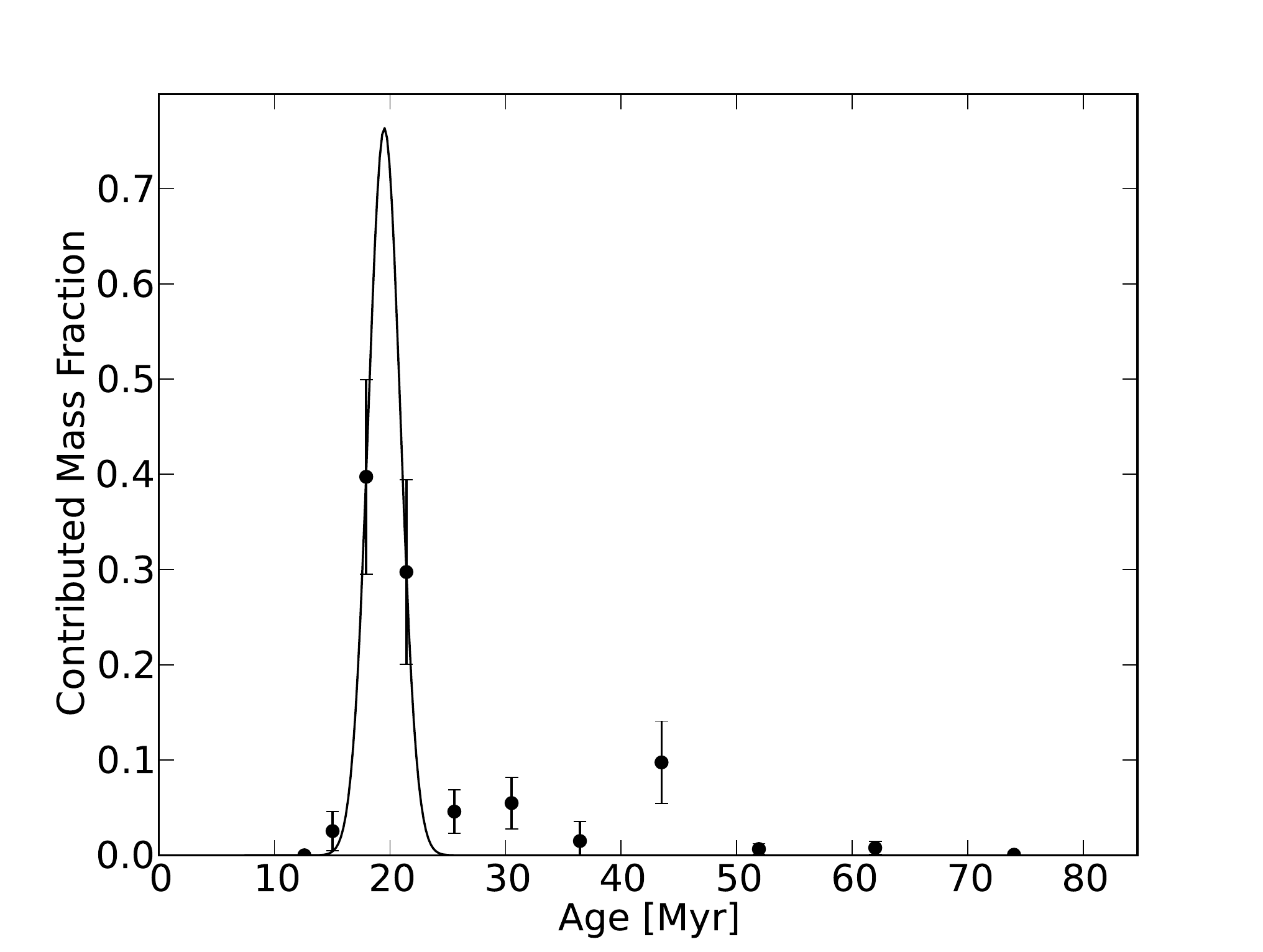}}
   \caption{Results for the fitting of the SFH of NGC 2004. The dots represent the results at individual ages and the solid line shows the best Gaussian fit to the points with a peak at 19.5 Myr and a standard deviation of 1.4 Myr.}
              \label{fig:ngc2004_sfh_fit}
    \end{figure}

We used for the fitting of the SFH of NGC 2004 the Parsec 1.1 isochrones. The result is shown in Figure \ref{fig:ngc2004_sfh_fit}. As expected, the peak of the star formation is at around 20 Myr, with a dispersion of 1.4 Myr. Besides the prominent peak there are also points with a non-zero star formation at higher ages. However, we can rule out at a 2$\sigma$ level star formation at these ages mainly because of two reasons. First, a similar behavior is seen in the tests with artificial coeval clusters (see Section \ref{sec:art_cluster_test}) which is indicative that the non-zero values at higher ages are introduced by the fitting itself. Second, a forming cluster that undergoes an extended period of low star formation followed by a burst of high star formation is a un-physical scenario (e.g. \citealt{Cabrera-Ziri14}).
The usage of \citet{Marigo08} isochrones does not change the overall result.

\subsection{NGC 2100}

\begin{figure}  
  \resizebox{\hsize}{!}{\includegraphics{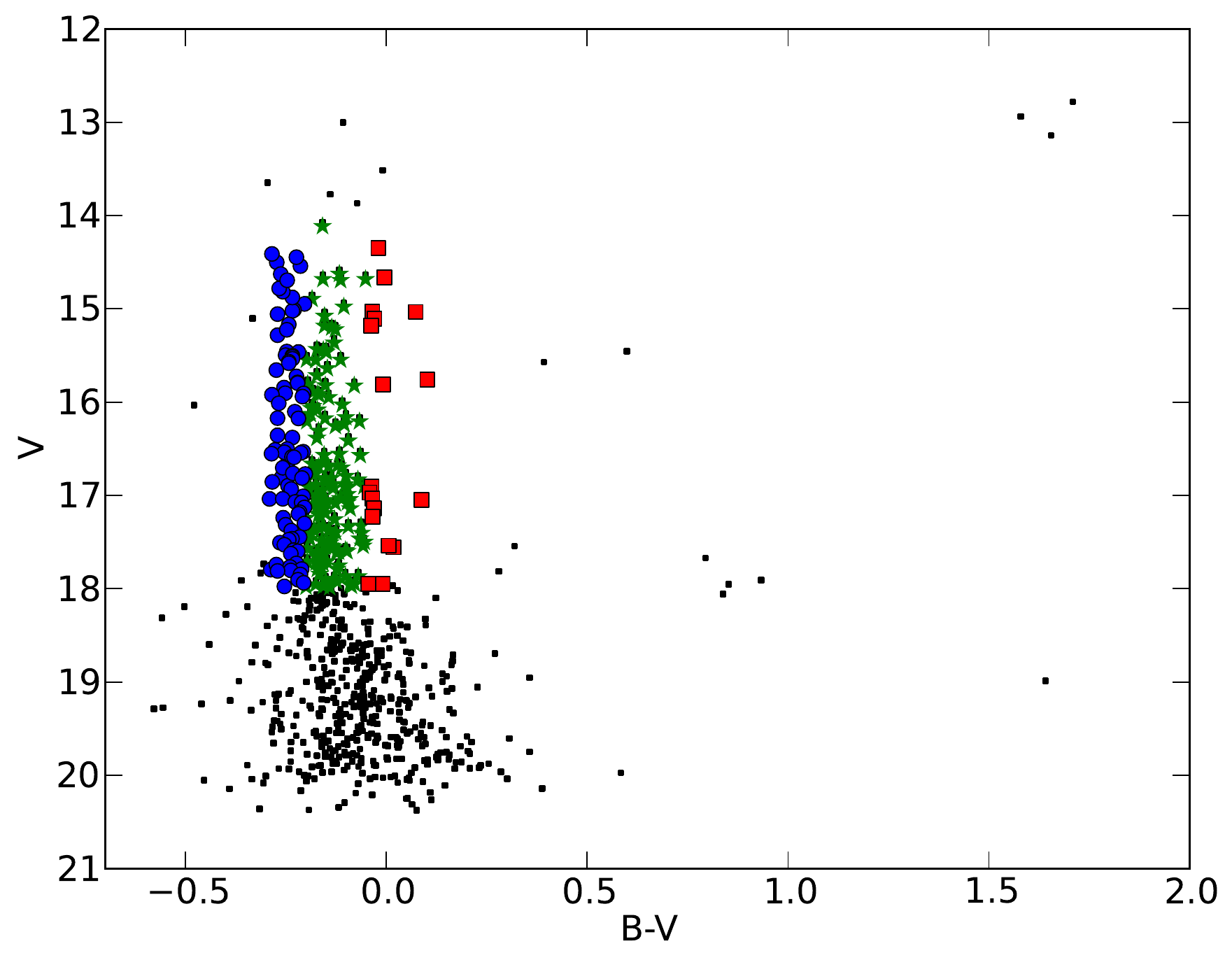}}
   \caption{CMD of the cluster NGC 2100. We used the stars that are marked by circles (in blue), the asterisk symbols (in green) and squares (in red) to determine if differential reddening could be responsible for the color spread in the main sequence.}
              \label{fig:NGC2100_cmd_red_blue_MS}
    \end{figure}

According to previous studies, NGC 2100 is the youngest cluster in our sample of YMCs (see Table \ref{tab:Cluster_Param}). By superimposing model isochrones over the cluster CMD we find an extinction of $E(B-V)$ = 0.24 for this cluster, which agrees with the value found by \citet{Keller2000}. Moreover, we determined a metallicity of 0.007 and a distance modulus of 18.5 mag. Figure \ref{fig:NGC2100_cmd_red_blue_MS} shows the CMD of the cluster. The MS is located at a $B-V$ color of $\sim -$0.1 and extends up to a magnitude of $\sim$14 (in apparent V magnitude). Besides the MS there are three bright and red stars at $V \sim$13 and $B-V \sim$1.7 which we believe are associated with the cluster. The other fainter red stars are most likely field stars that were not removed completely. NGC 2100 shows a broad MS with several stars scattered to redder colors. This feature could have been caused by different things: 1) Field stars, 2) Signs of a possible age spread inside the cluster population or 3) Differential reddening across the cluster. 

\begin{figure}
   \resizebox{\hsize}{!}{\includegraphics{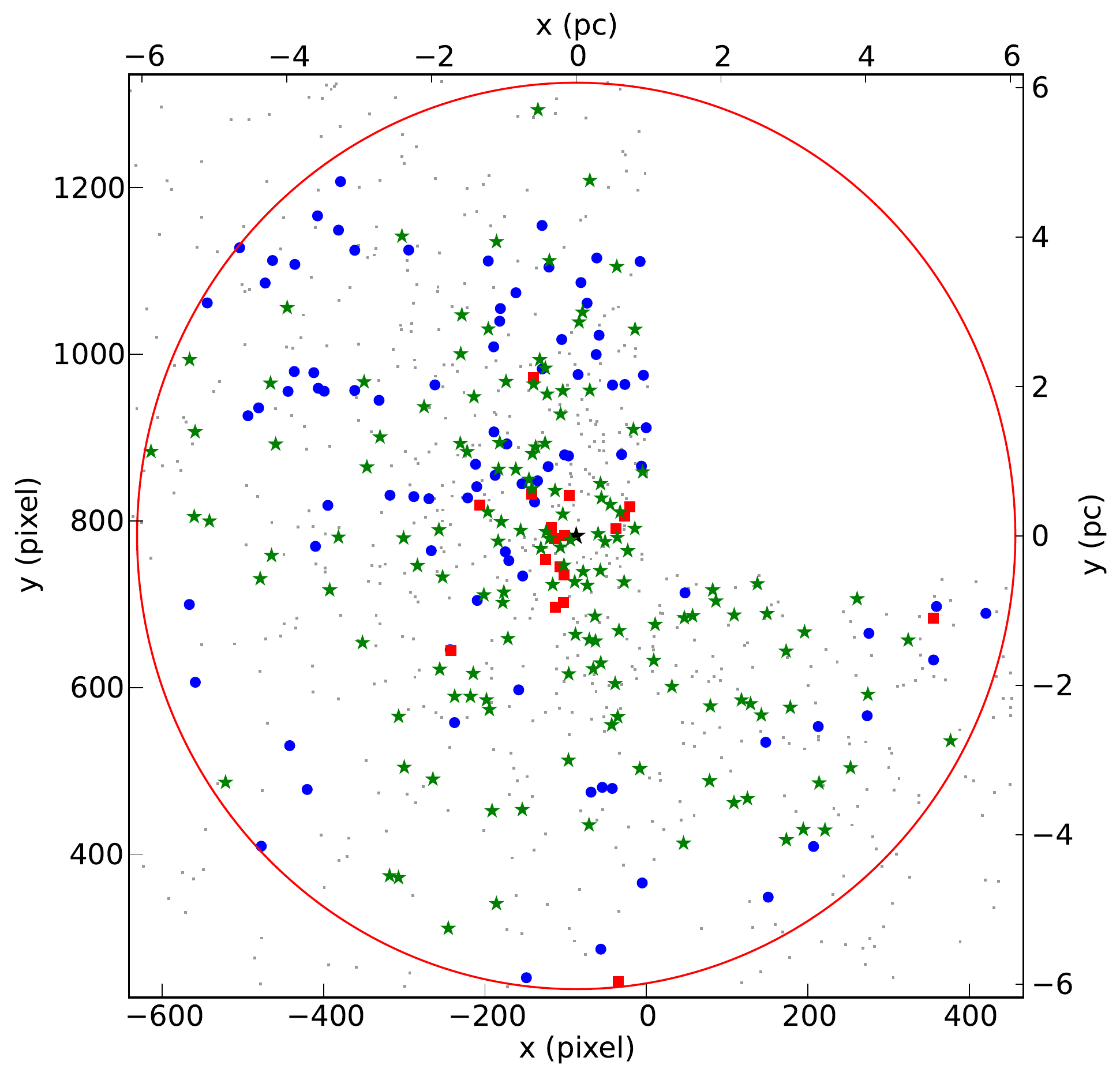}}
   \caption{Spatial positions of the red, green and blue stars marked in Figure \ref{fig:NGC2100_cmd_red_blue_MS}. The red stars preferentially populate the very central region of the cluster whereas the blue stars are found everywhere but the central parts. This might suggest that NGC 2100 is affected by differential reddening.}
              \label{fig:NGC2100_xy_plot_red_blue_MS}
    \end{figure}

To investigate which effect causes this broadening we divided the MS in to three parts, one at the blue, one at the red side, and the region in between the two, with the same magnitude cuts well above the completeness limit (see Figure \ref{fig:NGC2100_cmd_red_blue_MS} the circles (in blue), squares (in red) and the asterisks (in green)) and plotted their spatial positions (see Figure \ref{fig:NGC2100_xy_plot_red_blue_MS}). If the red stars were stars not associated with the cluster we would expect them to be distributed evenly over the selected region. They should also appear in the CMD of the region that we used to subtract the field population from the cluster population. However, there is no indication for them in the field CMD (see Figure \ref{fig:Cluster_cmds}). On the other hand if the stars would belong to a somewhat older population they should be distributed in a similar manner as the blue stars, i.e. less concentrated towards the center. But what we actually see is that the     
red stars preferentially populate the central region of the cluster. The blue stars avoid the central and south-west parts of the cluster and are mostly found in the north-west. The green stars that populate the intermediate part of the MS are centrally condensed and follow the overall distribution of the cluster stars. This all is a strong indication that the stars in NGC 2100 are affected by differential reddening with higher extinction values towards the central parts of the cluster and lower extinction towards the north-east. The fact that the extinction is not constant across the cluster will introduce an additional scatter in the SFH of NGC 2100. In our analysis we have to consider this effect.

\begin{figure}  
  \resizebox{\hsize}{!}{\includegraphics{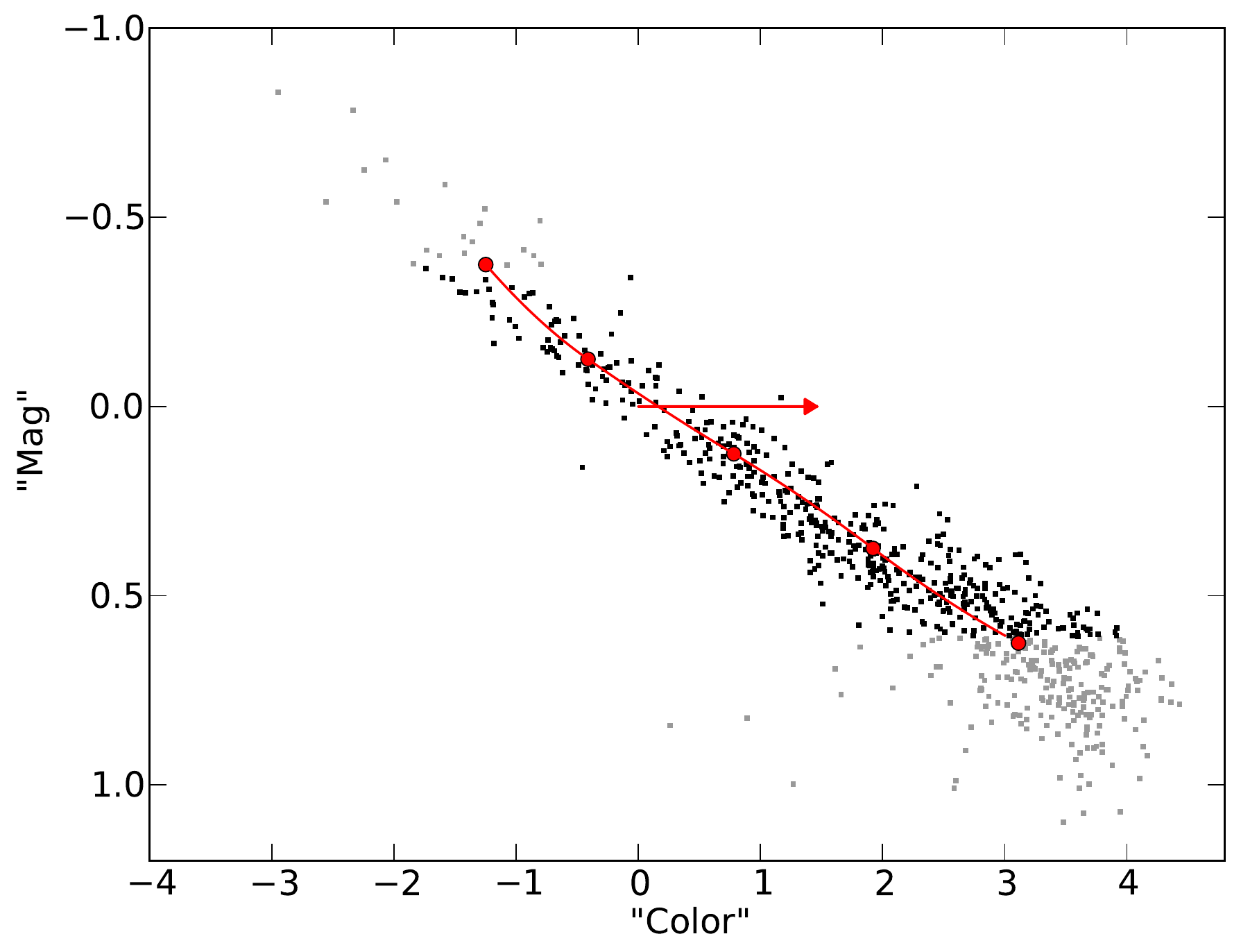}}
   \caption{CMD of NGC 2100 rotated such that the extinction vector (red arrow) is horizontal. The line (in red) along the MS is the fiducial line as it is determined in the text. The stars that are used to analyze the extinction variations across the cluster are colored black.}
              \label{fig:NGC2100_rot_cmd}
    \end{figure}

\begin{figure}  
  \resizebox{\hsize}{!}{\includegraphics{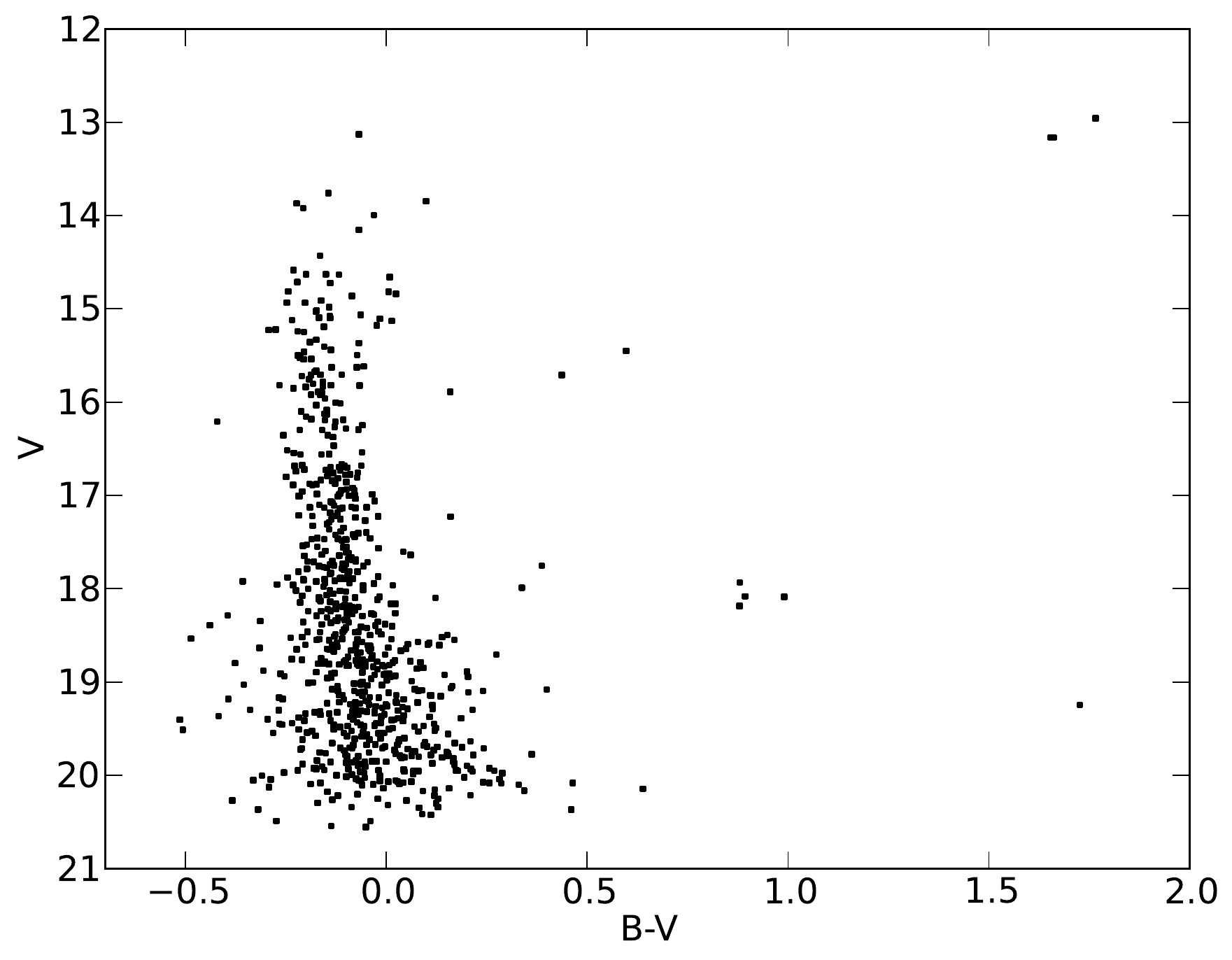}}
   \caption{CMD of NGC 2100 after correcting it for differential extinction.}
              \label{fig:NGC2100_cmd_diff_ext_corr}
    \end{figure}

Therefore, before we fit the SFH of the cluster, we take an extra step here and correct the CMD of NGC 2100 for differential extinction. We adopt the method from \citet{Milone12}, which is explained in great detail in that paper. Here we give a short description of the basic steps that we have performed: First, we define the reddening vector in the CMD of NGC 2100 and rotate the CMD around an arbitrarily chosen point in the CMD such that the reddening vector is horizontal in the new "color"-"magnitude" diagram (see Figure \ref{fig:NGC2100_rot_cmd}). We will refer to the new axes as "color" and "magnitude". In the second step we define a fiducial line along the MS in the rotated CMD by calculating the median "color" of the stars in 0.25 "magnitude" bins and quadratically interpolating between them. Then we choose all stars in the real magnitude interval of 15.5$\lesssim B \mathrm{[mag]} \lesssim$18.5 as reference stars that are used to determine the extinction differences across NGC 2100. For each reference star we identified the closest spatial neighbors and calculated the median "color" distance to the fiducial line of these neighbors (not including the respective star itself). Depending on where in the cluster the star is located (center or outskirts) we used 15 or 20 closest neighbors. As the extinction vector is parallel to the "color" axis in the rotated CMD the median displacement of a star's neighbors is equivalent to a shift caused by extinction. We then divided the cluster field in annulus segments in which the extinction has approximately the same level and calculated the median reddening in each segment. As a last step we corrected all stars in the cluster CMD for this median extinction, depending on its position in the cluster.  
We performed the above procedure two times in a sequence. A third correction would not have changed the final result by much. As we only have a small number of stars ($\sim$400) in this cluster that we used for the determination of extinction variations we can only correct on large scales and might miss possible variations on smaller scales. Therefore the corrected CMD might still be affected by differential extinction to a certain degree.
Figure \ref{fig:NGC2100_cmd_diff_ext_corr} shows the CMD of NGC 2100 after it has been corrected for differential extinction. The MS is now better defined, especially at intermediate magnitudes, whereas the upper part still is broadened. We also note that, in order to match theoretical models, we have to adjust the mean extinction from $E(B-V)$ = 0.24 to 0.17.

\begin{figure}
\centering
\resizebox{\hsize}{!}
{\includegraphics{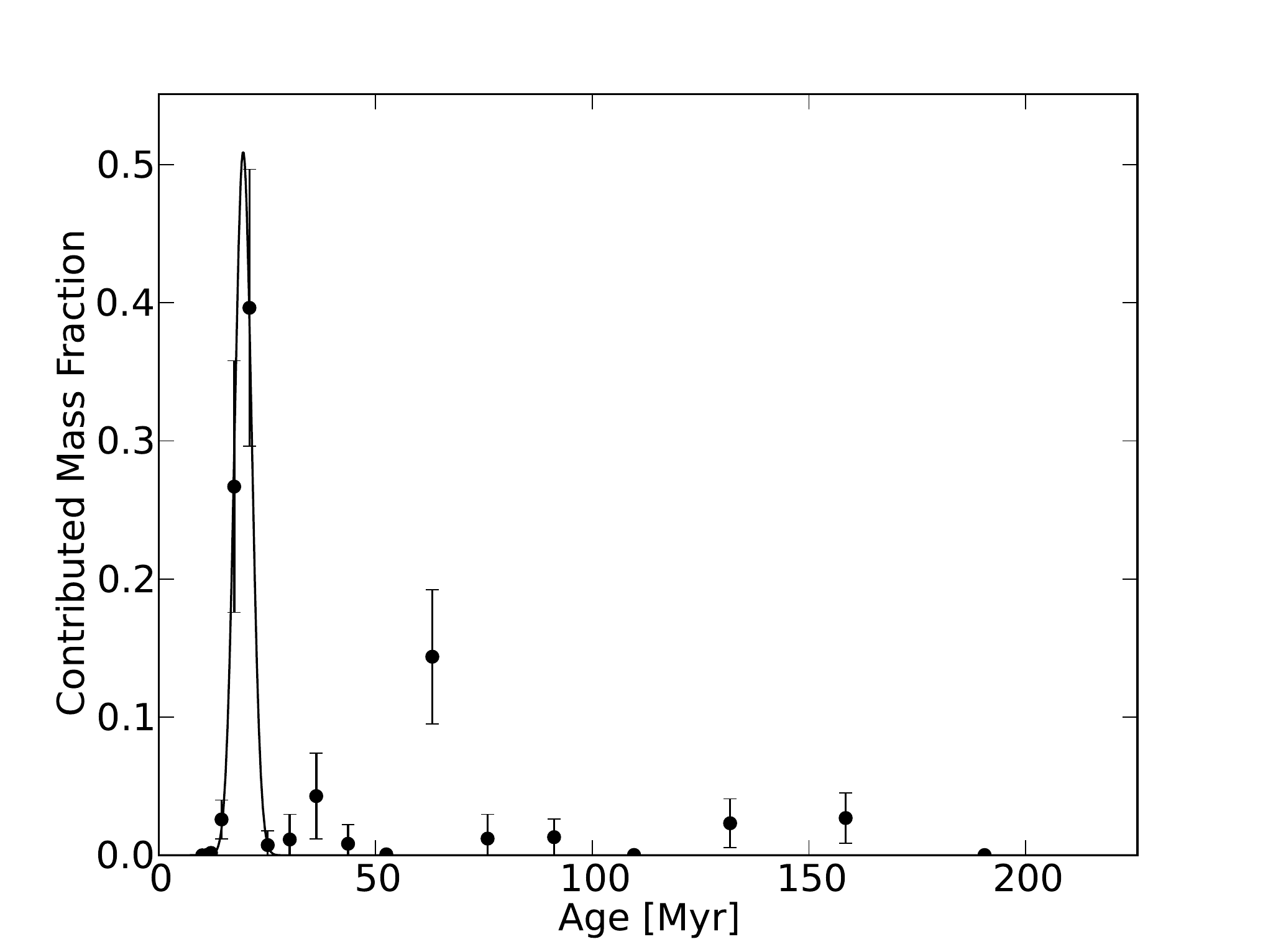}}
\caption{Results of the SFH fit of NGC 2100. The dots represent the results at individual ages and the solid line shows the best Gaussian fit to the points with a peak at 19.5 Myr and a standard deviation of 2.0 Myr.}
\label{fig:ngc2100_sfh_fit}
\end{figure}

\begin{figure*}
\centering
\includegraphics[width=19cm]{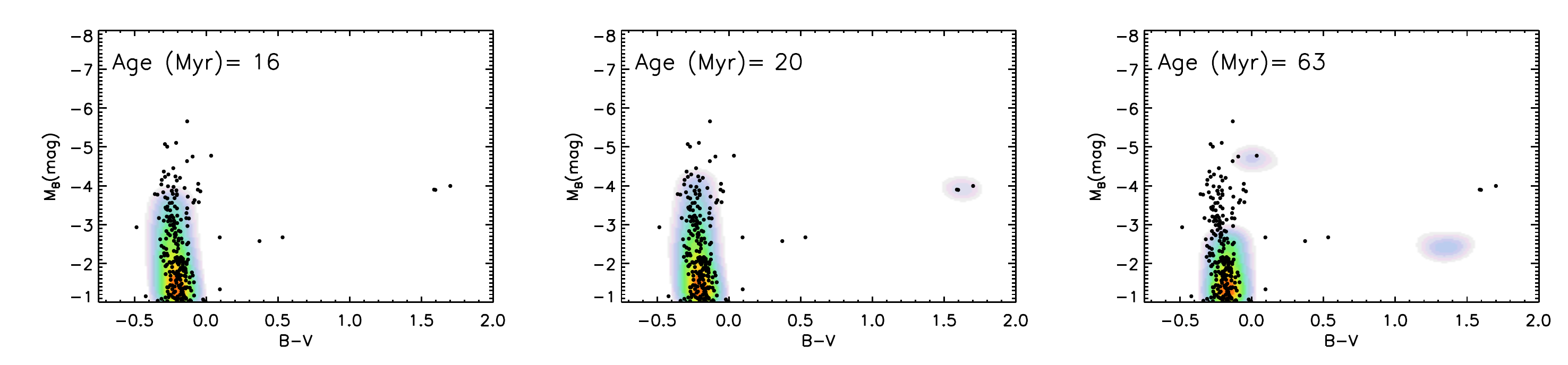}
\caption{CMD of NGC 2100 after correction for differential extinction. Overplotted are model Hess diagrams at three different ages in the range from 16 to 63 Myr.}
\label{fig:ngc2100_hess}
\end{figure*}

Figure \ref{fig:ngc2100_sfh_fit} displays the results of the SFH fit of NGC 2100 together with a Gaussian that matches best the highest peak. The star formation shows two peaks, a larger one with a maximum at about 20 Myr followed by a smaller peak at about 60 Myr. Additionally there are some non-zero values between 90 and 160 Myr. To quantify the reliability of the fitting we compare the results with theoretical Hess diagrams. These diagrams at three different ages, together with the corrected CMD of NGC 2100, are shown in Figure \ref{fig:ngc2100_hess}.
We see that at an age of 20 Myr, the Hess diagram matches best the position of the evolved red stars and the top of the MS.  
At a younger age of about 16 Myr (upper left) the three blue loop stars cannot be explained as there are no stars that evolved off from the MS to redder colors. This is also caused by the usage of the Parsec 1.1 isochrone set that contains only stars up to 12 M$_{\sun}$. As in the previous clusters there are a few bright stars above the MS that cannot be explained by the models.

\begin{figure}
\centering
\includegraphics[width=6.5cm]{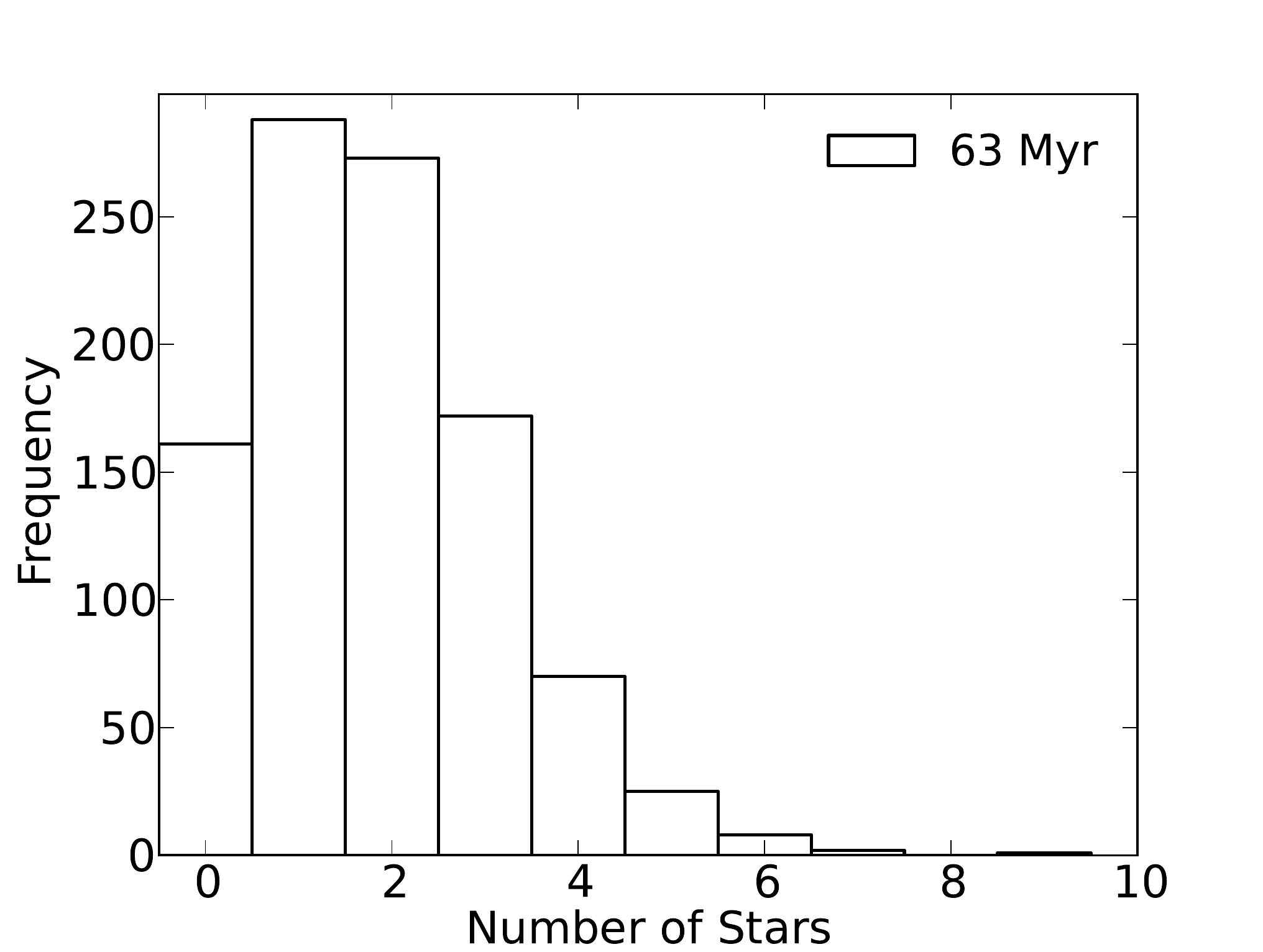}
\caption{Occurrence rate of post-MS stars at an age of 63 Myr in a simulation of 1000 artificial clusters constructed from the fitted SFH of NGC 2100.}
\label{fig:ngc2100_stars_hist}
\end{figure}

The fitting of the SFH of NGC 2100 shows, in addition to the highest peak at 20 Myr, a second peak at 63 Myr. The third panel of Figure \ref{fig:ngc2100_hess} shows the CMD of NGC 2100 with a Hess diagram at 63 Myr superimposed. The fitted SFH at this age is most likely due to the three stars at $B-V \sim$0.1 and $B \sim$ $-$5 mag on the red side of the MS. At the same time the models predict blue loop stars to be at $B-V \sim$1.4 and $B \sim$ $-$2.5 mag where no stars are observed. To check if the peak at 63 Myr is compatible with the observations from a statistical point of view, we did the same Monte Carlo test as we already did with NGC 1847 and created synthetic clusters with a SFH that is the same as predicted for NGC 2100 by the fitting. Again, the number of stars in the artificial clusters are normalized to the number of stars brighter than an absolute $V$ band magnitude of 0.0 to avoid incompleteness effects. In these simulations, we searched for blue loop stars at an age of 63 Myr with a $B-V$ color $>$0.2 in order to not get confused with the evolved stars above the MS. Figure \ref{fig:ngc2100_stars_hist} shows the occurrence rate of post-MS stars at an age of 63 Myr from the simulations. In 84\% of the cases the test predicts at least one post-MS star. Therefore we can almost certainly rule out this age.

Beyond the two peaks, the fitting of the SFH also yields a very low level of star formation at ages between 90 and 160 Myr. But this result, as discussed in Section \ref{sec:ngc2004}, is most likely not real.

\begin{table} 
\caption{Summary of the results from the SFH fit to the highest peak\label{tab:Fitting_results}}
\centering
\begin{tabular}{l c c} 
\hline\hline
\noalign{\smallskip}
Cluster & Fitted SFH & Fitted SFH of the corresponding\\
&of highest peak & highest peak in artificial cluster \\
\noalign{\smallskip}
\hline
\noalign{\smallskip}
NGC 1831 & 924$\pm$126 Myr & 949$\pm$57 Myr\\
NGC 1847 & 56.7$\pm$4.8 Myr & 60.0$\pm$1.0 Myr\\
NGC 1850 & 93.4$\pm$18.3 Myr & 96.2$\pm$5.7 Myr\\
NGC 2004 & 19.5$\pm$ 1.4 Myr & 20.1$\pm$1.5 Myr\\
NGC 2100 & 19.5$\pm$2.0 Myr & 20.6$\pm$1.6 Myr\\
NGC 2136 & 123.3$\pm$22.6 Myr & 130.2$\pm$9.6 Myr\\
NGC 2157 & 98.3$\pm$13.2 Myr & 102.2$\pm$8.4 Myr\\
NGC 2249 & 1110$\pm$139 Myr & 1235$\pm$111 Myr\\
\hline
\noalign{\smallskip}
NGC 1856 & 281$\pm$33.6 Myr & $--$\\
NGC 1866 & 177$\pm$18.4 Myr& $--$\\
\noalign{\smallskip}
\hline
\end{tabular}\\
\tablefoot{The last two clusters (NGC 1856 and NGC 1866) are not studied in this work. They were already analyzed by \citet{BastianSilva13}.}

\end{table}

\section{Discussion and Conclusions\label{sec:disc}}

\subsection{Discussion of the Results}

A common feature in the CMDs of intermediate age (1-2 Gyr) LMC clusters is an extended or bifurcated MSTO which could be interpreted as an age spread of the order 100-500 Myr (e.g. \citealt{Goudfrooij11a,Goudfrooij11b}). In this study we test this hypothesis. An extended history of star formation in intermediate age clusters implies that an age spread of the same order should be present in young ($<$1 Gyr) clusters with similar properties as well. We analyzed a sample of eight young massive LMC clusters in continuation of the study by \citet{BastianSilva13} and searched for potential spreads in ages inside them. We compared the CMDs of these clusters with model Hess diagrams and fitted the SFH of all clusters using the \textit{FITSFH} code. The results of the fitting of the highest peak are summarized in Table \ref{tab:Fitting_results}.
Most of the clusters in our sample have a single and well defined peak in the fitted SFH. But there are also some clusters which show periods of low extended star formation, like NGC 1850, or a second burst, like NGC 1847 or NGC 2100, in addition to the main star formation peak.
Therefore we performed Monte Carlo simulations of artificial clusters with the same fitted star formation as the real ones to assess the reliability of the SFH fit. For NGC 1850 we were able to rule out the fitted star formation at a few 100 Myr. However, using this statistical tool, we cannot make any final statement if the younger star formation in NGC 1847 is real or not, although there are physical effects (rotation, blending, binaries) that would cause the MS extend to brighter magnitudes than would be expected from a nominal isochrone at this age. \citet{BastianStrader14} studied the gas and dust content in a sample of 13 LMC and SMC clusters, including NGC 1847, with masses greater than $10^4\mathrm{M_{\sun}}$ and ages between 30 and 300 Myr. They found no evidence for any gas or dust left inside the clusters, even in the most youngest ones. They conclude that clusters are very efficient in removing their gas content right after their formation. This result suggests that the younger period of star formation might not be real. However, the overall age spread, also including the younger burst, is smaller than $\sim$45 Myr (if it is real) which is inconsistent with the proposed age spread of 100-500 Myr for intermediate age clusters.

NGC 2100 is a special case in our sample of young clusters as its CMD is affected by differential extinction across the cluster field. We corrected the photometry for the varying reddening using the method described in \citet{Milone12}. We were only able to account for large scale variations in our correction because of the limited number of stars available. Therefore, there might be some differences in extinction on smaller scales left that introduces some scatter in the final results. The SFH fit of this cluster yields two peaks, one larger at 20 Myr and a smaller one at 63 Myr. Monte Carlo simulations of artificial clusters showed that the second peak is most likely not real.

We also note that the MS of some of the clusters extends to brighter magnitudes than would be expected from the position of the post-MS stars. In the case of NGC 1847, as already discussed, the bright stars introduce an additional younger peak in the fitted SFH. As we used Parsec 1.1 isochrones that cover only stars with masses smaller than 12 M$_{\sun}$ which corresponds to the turn-off mass at $\sim$16 Myr, we additionally compared the CMD of NGC 2004 with the isochrone set from \citet{Marigo08} that includes stellar masses up to 100 M$_{\sun}$. We found that an age of 12.5 Myr is required to explain the brightest MS stars. At this point, we cannot entirely rule out that the discrepancy between the MSTO and the evolved stars is due to an age spread. However, we can restrict the potential age spread to be $\lesssim$8 Myr as we see no sign of recent star formation, nor for ages older than $\sim$28 Myr.

By fitting the SFHs to the LMC clusters in our sample we get age spreads (given by the dispersion of the best-fitting Gaussians to the highest star formation peak) that are all upper limits of the real age spreads that might be present in the clusters. We find that the age dispersion that we got for the clusters is increasing with increasing age of the corresponding age of the cluster. The youngest clusters in our sample have age spreads smaller than a few Myr, whereas the two oldest (NGC 1831 and NGC 2249) have a fitted dispersion in age of the order of 120-140 Myr. The lower four panels of Figure \ref{fig:artificial_cluster} show the results of the SFH fits of four simulated clusters. We see the same trend of increasing "age spread" in the results for the simulated coeval clusters. A comparison with the results of our cluster sample shows that, on the one hand, the age dispersion is comparable for the youngest age. On the other hand, the older clusters show spreads in age that are larger (2 to 4 times) than we would expect from our simulations. This spread therefore must have resulted from other effects. The apparent larger age range might be introduced by an underestimation of the photometric errors, interacting binaries in the cluster and/or rotating stars as already proposed by e.g. \citet{Larsen11}.

We determined the ages of NGC 1831 and NGC 2249 to be $\sim$900 Myr and $\sim$1.1 Gyr, respectively. Therefore they can already be classified as intermediate age LMC clusters. Moreover, the age spread larger than 100 Myr that we found for both clusters would support the interpretation of the extended MSTO by \citet{Goudfrooij11a,Goudfrooij11b}. However, this is in conflict with the results we got for the younger clusters in our sample and in the sample of \citet{BastianSilva13}. Their age dispersion is much smaller as would be expected from the proposed age spreads in the intermediate age LMC clusters. 
In addition, as already mentioned, all age spreads are upper limits and there are non-negligible extrinsic (photometric errors, differential extinction) and intrinsic (interacting binaries, rotating stars) contributions that increase the apparent spread. Therefore the fitted extended SFH of the two intermediate age clusters in our sample does not support the \citet{Goudfrooij11a,Goudfrooij11b} scenario.

\citet{Goudfrooij11b} suggest that clusters that have an escape velocity greater than a critical value of 10 km/s at the beginning of their lifetimes should be capable to host multiple generations of stars. Table \ref{tab:Cluster_Param} lists the escape velocities of all clusters in our data set plus the clusters from \citet{BastianSilva13}. We estimate the value of $v_{esc}$ the clusters had when they were 10 Myr old using the method presented in \citet{Goudfrooij11b} that accounts for mass loss. We do not assume any evolution in radius. Whereas the value of the younger clusters does not change by much, the escape velocity of the older clusters ($>$100 Myr) will be larger by a factor between 1.26 and 1.4. After applying this correction to the clusters, NGC 1850, NGC 1856 and NGC 1866 have escape velocities that are larger than 10 km/s. None of them shows signs of an extended history of star formation disagreeing with the predictions made by \citet{Goudfrooij11b}. Another criterion for an extended SFH is given by \citet{Conroy11} who stated that clusters more massive than $10^4 \mathrm{M_{\sun}}$ should have age spreads. However, all our sample clusters are well above this threshold.

Analyzing the CMDs of young and massive LMC clusters we found a new age for the main cluster of the binary system NGC 1850. Our result of about 100 Myr is significantly higher than the commonly used value of  $\sim$30-50 Myr. This increase in age also affects the mass of the cluster which increases by a factor of two, from 7.2$\cdot 10^4\mathrm{M_{\sun}}$ to 1.4$\cdot 10^5\mathrm{M_{\sun}}$. This new mass is in agreement with the predicted mass-to-light ratio for a 100 Myr population.

\subsection{Comparison with the Intermediate Age Clusters}

In Table~\ref{tab:Cluster_Param} we show the estimated current masses and escape velocities of the clusters in our sample.  Comparison with Table~2 of \citet{Goudfrooij11b} shows the clusters studied here have masses and escape velocities in the same range as the intermediate age clusters that display the extended MSTO phenomenon. However, \citet{Goudfrooij11b} have suggested that large ``correction factors" need to be applied to the intermediate age clusters in order to estimate their initial masses and escape velocities.  The scenario advocated by these authors is that a large cluster was formed in a near-instantaneous burst, and that this cluster was able to retain the ejecta from some of the evolved stars within it, as well as accrete new gas from its surroundings, in order to form a second generation of stars (which lasts for $300-600$~Myr). During the formation of the second generation, the majority ($>$80-90\%) of the first generation stars are lost (i.e. they are no longer in the present day cluster). For this, the initial massive cluster is assumed to be 1) tidally limited, 2) in a strong tidal field (suitable for the inner regions of the Milky Way, not tailored to that expected in the LMC/SMC) and 3) mass segregated (which accelerates mass loss by causing a cluster to expand over its Jacobi radius).  The initial cluster then loses the vast majority of its initial population of stars, and what is left today is only the second generation stars.  It is unclear in this model why the second generation of stars would show a Gaussian type age distribution as observed.  When applying these correction factors, \citet{Goudfrooij11b} suggest that a cluster must have an escape velocity in excess of $10-15$~km/s in order to form a second generation of stars.

However, given the relatively weak tidal fields in the LMC/SMC, and the fact the intermediate age (and older) clusters do not appear to be tidally limited (e.g., \citealt{Glatt09}), it is more likely that the intermediate age clusters have not undergone strong mass loss, hence their current masses and escape velocities are similar to their initial values. In this interpretation, we see that there is no relation between the mass or escape velocity of clusters that do or do not display the extended MSTO feature (i.e. potential evidence for an age spread).

Hence, we conclude the scenario put forward by \citet{Goudfrooij11b}, that a minimum escape velocity is required for clusters to undergo extended secondary star formation events, is highly dependent on the adopted ``correction factors".  If standard ``correction factors" are adopted for the LMC/SMC, there is no relation between the escape velocity and the existence of age spreads or the extended MSTO feature.

\subsection{Comparison with other YMCs in the LMC}

Our findings are in line with previous studies of young massive LMC clusters. \citet{Milone13} ruled out a potential age spread present in NGC 1844 that shows broadened MS features. They derived an age of NGC 1844 of $\sim$150~Myr. \citet{Li13} proposed that NGC 1805 (44.7 Myr, 4.7$\cdot10^3\mathrm{M_{\sun}}$) and NGC 1818 (17.8 Myr, 2.6$\cdot10^4\mathrm{M_{\sun}}$) are simple stellar populations with no evident age spreads. These results are all in disagreement with the findings e.g. by \citet{Rubele13} and \citet{Goudfrooij11a,Goudfrooij11b} who explained the extended MSTO feature observed in many intermediate age LMC clusters with an age spread of 100 - 500 Myr inside the cluster. As a solution to this discrepancy we propose that rotating stars are (at least partially) responsible for the anomalous features observed in stellar clusters as it is already suggested by e.g. \citet{BastianDeMink09} and \citet{Yang13}. We will test this hypothesis in an upcoming paper [Niederhofer et al., in prep.] where we  will analyze observed CMDs of YMCs using the new population synthesis code (SYCLIST) of the Geneva group \citep{Georgy14} that includes all effects of stellar rotation.


\begin{acknowledgements} 
This research was supported by the DFG cluster of excellence "Origin and Structure of the Universe"\\
We thank the anonymous referee for useful comments that helped to improve the manuscript.

\end{acknowledgements}

\end{document}